\documentclass[a4paper,11pt]{article}
\pdfoutput=1 					

\usepackage{jinstpub} 

\usepackage{graphicx,subfig,caption}

\usepackage[all]{hypcap}

\usepackage{siunitx}

\usepackage{booktabs}

\title{\boldmath The novel properties of SF$_6$ for directional dark matter experiments}

\author[a,1]{N. S. Phan,\note{Corresponding author.}}
\author[a]{R. Lafler,}
\author[a]{R. J. Lauer,}
\author[a]{E. R. Lee,}
\author[a]{D. Loomba,}
\author[a]{J. A. J. Matthews,}
\author[a]{and E. H. Miller}

\affiliation[a]{Department of Physics and Astronomy, University of New Mexico, NM 87131, USA}

\emailAdd{nphan@unm.edu}

\abstract{SF$_{6}$ is an inert and electronegative gas that has a long history of use in high voltage insulation and numerous other industrial applications.  Although SF$_{6}$ is used as a trace component to introduce stability in tracking chambers, its highly electronegative properties have limited its use in tracking detectors.  In this work we present a series of measurements with SF$_{6}$ as the primary gas in a low pressure Time Projection Chamber (TPC), with a thick GEM used as the avalanche and readout device. The first results of an $^{55}$Fe energy spectrum in SF$_{6}$ are presented.  Measurements of the mobility and longitudinal diffusion confirm the negative ion drift of SF$_{6}$.  However, the observed waveforms have a peculiar but interesting structure that indicates multiple drift species and a dependence on the reduced field ($E/p$), as well as on the level of water vapor contamination. The discovery of a distinct secondary peak in the waveform, together with its identification and use for fiducializing events in the TPC, are also presented.  Our measurements demonstrate that SF$_{6}$ is an ideal gas for directional dark matter detection.  In particular, the high fluorine content is desirable for spin-dependent sensitivity, negative ion drift ensures low diffusion over large drift distances, and the multiple species of charge carriers allow for full detector fiducialization.}


\begin{document}
\maketitle
\flushbottom


\section{Introduction}
\label{sec:intro}

Sulfur hexafluoride (SF$_6$) is an inert, odorless, and colorless gas commonly known as an electron scavenger because of its large electron attachment cross-section \cite{Crompton1983, Smith1984, Petrovic1985, Miller1994, Datskos1993, Spanel1995, Braun2005, Viggiano2007, Merlino2008}.  The high electron affinity coupled with its non-toxicity and non-flammability make it suitable for use in many practical applications, including as a gaseous dielectric insulator in high voltage power devices, plasma etching of silicon and Ga-As based semiconductors, thermal and sound insulation, magnesium casting, and aluminum recycling (Refs.~\cite{LGC1997, LGC2000} provide an extensive review of the properties and applications of SF$_6$).  In particle detectors, SF$_6$ has been used as a quencher in Resistive Plate Chambers (RPCs) operated in both avalanche and streamer modes, enabling more stable operation by suppressing streamer formation in the former, and reducing the energy of discharges and allowing lower voltage operation in the latter \cite{camarri1998, aielli2002}.  As a result of its many diverse commercial and research applications, SF$_6$ is one of the most extensively studied gases \cite{LGC1997}. 
 
Nevertheless, with the exception of RPCs, studies of SF$_6$ in conditions applicable to particle physics detectors are scarce. Although SF$_6$ was considered as a negative ion gas in rare searches \cite{nygrenBB}, the high electron affinity was deemed a barrier for stripping the electron from the negative ion in the avalanche region, a necessary first step for initiating gas gain amplification.  However, with the advent of Micro-patterned Gas Detectors (MPGDs), which have flexible geometries that can sustain high electric fields in the avalanche region even at low pressures, the potential for achieving gas gain in SF$_6$ has been realized, as shown in this work.  Demonstrating this for low energy event detection has opened up the possibility for its use in a variety of experiments, such as directional dark matter searches.  Our work provides the first experimental evidence that SF$_6$ is in fact an excellent choice as a negative ion gas for TPC-based directional dark matter experiments.

Directional searches in TPCs require low pressures to lengthen recoil tracks, and low diffusion so they can be resolved, both of which are ideally suited to negative ion gases.  The idea of negative ion drift with carbon disulfide (CS$_2$) was first proposed by Martoff to circumvent the use of magnetic fields to achieve low diffusion in large TPCs \cite{martoff2000}.  Negative ion TPCs were first successfully demonstrated with CS$_2$-based gas mixtures by DRIFT, a directional dark matter experiment \cite{DRIFT2007,drift2014}.  At present DRIFT employs a mixture of 30:10:1 Torr CS$_2$:CF$_4$:O$_2$, which leverages the benefits of negative-ion CS$_2$ with the spin content of fluorine, an ideal target for spin-dependent (SD) interactions with WIMPs\footnote{For SD dark matter searches neither $^{12}$C or $^{32}$S atoms have the nuclear spin content to be suitable detection targets, whereas $^{19}$F is excellent in this regard \cite{ellis}.}, and the capability to fiducialize the detector provided by O$_2$ \cite{SI2014}.
 This multi-component DRIFT gas mixture was tailored for directional DM searches where low diffusion, low backgrounds and the SD limit-setting capabilities are all essential. 
 
As demonstrated in this work, SF$_6$ has all of the benefits of the DRIFT gas mixture, along with additional advantages that make it more amenable to the underground environment.  We begin by discussing the motivation behind, and benefits of each component of the CS$_2$/CF$_4$/O$_2$ gas mix for directional dark matter experiments, and how these are matched by SF$_6$.  

In a detector with an electronegative gas, like CS$_2$, the free electrons produced by an ionization event are quickly captured, forming anions that drift in the thermal regime to the amplification and readout region.  In this regime, diffusion scales as $\sqrt{L/E}$, where $L$ is the drift distance and $E$ is the strength of the drift field, making it desirable to have high fields to minimize diffusion.  With this, good tracking resolution can be achieved over long drift distances, which are two necessary conditions for the high quality track reconstruction and large detection volumes required for directional dark matter and other rare event searches.  Like CS$_2$, which has an electron affinity of 0.55 eV \cite{cavanagh}, SF$_6$ is highly electronegative with electron affinity of 1.06 eV \cite{christophorou2001}.\footnote{The values quoted for SF$_6$ were recommended by Ref.~\cite{christophorou2001} based on results from Ref.~\cite{grimsrud} and Ref.~\cite{chen}, and the value for CS$_2$ is the most precise to date.  Note however that, similar to SF$_6$, the experimentally determined electron affinities of CS$_2$ have a large spread, ranging from $\sim 0.5 - 1.0$ eV \cite{NIST}.}  Thus, SF$_6$ should also behave like a negative ion gas, with similar drift properties to CS$_2$.

An additional advantage of electronegative gases is that they tend to display superior high voltage performance at low pressures over electron drift gases, such as CF$_4$ and N$_2$.  SF$_6$ is especially well suited in this regard, having a breakdown field strength that is about three times higher than air \cite{LGC1988} and N$_2$ \cite{LGC1982, LGC19841} at pressures below one atmosphere.

The CF$_4$ in the DRIFT gas mixture, as mentioned above, provides the fluorine target for SD WIMP interactions.  In this regard, with its high fluorine content, SF$_{6}$ has a clear advantage over CS$_2$/CF$_4$ mixtures for SD searches.  Thus, if the potential of SF$_6$ as a negative ion gas are borne out, there would be no need to sacrifice precious detection volume to the non spin-dependent CS$_2$, leading to a significant increase in the sensitivity to dark matter.

The motivation for O$_2$ in the DRIFT gas mixture came from the recent discovery that the combination CS$_2$/O$_2$ produces features in the signal waveform that allow event fiducialization \cite{SI2014}.  This enabled the ability to reject backgrounds from detector surfaces, a critical advance for gas TPCs used in rare searches.  With this, DRIFT demonstrated a $\sim$50 day, zero background limit that is currently the world's best for a directional experiment \cite{drift2014}.  We show in Section~\ref{sec:SF6properties} that the signal waveform in SF$_{6}$ also contains similar features that can be used for fiducialization (Section~\ref{sec:fiducial}). 
 
There are a number of other advantages of SF$_{6}$ over CS$_2$/CF$_4$O$_2$ mixtures.  One is the ability to purify via recirculation, which has not been demonstrated to satisfaction with any CS$_2$ mixture but should be straightforward with SF$_{6}$.  This would lower backgrounds and also lower costs and the manpower needed for transporting gas underground.  With respect to safe underground operations, SF$_6$ is non-toxic and non-flammable, whereas CS$_2$ is highly toxic and, with the addition of O$_2$, flammable and potentially explosive \cite{Wood1971}.  CS$_2$ also has a tendency to be absorbed into detector surfaces making operation and maintenance arduous.  Finally, SF$_6$ has an extremely high vapor pressure of 15,751 Torr at room temperature, compared to about 300 Torr for CS$_2$.

In order to realize the very appealing prospects of SF$_{6}$, the key features we need to demonstrate in this work are:

\begin{enumerate}

	\item Gas amplification with efficient stripping of the electron from SF$_6^{-}$ in the gain stage.

	\item  Gas gain and its dependance on pressure.  For example, if good gas gain can be achieved at high pressure, it would have implications for double-beta decay searches with SeF$_6$ (selenium hexafluoride), which has a similar molecular structure to SF$_{6}$ \cite{nygrenParisTPC}.
	
	\item  Low thermal diffusion in SF$_6$, as expected from a negative ion gas, and how it compares to CS$_2$.
	
	\item Features in the signal waveforms that could be used to fiducialize events along the TPC's drift direction.
	
\end{enumerate}


\section{Experimental apparatus and method}
\label{sec:expsetup}

\begin{figure*}[]
\centering
\subfloat[Acrylic cylindrical detector]{
	\includegraphics[width=0.6\textwidth]{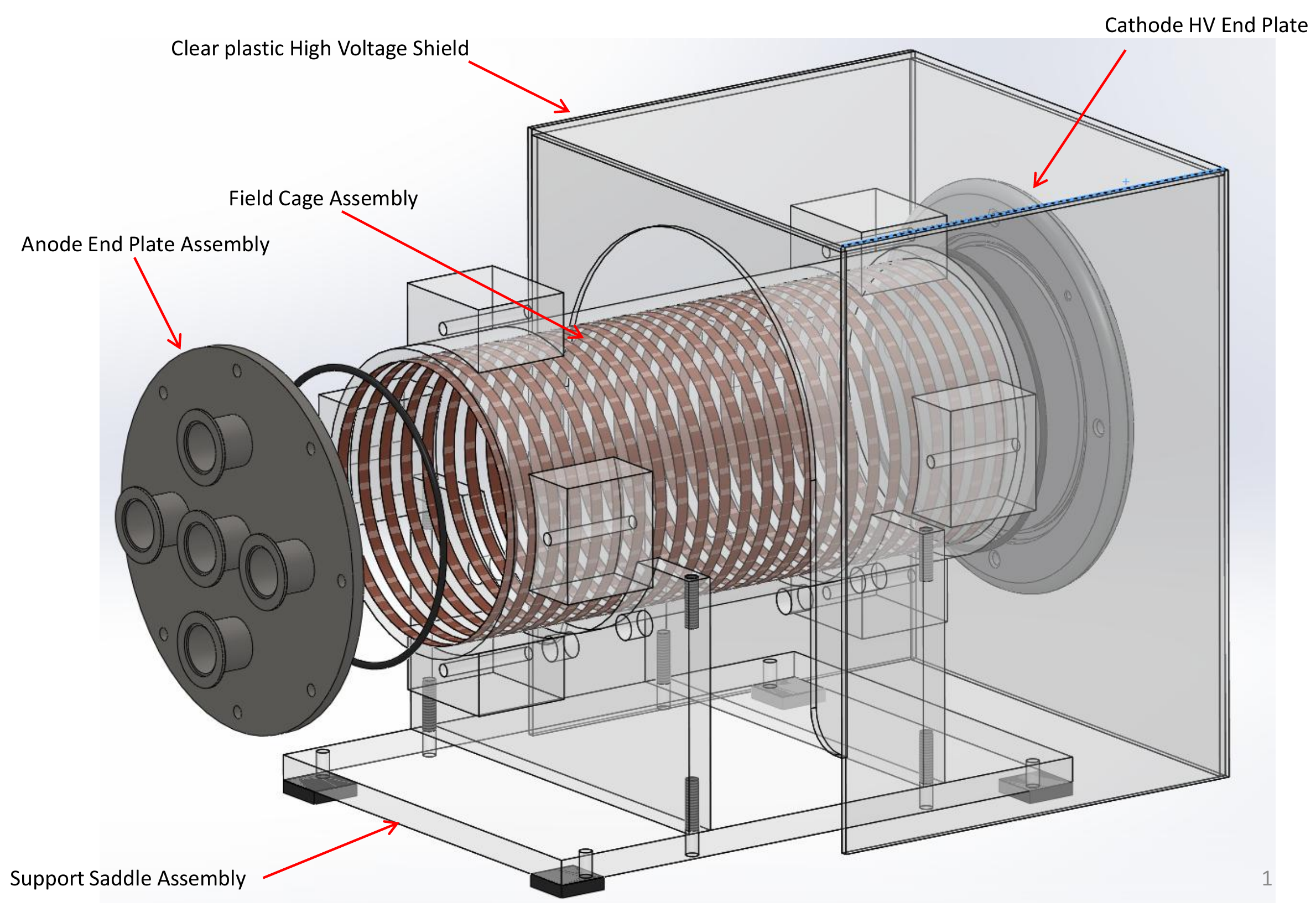}
\label{fig:expsetup}}
\qquad
\subfloat[Inner view of anode end plate]{
	\includegraphics[width=0.55\textwidth]{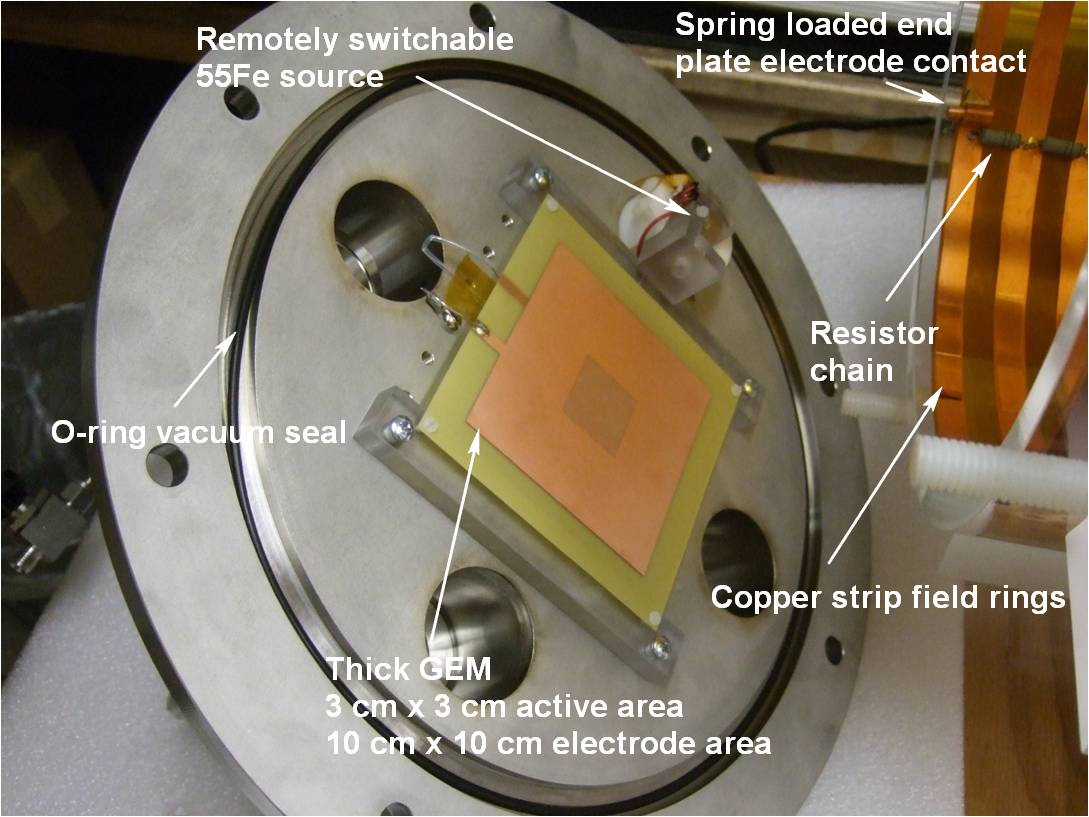}
\label{fig:vesselinnerlid}}
\caption{ (a) A schematic of the detector showing its primary cylindrical acrylic body, field cage, aluminum end plates, support saddle, and high voltage shield.  The laser (not shown) sits near the anode plate and fires pulses through a quartz window onto the cathode to create photoelectrons at a known location.  (b) A photograph of the inner side of the anode plate which shows the O-ring, switchable $^{55}$Fe source, and THGEM.}
\label{fig:vesselsetup}
\end{figure*}

\subsection{Acrylic detector}
\label{sec:detector}

The TPC detector used to make measurements for this work (Figure~\ref{fig:vesselsetup}) consisted of a 60 cm long acrylic cylinder with an inner diameter of 30.5 cm.  The two ends of the detector were made from aluminum plates, one serving as the cathode that could be powered up to a maximum voltage of ${-60}$ kV, and the other as the grounded anode.  The acrylic TPC with its aluminum end-caps also served as the vacuum vessel.  The field rings were made from a kapton PC flex board with 1.3 cm wide copper strips placed at a pitch of 2.54 cm and connected to 23 (56 M$\Omega$) resistors.  Gas amplification was provided by a single 0.4 mm thick GEM (THGEM) that was custom fabricated at CERN with an active area of $3 \times 3$ cm$^2$.  The THGEM had a hole pitch of $\sim$0.5 mm and hole diameter of $\sim$0.3 mm, with an annular region of thickness 0.05 mm etched around each hole to eliminate burring from the drilling process.  The THGEM was mounted on two acrylic bars attached to the anode plate.  The surface of the THGEM facing the cathode was grounded to the anode plate while the other surface was held at high voltage ($610-1020$ V).  Signals were read out from the high voltage surface with an ORTEC 142 charge sensitive preamplifier, which had a 20 ns rise-time (at zero capacitance) and a 100 \si{\micro\second} decay time constant.  

\subsection{Charge generation}
\label{sec:chargegeneration}

Ionization was introduced into the gas volume either with an internally mounted and remotely switchable $^{55}$Fe 5.9 keV X-ray source (Figure~\ref{fig:vesselinnerlid}), or by a system using a Stanford Research Systems (SRS) NL100 337.1 nm pulsed nitrogen laser, which was used to produce photoelectrons by illuminating the aluminum cathode.  The NL100 laser had a FWHM pulse width of 3.5 ns, a pulse energy of 170 mJ, and a peak power of 45 kW.  The spot size in the longitudinal, or drift, dimension was essentially a delta function, whereas the projected spot size in the X and Y (lateral) dimensions was a $1 \times 3$ mm$^2$ rectangle.  Measurements of transverse diffusion require an instrumented XY readout, which is the subject of future work.

\subsection{Operation and data acquisition}
\label{sec:opDAQ}

After the vacuum vessel was sealed, a long pump-down with an Edwards XDS10 dry scroll vacuum pump (base pressure $<$ 0.1 Torr) was conducted to minimize out-gassing from the acrylic cylinder and other components inside the detector.  The vessel was then back-filled with approximately 200 Torr of SF$_{6}$ gas (99.999\% purity), and flushed.  This was done to dilute any residual out-gassed contaminants that the vacuum pump was not able to remove.  The vessel was once again back-filled with gas to approximately 200 Torr and slowly pumped down to the final operating pressure, with a precision of 0.05 Torr.  During this slow pump down, both the cathode and GEM were ramped up to operating voltages.  This procedure assured a minimum time between the introduction of fresh gas into the detector and the start of data acquisition. 

As the various measurements of SF$_{6}$ properties were performed as a function of the operating pressure and drift field, these were changed between each setting.  This was done by raising the pressure back up to 200 Torr and, as before, slowly pumping down to the new pressure setting while concurrently setting the new cathode voltage.  This procedure was repeated between each set of measurements, and its importance will be explained in Section~\ref{sec:watervapor} where the presence and effects of water vapor are discussed.  

Although the focus of this paper is on SF$_{6}$, for comparative purposes we also present measurements of CS$_{2}$ properties made using the same setup.  For this gas, the operating procedure was different than the one used for the SF$_6$.  After the long pump down, the detector was back-filled to the operating pressure and all sets of measurements were taken without a pressure raise and pump down between each setting.  When the cathode was brought to full operating voltage, a spark-down period of $30-60$ minutes allowed micro-sparks due to the acrylic charging-up to subside before powering up the THGEM to full voltage.

All measurement waveforms were acquired with a Tektronix TDS 3054C digital oscilloscope and National Instruments data acquisition software, where every triggered event was read out and saved to file for analysis.  The saved files contain the voltage signals from the ORTEC charge sensitive preamplifier, which integrated the charge collected by the THGEM readout surface with a rise time of $\sim$100 ns, and an exponential decay time constant of $\tau = 100$ \si{\micro\second}.  The current, $I(t)$, entering the preamplifier is related to the detected voltage signal, $V(t)$, by 
\begin{equation}\label{eq:It}
I(t) \propto \frac{dV}{dt}- \left( -\frac{V}{\tau}\right),
\end{equation}
where the second term is for removing the decay tail.  We used Equation~\ref{eq:It} to compute $I(t)$ from our measurements of $V(t)$.  After the conversion, pulses were smoothed with a Gaussian filter to suppress high frequency noise and to improve signal to noise.  We then extracted the drift speed, diffusion, and other quantities from these processed waveforms.


\section{SF$_6$ waveforms}
\label{sec:SF6properties}

\subsection{Capture and transport in SF$_6$}
\label{sec:transport}

Measurements made under differing conditions have shown that electron capture by the electronegative SF$_{6}$ molecule occurs rapidly \cite{Crompton1983, Smith1984, Petrovic1985, Miller1994, Datskos1993, Spanel1995, Braun2005, Viggiano2007, Merlino2008} with the immediate product being SF$_{6}^{-*}$, a metastable excited state of the anion, SF$_{6}^{-}$.  The latter forms subsequently from the collisional or radiative stabilization of the excited state \cite{LGC2000}.  The electron capture cross-sections by SF$_{6}$ are very large \cite{Crompton1983, Smith1984, Petrovic1985, Miller1994, Datskos1993, Spanel1995, Braun2005, Viggiano2007, Merlino2008} and estimates of the capture mean-free-path are about a micron at the pressures and drift fields of our experiments.  This assumes that the electrons produced by the laser illumination of the cathode have near zero kinetic energies, where the capture cross-sections peak.  The metastable SF$_{6}^{-*}$ leads to subsequent products besides SF$_{6}^{-}$, whose relative abundances depend on the lifetime of SF$_{6}^{-*}$, the electron energy, gas pressure, temperature, and drift field:
\begin{equation} \label{eq:attachment}
\mathrm{SF_6 + e^- \rightarrow SF_6^{-*}   } \qquad	\text{	(attachment, metastable) }
\end{equation}

\begin{equation} \label{eq:autodetachment}
\mathrm{SF_6^{-*} \rightarrow SF_6 + e^- 		}	\qquad	\text{	(auto-detachment) }
\end{equation}

\begin{equation} \label{eq:stabilization}
\mathrm{ SF_6^{-*} + SF_6 \rightarrow SF_6^- + SF_6 	}		\qquad	\textrm{(collisional stabilization)  }
\end{equation}

\begin{equation} \label{eq:autodissociation}
\mathrm{ SF_6^{-*} \rightarrow SF_5^- + F 		}		\qquad\textrm{(auto-dissociation)   } 
\end{equation}

Thus, after the quick electron capture leading to SF$_{6}^{-*}$, the auto-detachment reaction (\ref{eq:autodetachment}) will compete with collisional stabilization, reaction (\ref{eq:stabilization}), and auto-dissociation, reaction (\ref{eq:autodissociation}).  To determine whether auto-detachment plays a significant role in our experiment, which could lead to a significant distortion of the waveform, we consider bounds on the lifetimes of these reactions.

Measurements of lifetimes for auto-detachment have a broad range, from $\sim$10 \si{\micro\second} to one ms, depending on the experimental technique used.  Under collision-free conditions, time-of-flight (TOF) mass spectrometric experiments indicate the lifetime is between $10 - 68$ \si{\micro\second} \cite{Edelson, Compton, Harland, LGC1971, LGC1978, LGC1984}.  Measurements made with ion cyclotron resonance (ICR) experiments, however, give lifetimes in the ms range \cite {Henis, Odom, Foster}.  The difference in measured lifetimes between the two techniques reflect different electron energies, with those in ICR experiments typically much lower than in TOF experiments \cite{LGC2000}, and closer to the energies in our experiment.

The lifetime for collisional stabilization (\ref{eq:stabilization}) depends on the cross-section and collision rate.  The former is large, and the latter can be estimated by considering the collision mean-free path, $\lambda$, for SF$_{6}^{-*}$ in SF$_{6}$. Assuming that this is similar to that of SF$_{6}^{-}$ in SF$_{6}$, we can use:
\begin{equation} \label{eq:lambda}
\lambda = \frac{\left(3MkT\right)^{1/2}v_d}{eE}
\end{equation}
\cite{McDaniel}, where $T= 296$ K, $M$ is the mass of the SF$_6$ molecule, $v_d$ is the drift speed, and $E$ is the drift field. Using our measured drift speeds (see Section~\ref{sec:mobility}) we estimate $\lambda\sim 0.1-1$ \si{\micro\meter}, implying a collisional mean-free time of $\sim 1-10$ ns.  This is many orders of magnitude less than the lifetimes for auto-detachment, indicating that the latter process should be inconsequential in our experiment.  This is confirmed by our waveforms shown in Section~\ref{sec:waveforms}.
 
Besides reactions (\ref{eq:stabilization}) and (\ref{eq:autodissociation}), which lead to the production of SF$_6^-$ and SF$_5^-$, other processes occurring at either the site of initial ionization or during drift to the anode can lead to additional negative ion species.  For example, the metastable SF$_6^{-*}$ produced initially can also lead to F$^-$ and SF$_4^-$ (e.g., via auto-dissociation \cite{LGC2000}), although at much lower probabilities; reactions producing these species have much lower production cross-sections and require much higher electron energies than those for SF$_6^-$ and SF$_5^-$ \cite{Lehmann, Rao, Kline, LGC2001}.  Therefore, in our experiment we expect the initial charge carriers to be dominated by SF$_5^-$ and SF$_6^-$, with their relative contributions estimated from production cross-sections.

The cross-section for reaction (\ref{eq:stabilization}) is peaked at zero electron energy \cite{LGC2001, Hotop, Klar, Ling}, falling by a factor of about $100$ at 0.1 eV \cite{Kline, LGC2001, Hunter}, whereas that for reaction (\ref{eq:autodissociation}) has a peak at 0 eV \cite{Hunter} and a smaller one at $\sim$0.38 eV \cite{Kline,  LGC2001, Hunter}.  At 0 eV, the SF$_{6}^-$ cross-section is larger by a factor $1000$ than that for SF$_{5}^-$, but only a factor $\sim$30 at 0.1 eV because the SF$_{6}^-$ cross-section falls much more rapidly with energy than that of SF$_{5}^-$.  For the low electron energies expected in our experiments, however, SF$_{6}^-$ should be the dominant charge carrier arriving at the anode.  Because of the higher mobility of SF$_{5}^-$ (\cite{Patterson, Urquijo, Fleming}, and see Section~\ref{sec:mobility} below) we should detect two peaks in the signal waveform, with the faster SF$_{5}^-$ arriving earlier in time.  This is the basis for fiducialization, and is discussed in detail in Section~\ref{sec:fiducial}.  

A number of possible reactions involving the drifting SF$_{5}^-$ and SF$_{6}^-$ with the neutral gas could, however, complicate this simple picture.  At low drift fields, neutral, electron-hungry SF$_{6}$ molecules will form clusters around the negative ions \cite{Patterson}.  Clusters of SF$_{6}^{-}$(SF$_{6}$)$_{n}$ and SF$_{5}^{-}$(SF$_{6}$)$_{n}$ ($n = 1, 2, 3, ...$) have been observed but with mobilities less than those of SF$_{5}^{-}$ and SF$_{6}^{-}$ \cite{Patterson}.  This phenomena could therefore partly explain the long tail observed on the slow side of the SF$_{6}^{-}$ peak in our low reduced field waveforms (Figure~\ref{fig:avgwaveform40Torr5kV}).  

In addition to clustering, the drifting SF$_{5}^{-}$ and SF$_{6}^{-}$ could also interact with neutral SF$_{6}$ molecules or contaminants in the gas, leading to other species (see Section~\ref{sec:watervapor}).  These could appear as distinct features in our measured waveforms.  More important for us is the collisional detachment of energetically stable SF$_{5}^{-}$ and SF$_{6}^{-}$ via the following reactions:
\begin{equation} \label{eq:SF5coll}
\mathrm{ SF_5^{-} + SF_6 \rightarrow SF_5 + SF_6  + e^{-}	}		\qquad	\textrm{(collisional detachment)  }
\end{equation}

\begin{equation} \label{eq:SF6coll}
\mathrm{ SF_6^{-} + SF_6 \rightarrow SF_6 + SF_6  + e^{-}}		\qquad\textrm{(collisional detachment).   } 
\end{equation}
Such processes would be followed by re-attachment via reaction (\ref{eq:attachment}), and the subsequent reactions (\ref{eq:stabilization}) and (\ref{eq:autodissociation}) that lead back to SF$_{5}^{-}$ or SF$_{6}^{-}$.  The attachment/detachment of the electron could result in a smeared waveform due to the different drift speeds of the charge carriers.  However, the probability of detachment via reactions (\ref{eq:SF5coll}) and (\ref{eq:SF6coll}) is very small for center-of-mass energies $<$ 60 eV \cite{Wang}.  In comparison to the electron affinity of SF$_{5}$ ($2.7-3.7$ eV) \cite{Christodoulides} and SF$_{6}$ (1.06 eV), the threshold energy for detachment is much larger and is attributed to competing charge-transfer and collision-induced dissociation processes \cite{Wang, Olthoff, Champion}.  Nevertheless, there is evidence that energetically unstable states of SF$_6^{-}$ (i.e. SF$_6^{-*}$) can contribute to collisional detachment \cite{Wang, Olthoff}.  The relative contributions of these effects depend on the interaction energies at different reduced fields, but the detailed mechanisms is well beyond the scope of this work.

\begin{figure*}[]
\centering
\subfloat[$E= 172$ V$\cdot$cm$^{-1}$]{
	\includegraphics[width=0.32\textwidth]{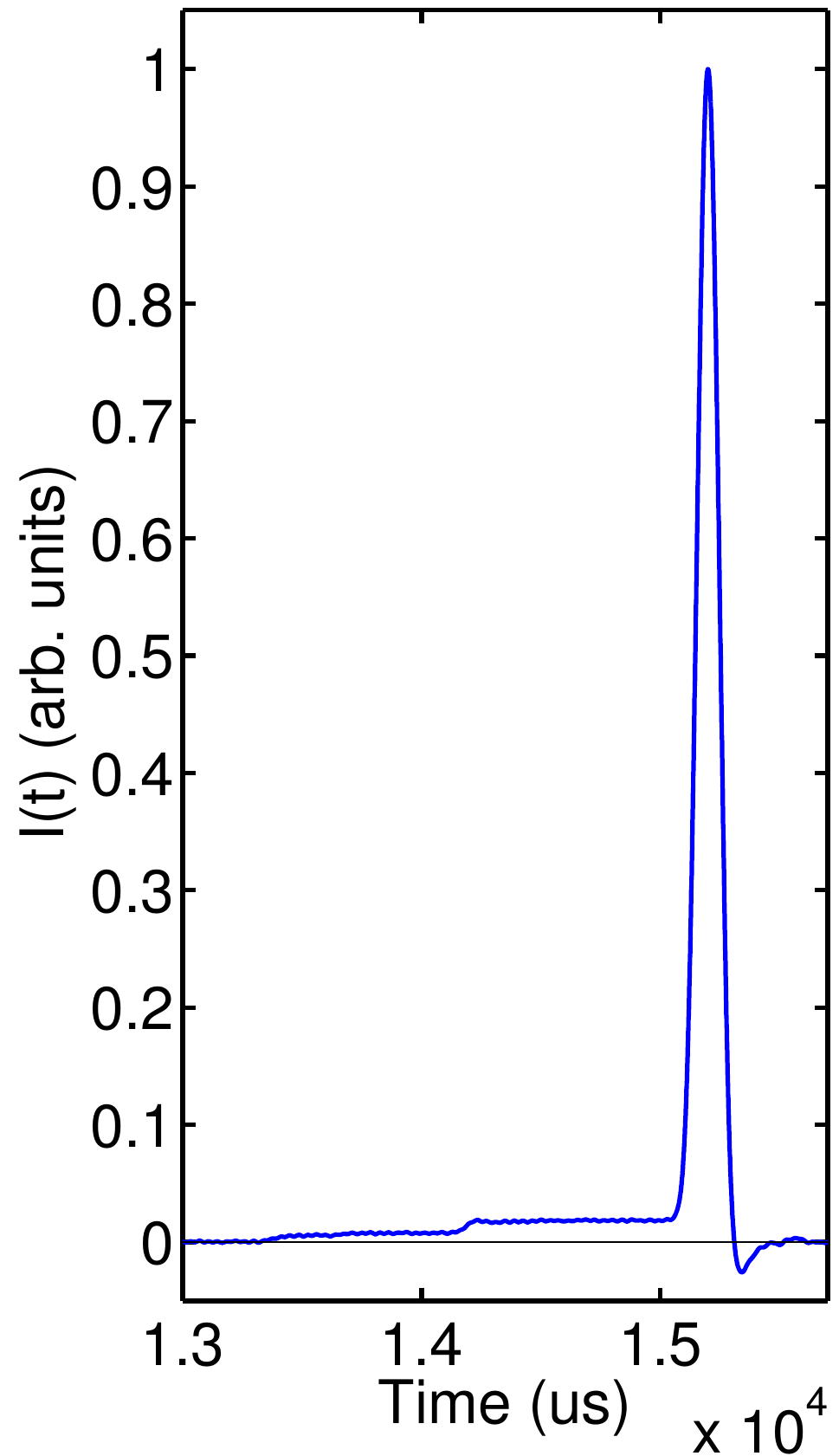}
\label{fig:10kV}}
\subfloat[$E= 343$ V$\cdot$cm$^{-1}$]{
	\includegraphics[width=0.32\textwidth]{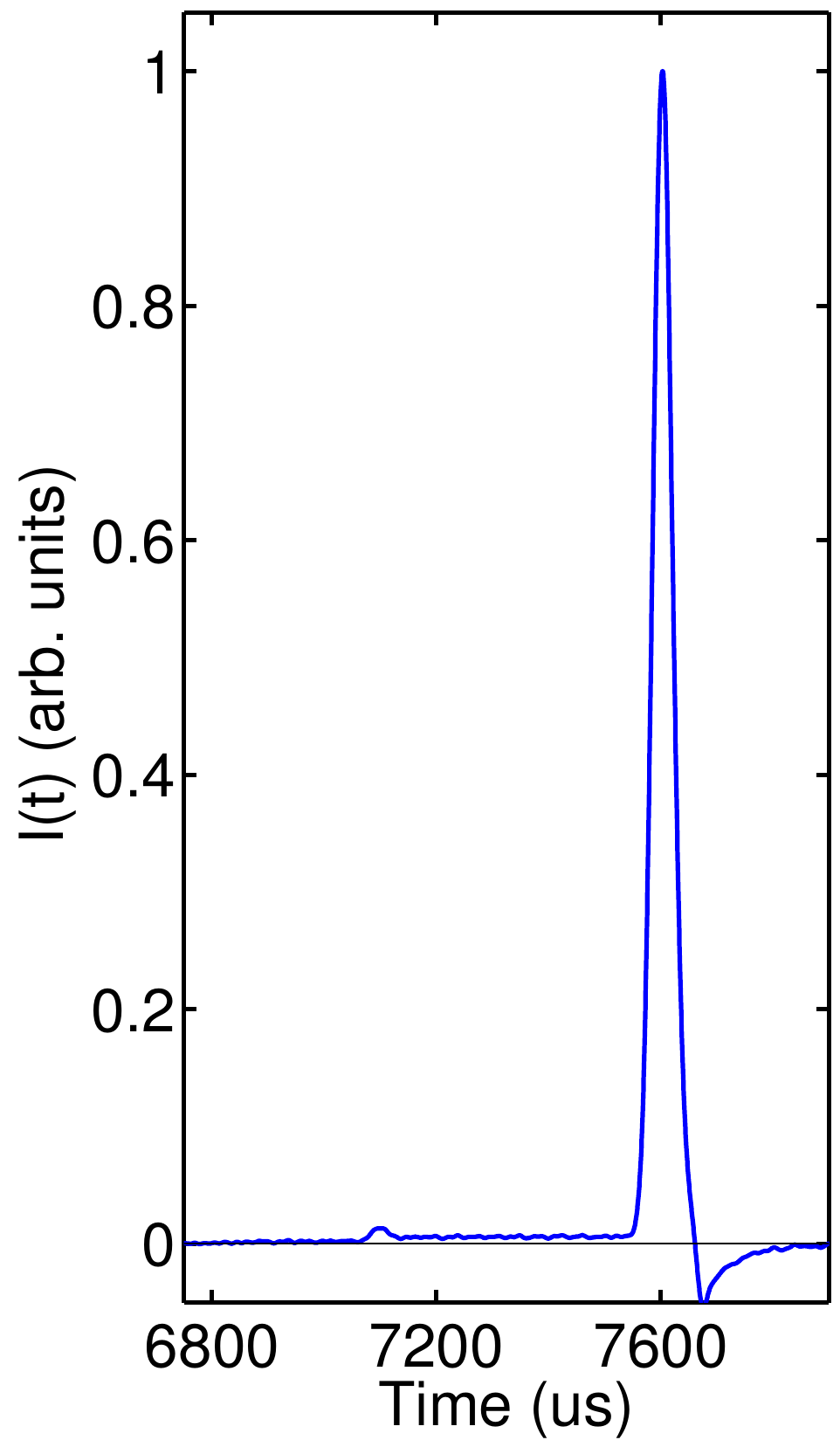}
\label{fig:20kV}}
\subfloat[$E= 515$ V$\cdot$cm$^{-1}$]{
	\includegraphics[width=0.32\textwidth]{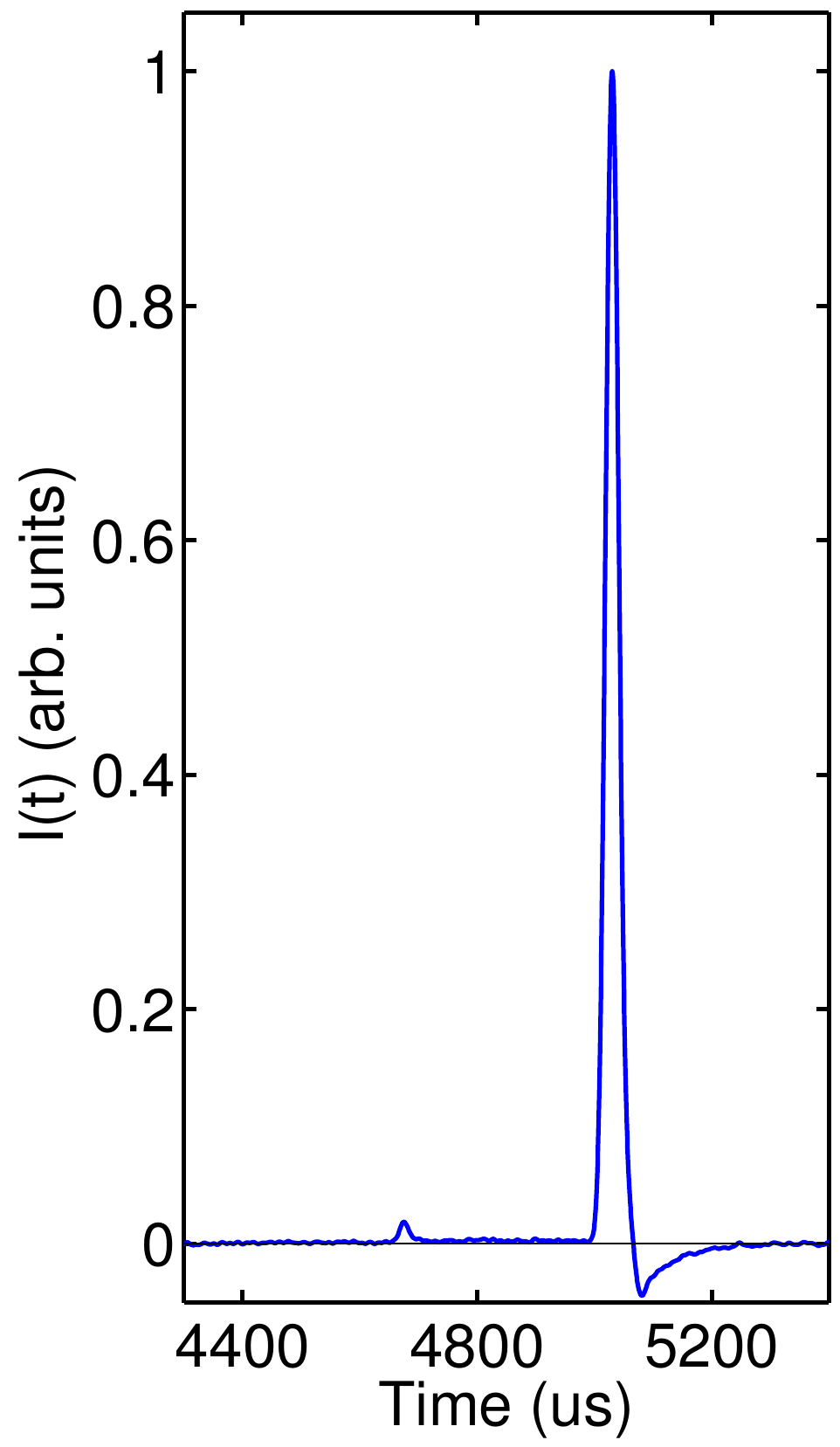}
\label{fig:30kV}}
\quad
\subfloat[$E= 686$ V$\cdot$cm$^{-1}$]{
	\includegraphics[width=0.32\textwidth]{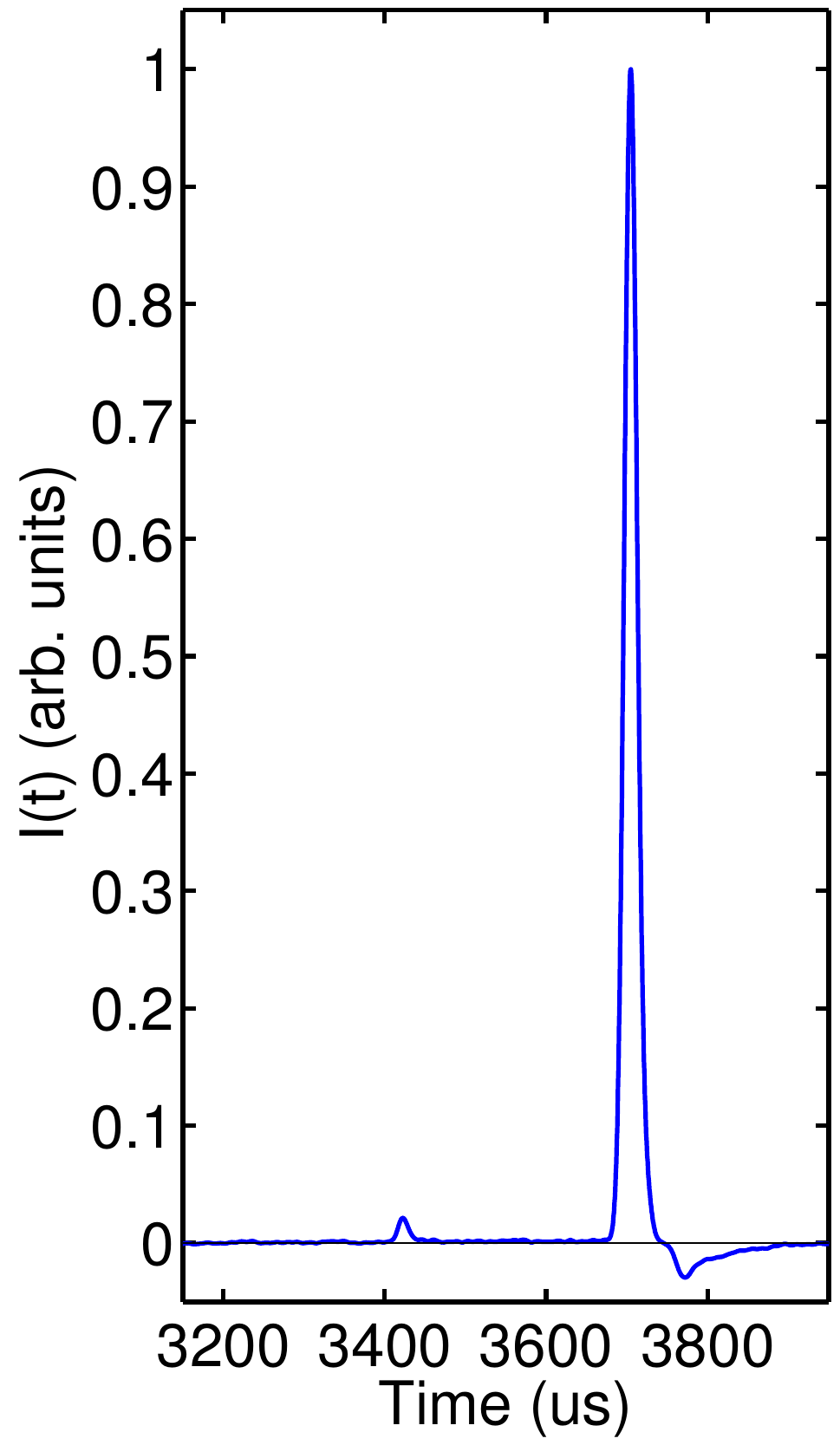}
\label{fig:40kV}}
\subfloat[$E= 858$ V$\cdot$cm$^{-1}$]{
	\includegraphics[width=0.32\textwidth]{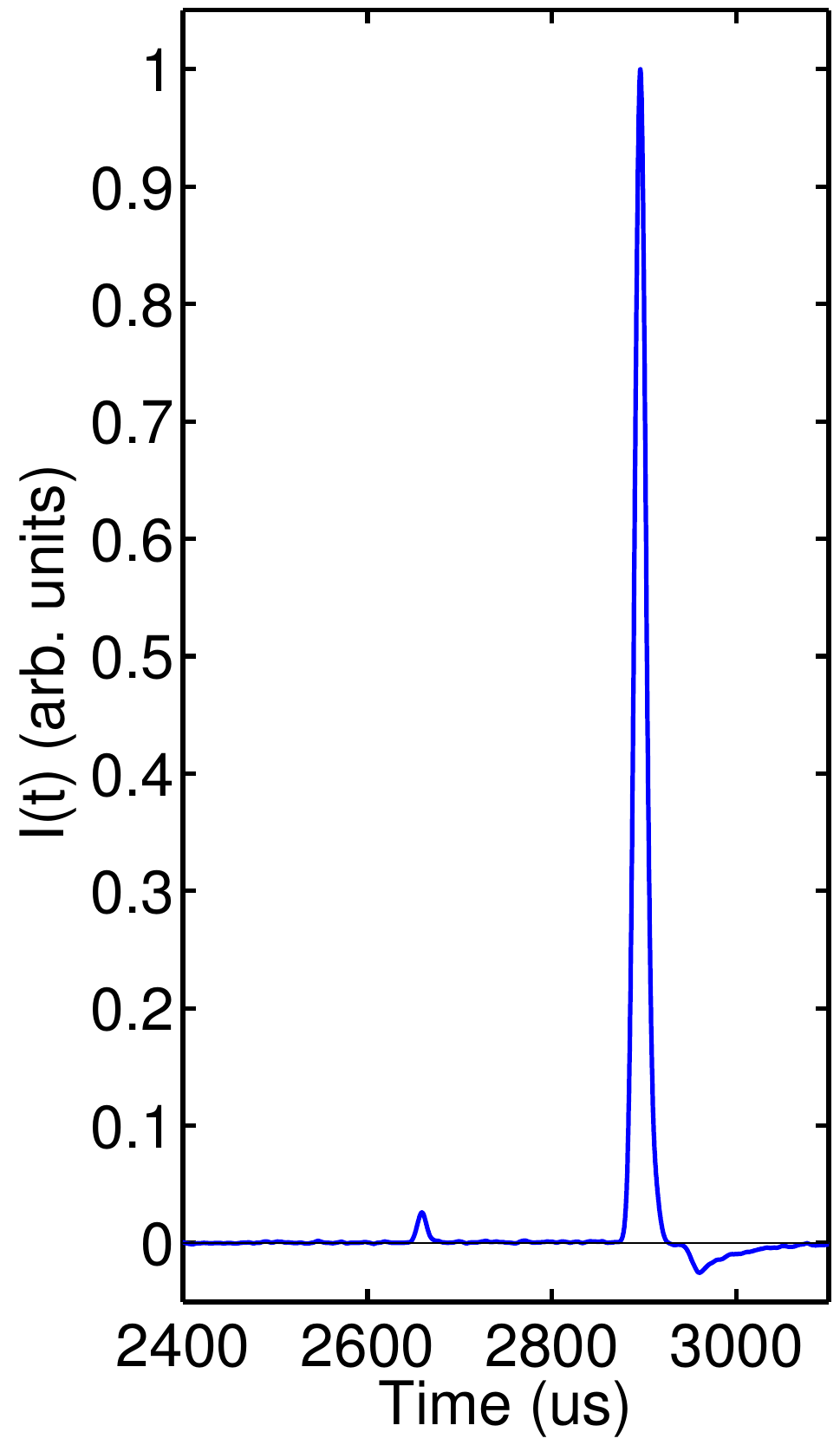}
\label{fig:50kV}}
\subfloat[$E= 1029$ V$\cdot$cm$^{-1}$]{
	\includegraphics[width=0.32\textwidth]{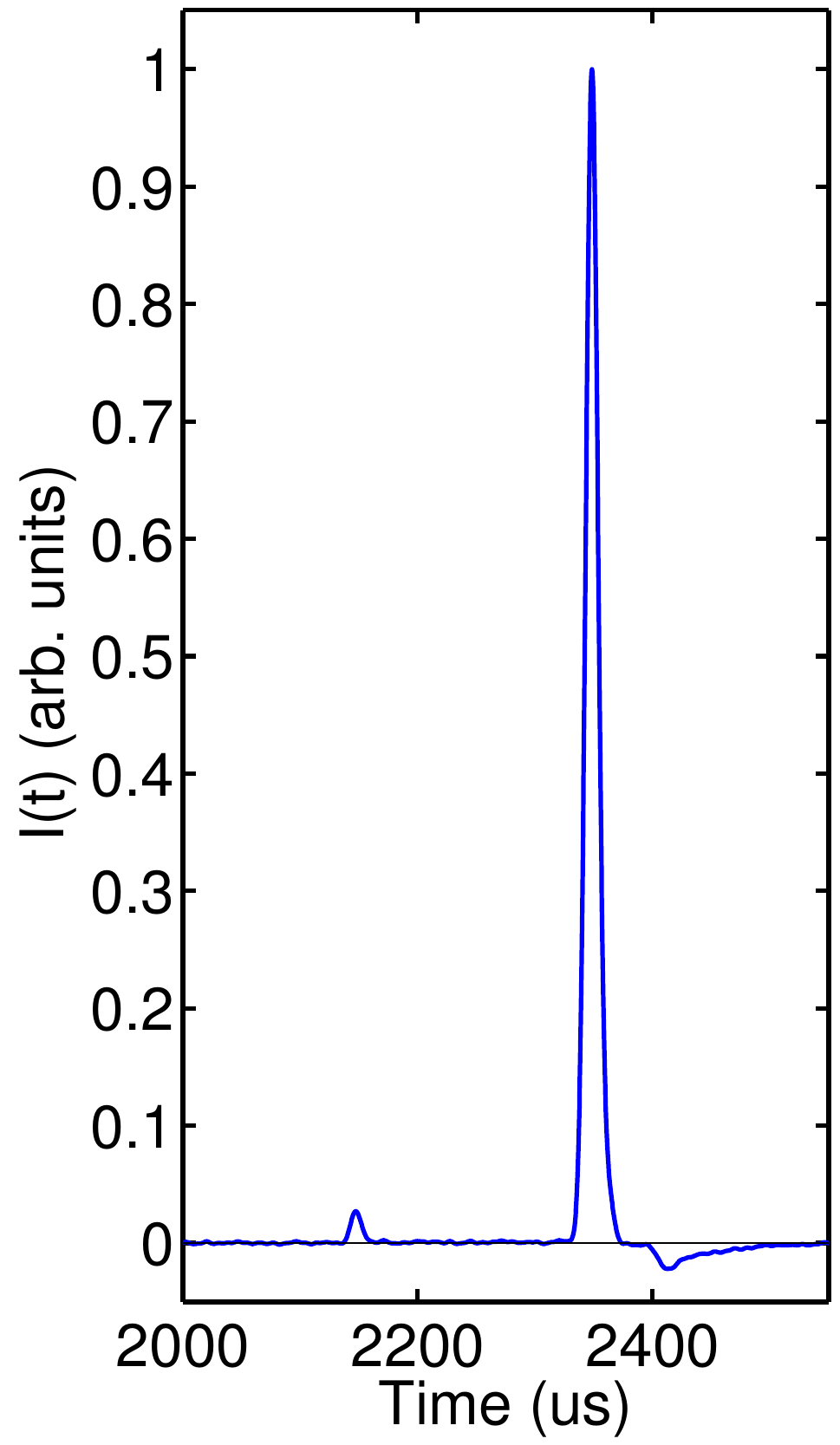}
\label{fig:60kV}}
\caption{(a) - (f) The average waveforms acquired in 20 Torr SF$_6$ at six different electric field strengths.  At low fields (a), there is an additional broad structure in addition to the two peaks.  This component appears to decrease in magnitude with increasing electric field and appears to vanish at the highest field (f).}
\label{fig:avgwaveforms}
\end{figure*}

\begin{figure*}[]
\centering
\subfloat[$E= 172$ V$\cdot$cm$^{-1}$]{
	\includegraphics[width=0.315\textwidth]{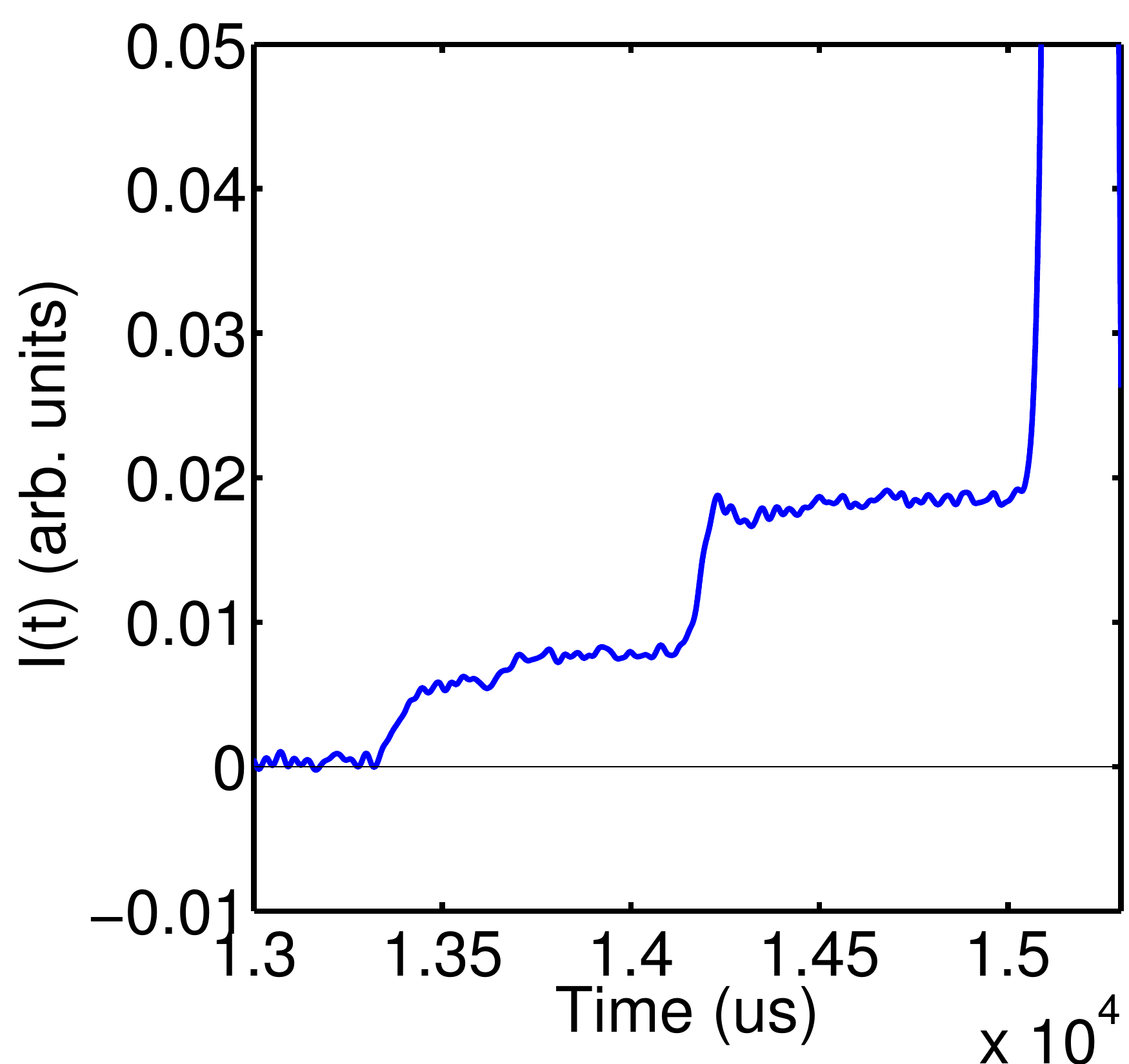}
\label{fig:10kVzoom}}
\subfloat[$E= 343$ V$\cdot$cm$^{-1}$]{
	\includegraphics[width=0.315\textwidth]{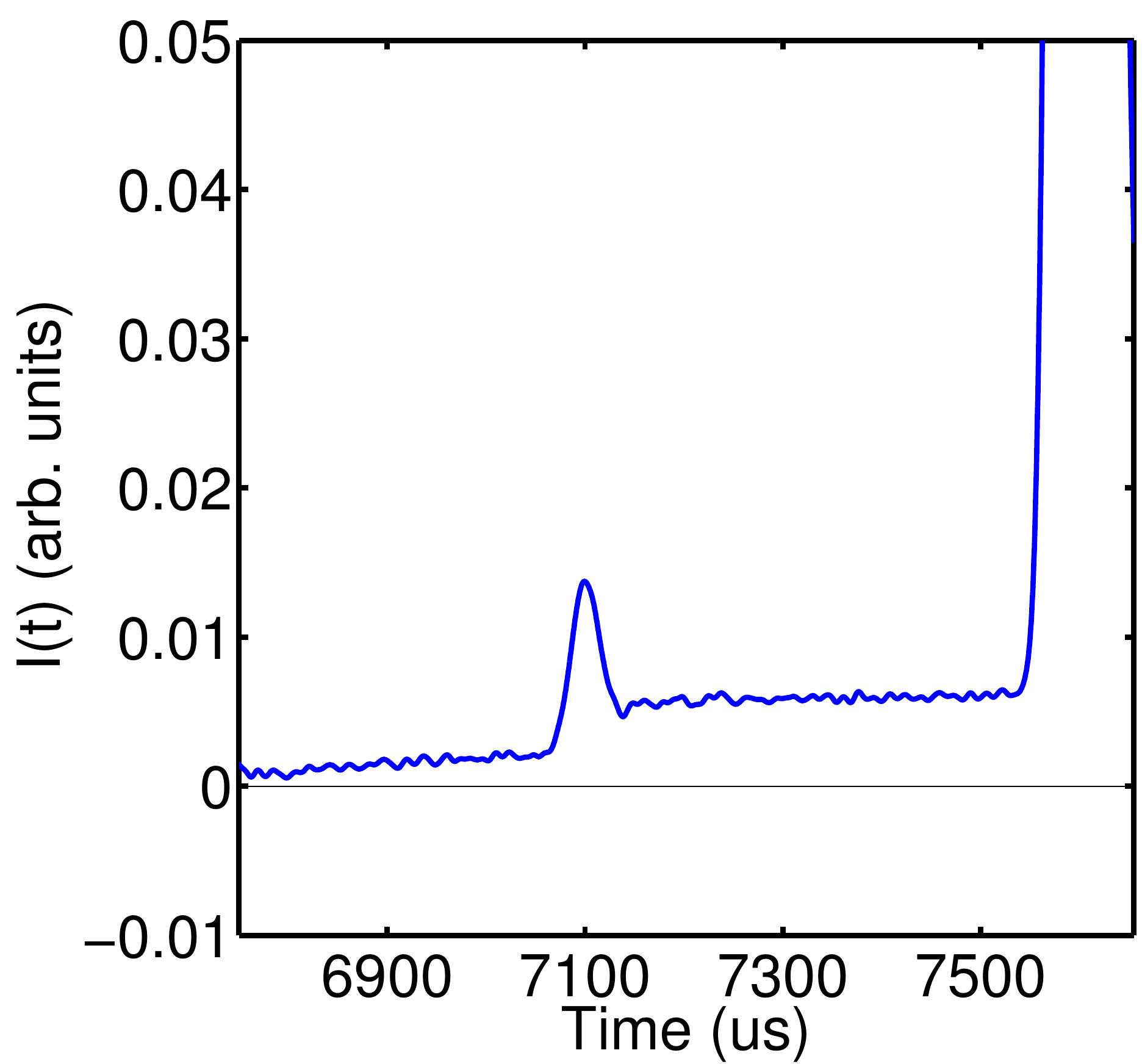}
\label{fig:20kVzoom}}
\subfloat[$E= 515$ V$\cdot$cm$^{-1}$]{
	\includegraphics[width=0.315\textwidth]{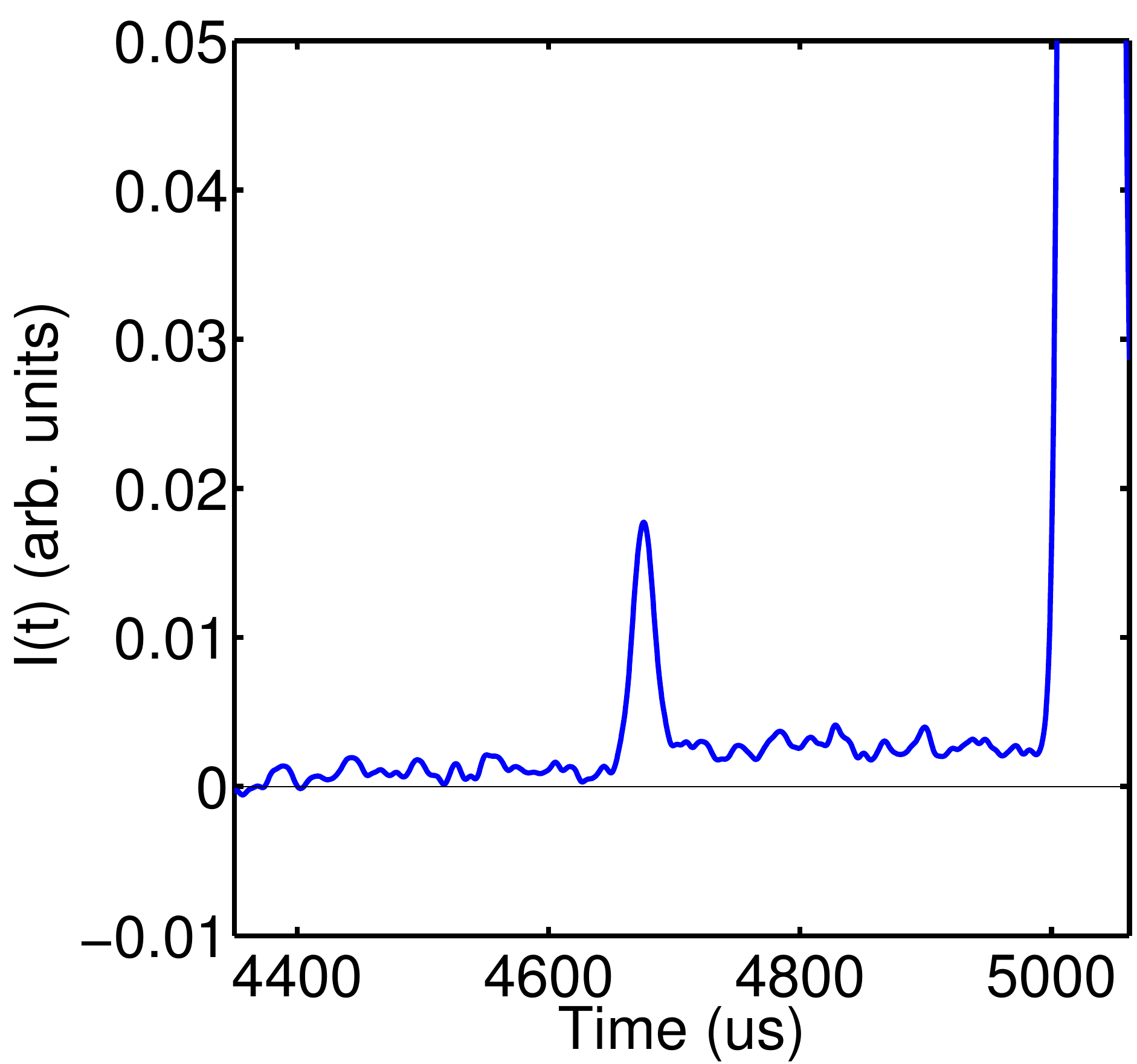}
\label{fig:30kVzoom}}
\quad
\subfloat[$E= 686$ V$\cdot$cm$^{-1}$]{
	\includegraphics[width=0.315\textwidth]{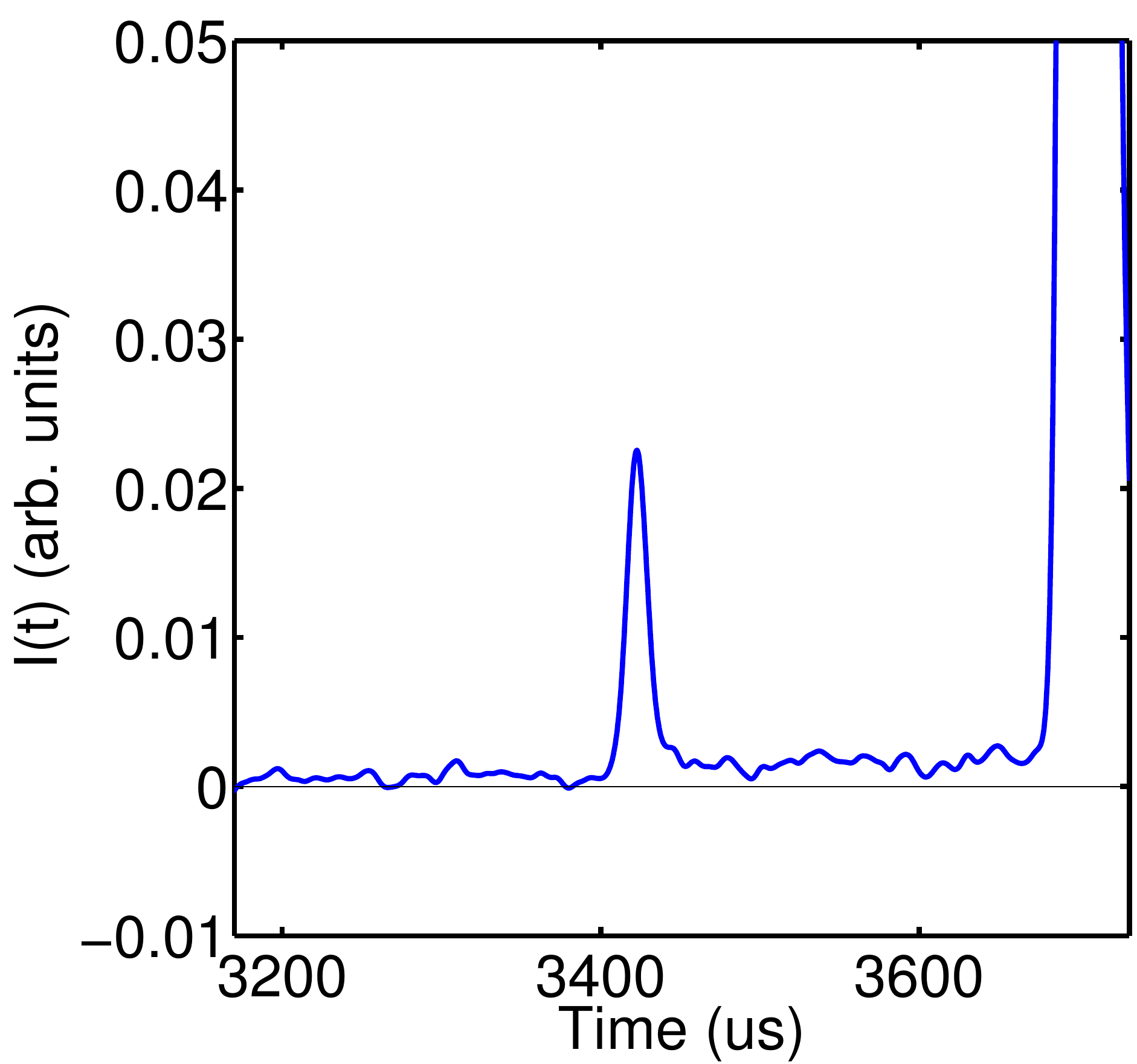}
\label{fig:40kVzoom}}
\subfloat[$E= 858$ V$\cdot$cm$^{-1}$]{
	\includegraphics[width=0.315\textwidth]{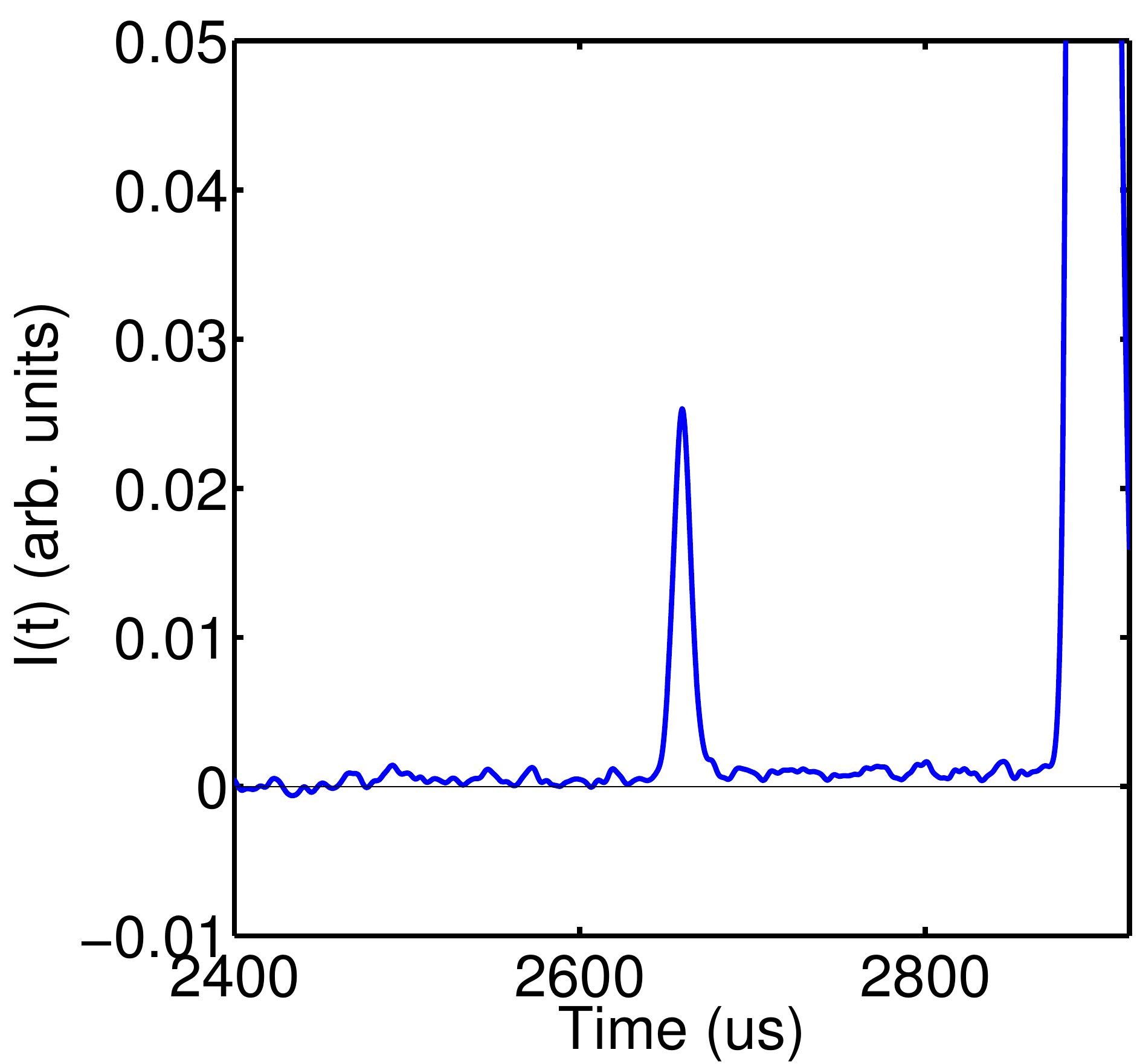}
\label{fig:50kVzoom}}
\subfloat[$E= 1029$ V$\cdot$cm$^{-1}$]{
	\includegraphics[width=0.315\textwidth]{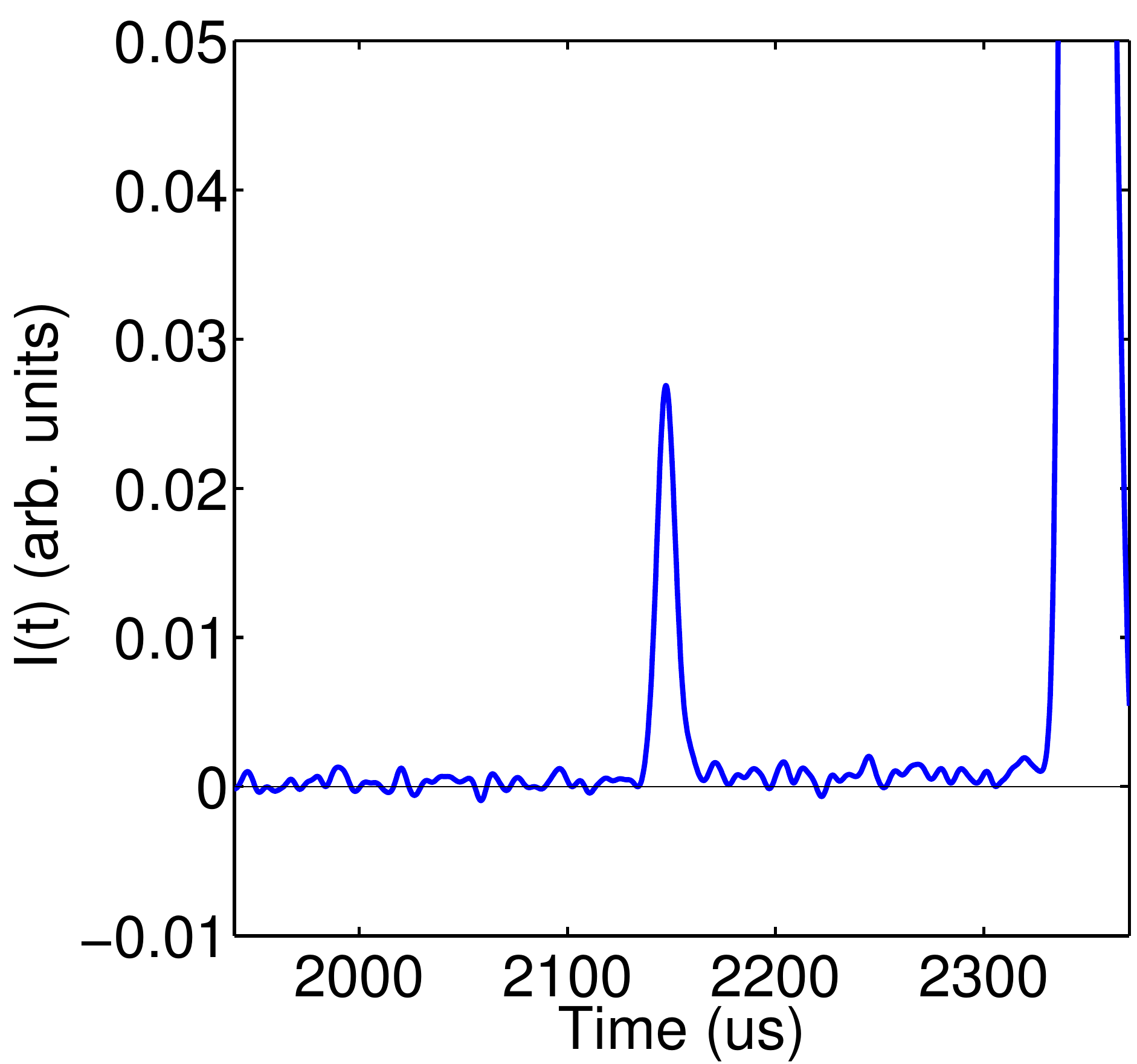}
\label{fig:60kVzoom}}
\caption{(a)-(f) The zoomed in views of the waveforms from Figure~\ref{fig:avgwaveforms}.  Charge outside of the peaks appears to decrease with increasing field strength while the SF$_{5}^{-}$ peak begins to emerge and grow in amplitude. }
\label{fig:avgwaveformszoom}
\end{figure*}

\subsection{Waveform features}
\label{sec:waveforms}

With an overview of the chemistry of electron drift and attachment in SF$_{6}$, we now turn to a detailed look at our data.  
Shown in Figures~\ref{fig:avgwaveforms} and \ref{fig:avgwaveformszoom} are the averaged current waveforms acquired in 20 Torr SF$_6$ ($N=6.522\times 10^{17}$ cm$^{-3}$ at $T=296$ K) for six different drift field strengths and gas gains up to a few 1000s (see Section \ref{sec:gasgain} for details on gas gain).  The averaging was done using one thousand individual waveforms, each acquired by illuminating the cathode with the nitrogen laser.  The laser also provided the initial trigger for the DAQ system. 

At low fields, the waveform consists of two peaks, one much smaller than the other, and a low amplitude broad component distributed outside the region of the two peaks.  The large main peak is SF$_{6}^{-}$ and the smaller secondary peak arriving earlier is SF$_{5}^{-}$.  The non-peak component does not appear continuous but displays a step in amplitude at the location of the smaller peak, and a second step to the baseline at an earlier time.  With increasing field strength, this non-peak component gradually subsides until it is barely discernible at $E = 1029$ V$\cdot$cm$^{-1}$ (Figure~\ref{fig:60kVzoom}) leaving just the two sharp peaks.  The origin of this component is water vapor contamination from out-gassing in the acrylic vessel, and is the subject of Section~\ref{sec:watervapor}.

\begin{figure*}[]
\centering
\subfloat[40 Torr, $E= 86$ V$\cdot$cm$^{-1}$]{
	\includegraphics[width=0.30\textwidth]{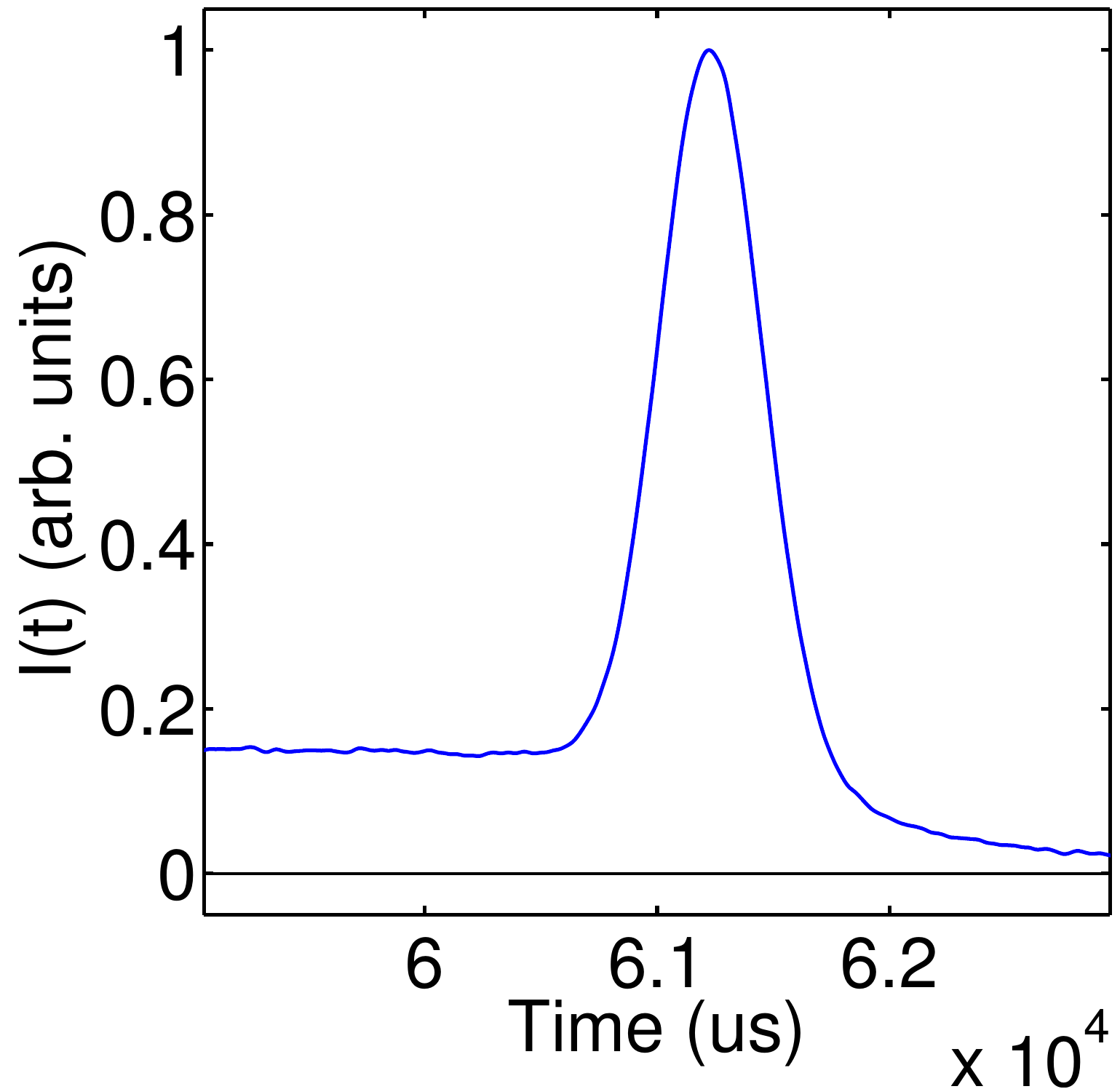}
\label{fig:avgwaveform40Torr5kV}}
\subfloat[30 Torr, $E= 86$ V$\cdot$cm$^{-1}$]{
	\includegraphics[width=0.30\textwidth]{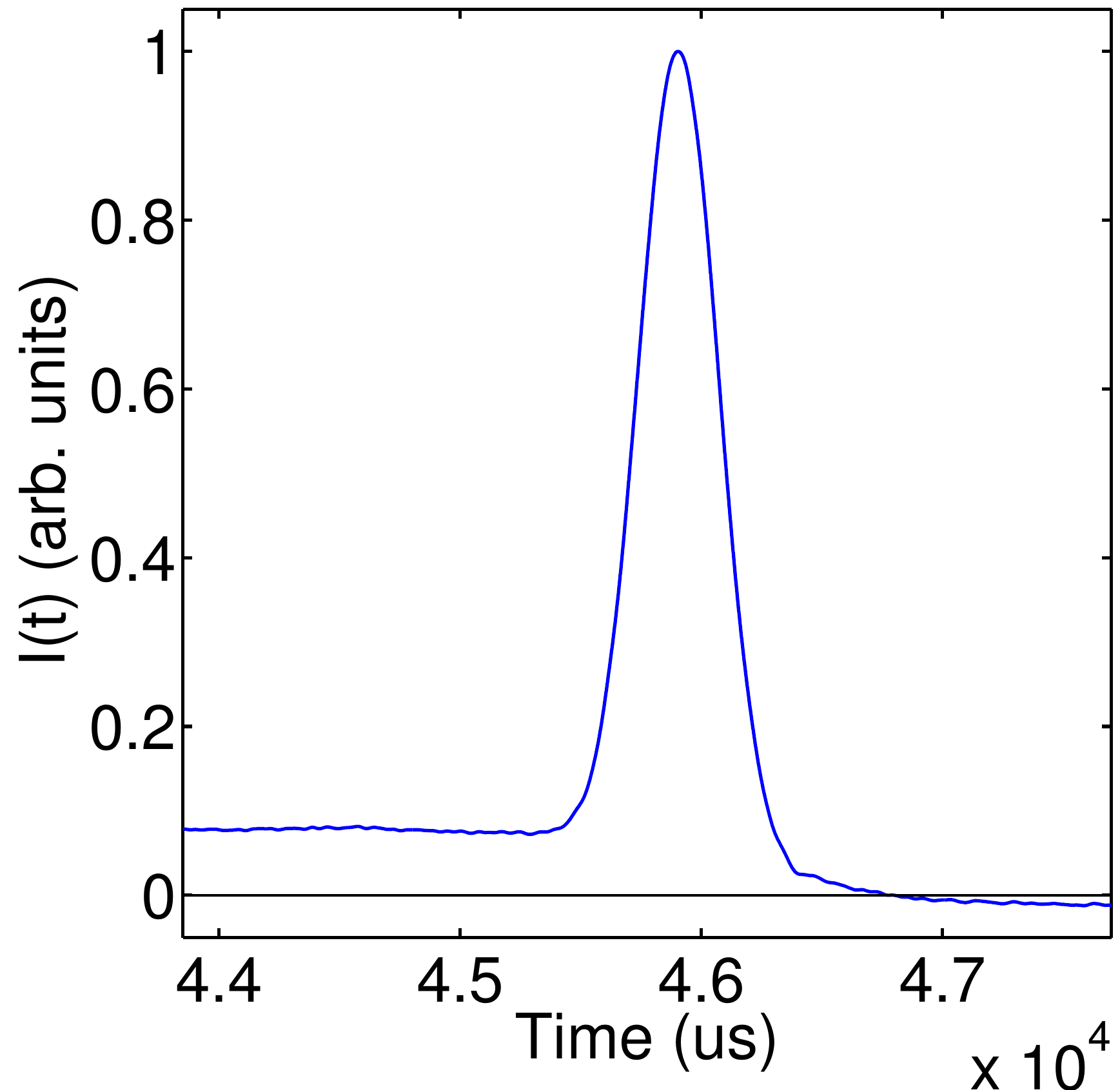}
\label{fig:avgwaverform30Torr5kV}}
\quad
\subfloat[20 Torr, $E= 86$ V$\cdot$cm$^{-1}$]{
	\includegraphics[width=0.30\textwidth]{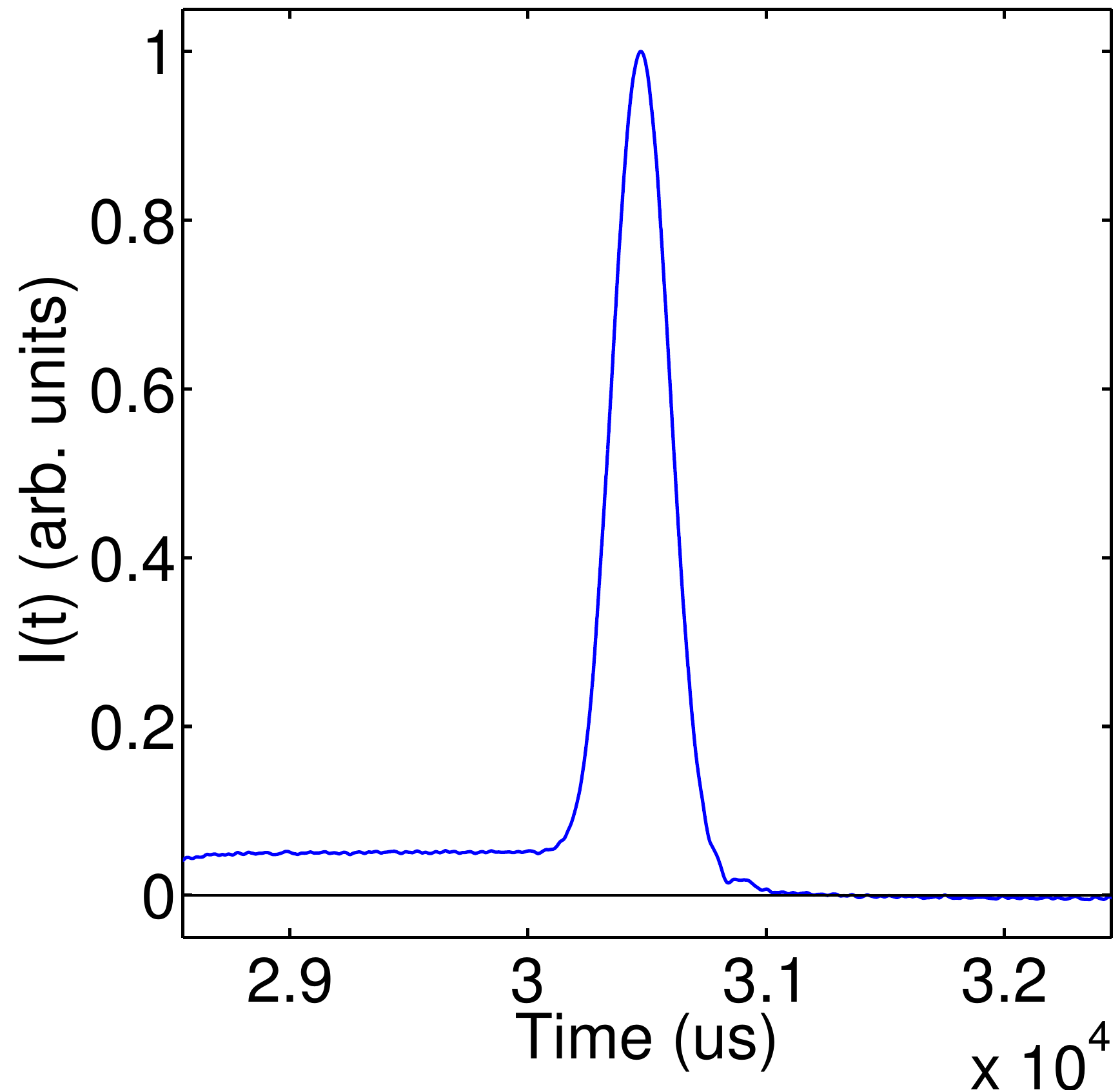}
\label{fig:avgwaveform20Torr5kV}}
\caption{(a)-(c) Average waveforms for 40, 30, and 20 Torr SF$_{6}$ at $E= 86$ V$\cdot$cm$^{-1}$.  Note the long tail on the right side of the peak in (a), which could be due to clustering at low reduced fields in SF$_{6}$.}
\label{fig:AvgWaveformsLowField}
\end{figure*}

The waveforms show a similar behavior as a function of inverse pressure, $1/p$.  Figure~\ref{fig:AvgWaveformsLowField} shows portions of waveforms taken at three pressures with a fixed drift field, $E = 86$ V$\cdot$cm$^{-1}$, the lowest used in our experiment.  The broad component decreases relative to the main SF$_{6}^{-}$ peak as the pressure is reduced, similar to what is observed with increasing drift field at fixed pressure.  This anti-correlation between the pressure and drift field would imply a reduced field ($E/p$ or $E/N$) dependence, but a detailed look at the data does not support this.  Comparing the waveforms in Figure~\ref{fig:30kVzoom} and Figure~\ref{fig:comp40Torr60kV} (blue curve), both at the same reduced field but different $E$ and $p$, we see clear differences in the amount of charge in the non-peak region (both waveforms are normalized with the SF$_6^-$ peak amplitude set to one).

Two other notable features seen on the right side of the SF$_{6}^{-}$ peak are the small negative amplitude dip and the long tail at low $E/p$.  As discussed in Section~\ref{sec:waveforms}, the latter could be due to SF$_{6}^{-}$(SF$_{6}$)$_{n}$ and SF$_{5}^{-}$(SF$_{6}$)$_{n}$ clusters that drift at a slower speed than the SF$_{6}^{-}$ anion.  The production and drift of SF$_{6}^{-}$(H$_{2}$O)$_{n}$ clusters, which is discussed in Section~\ref{sec:watervapor}, could also contribute to this tail.  But at higher reduced fields, the formation of such weakly bound clusters should be suppressed, which is supported by our higher $E/p$ data (Figures~\ref{fig:avgwaverform30Torr5kV} and \ref{fig:avgwaveform20Torr5kV}).  The second feature, the negative amplitude dip, is due to how the THGEM surfaces were electrically connected.  The surface facing the cathode was grounded to the aluminum anode end-cap, while the other readout surface is at positive high voltage.  As a result, the motion of the positive ions in the avalanche away from the readout induces a small positive signal, then the negative dip occurs as they approach the ground, which is capacitively coupled to the readout surface.

\subsection{Water vapor contamination}
\label{sec:watervapor}

\begin{figure*}[]
\centering
\subfloat[40 Torr, $E= 515$ V$\cdot$cm$^{-1}$]{
	\includegraphics[width=0.315\textwidth]{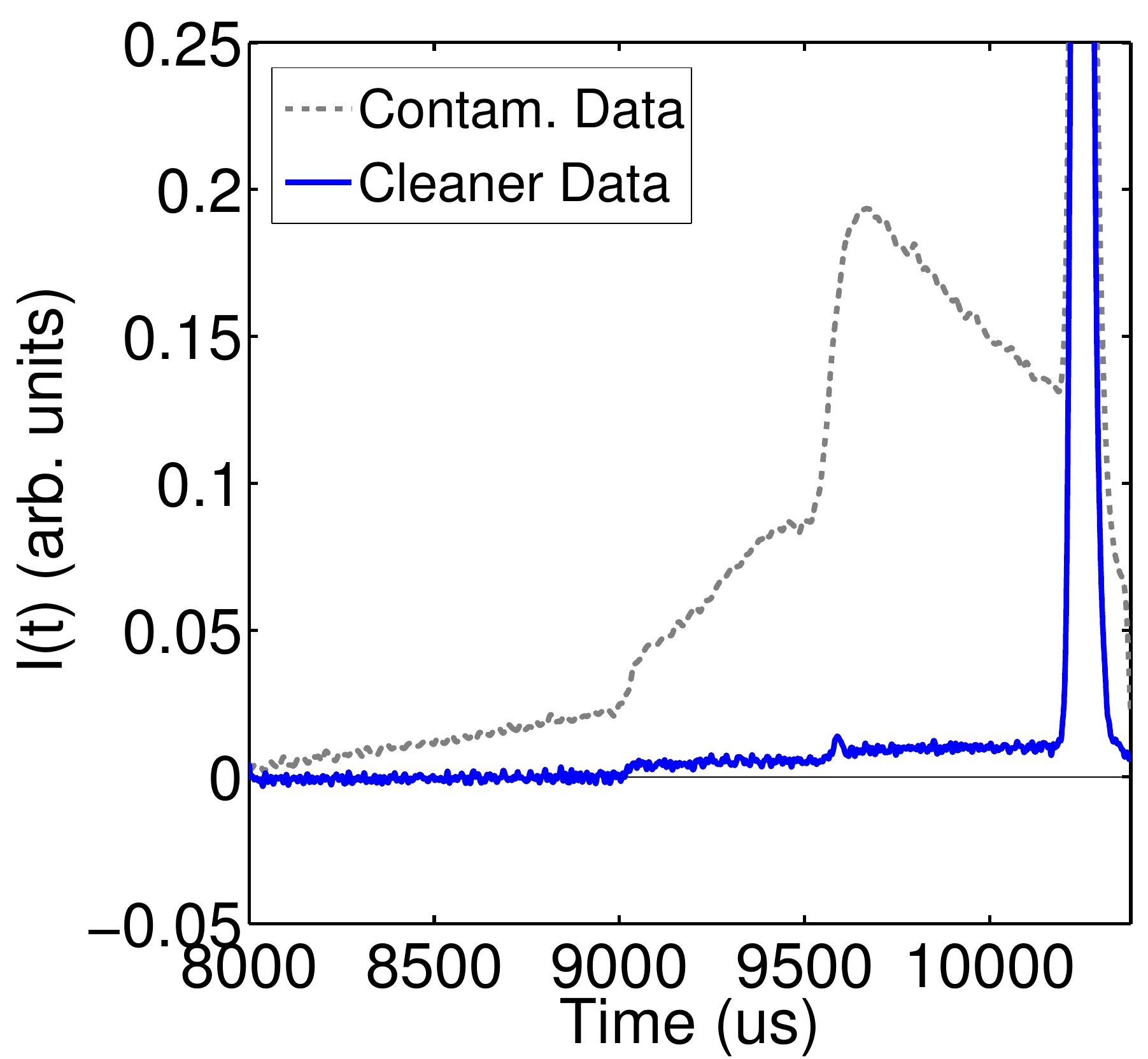}
\label{fig:comp40Torr30kV}}
\subfloat[40 Torr, $E= 1029$ V$\cdot$cm$^{-1}$]{
	\includegraphics[width=0.315\textwidth]{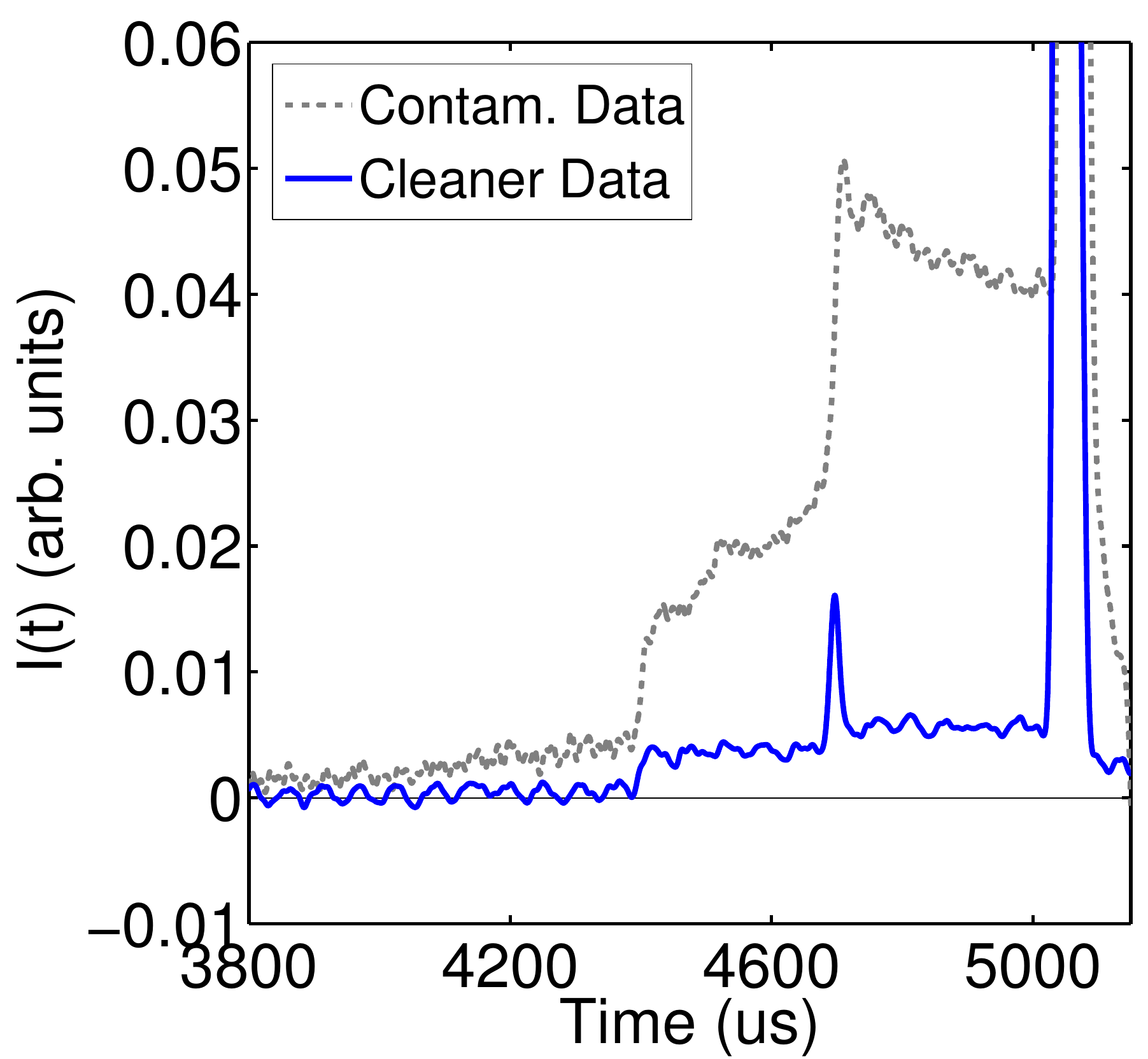}
\label{fig:comp40Torr60kV}}
\subfloat[20 Torr, $E= 1029$ V$\cdot$cm$^{-1}$]{
	\includegraphics[width=0.315\textwidth]{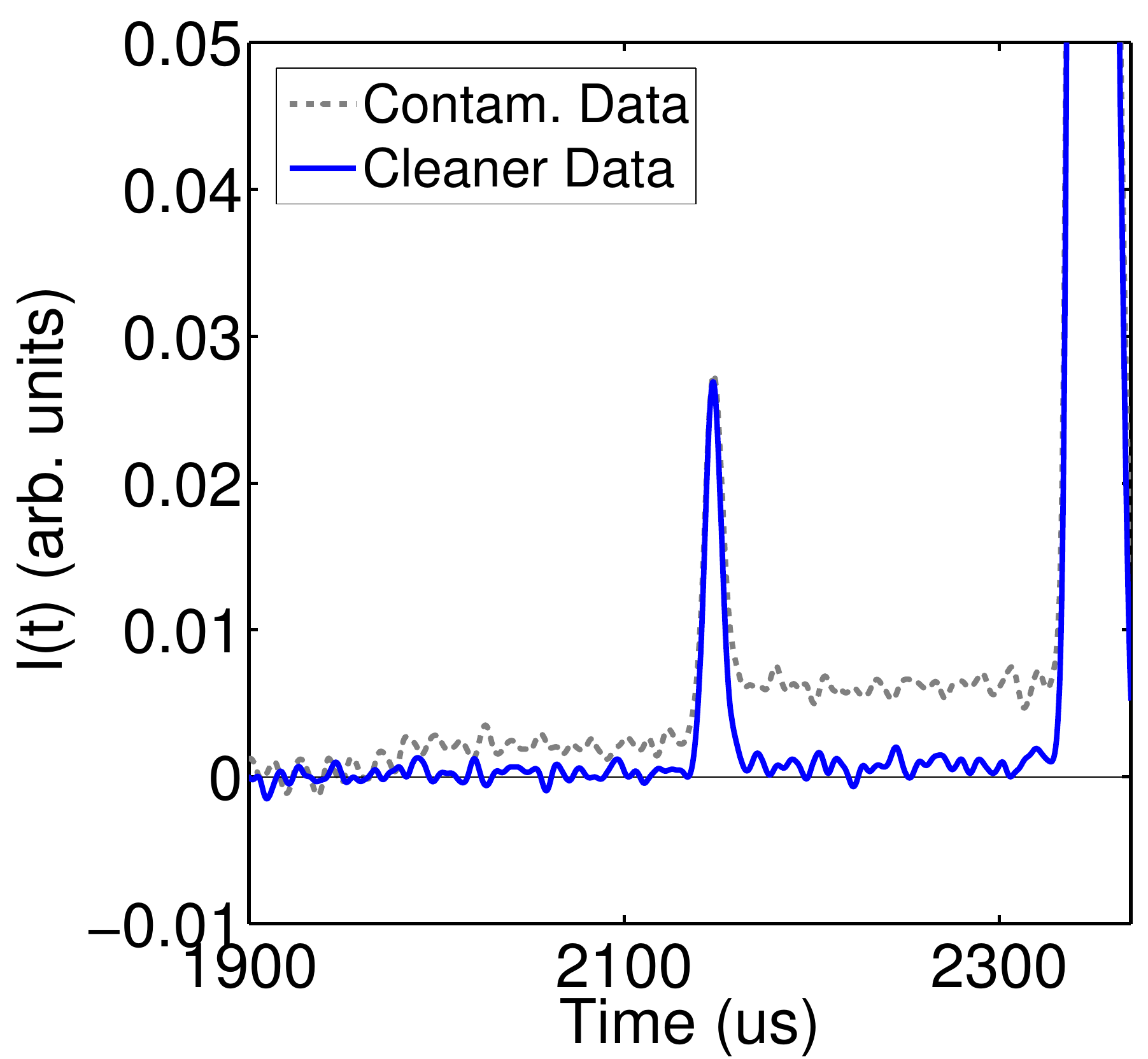}
\label{fig:comp20Torr60kV}}
\caption{(a)-(c) Comparison of waveform shapes for data with higher (dotted-gray) and lower (solid-blue) levels of water vapor contamination at several different reduced fields.  The primary SF$_{6}^{-}$ peak (outside the vertical range of the plots) has been normalized to one in every case. The effect can be considerable ($\sim$ 20\%) at lower reduced fields (a) but appears to diminish at a higher reduced field (c).}
\label{fig:CompCleanDirty}
\end{figure*}

The relative contribution of the broad component to the waveform discussed above (Figures~\ref{fig:avgwaveforms}, \ref{fig:avgwaveformszoom} and \ref{fig:AvgWaveformsLowField}) was found to depend on the length of the pump-out period prior to operation, and the subsequent rate of out-gassing as monitored with the baratron.  Given the propensity for plastics to absorb water vapor, the acrylic TPC vessel was an obvious source of contamination.  After numerous tests, which included separately adding small quantities of O$_2$ and water vapor into SF$_6$, we confirmed that the broad component was due to H$_2$O.
  
To reduce the out-gassing rate and dilute the concentration of contaminants, a long pump-down period (several days) followed by the flushing procedure outlined in Section~\ref{sec:opDAQ} was performed.  This greatly reduced the water vapor contamination as illustrated in Figure~\ref{fig:CompCleanDirty}, which shows waveforms of data taken before (dotted-gray) and after (solid-blue) this procedure was implemented.  The low contamination waveforms shown in solid-blue are from the same data shown in Figures~\ref{fig:avgwaveforms}-\ref{fig:AvgWaveformsLowField}.  Figure~\ref{fig:CompCleanDirty} shows the dramatic effects of water vapor on the waveforms and how strongly they depend on the reduced field, $E/p$.  The latter behavior provides an important clue to the detailed chemistry of water vapor interactions in SF$_6$, as discussed in detail below.  We can place an upper bound on the amount of water vapor contamination in these data using observations of the long term out-gassing rate.  By attributing the pressure rate-of-rise entirely to the out-gassing of water vapor, we estimate that the amount in the more contaminated data (dotted-gray waveforms in Figure~\ref{fig:CompCleanDirty}) was $<$1$\times$10$^{-1}$ Torr.  In the cleaner data (solid-blue curves in Figure~\ref{fig:CompCleanDirty}), where the detector had a much longer pump down period, we estimate that the amount of water vapor was $< $2 $\times$10$^{-3}$ Torr.  The additional step of flushing the vessel twice with SF$_6$ gas was also undertaken prior to data taking for the cleaner data.

While the effect of water vapor is quite significant, the physical mechanisms responsible for the observed features and their dependence on the reduced field are not fully understood.  Previous studies of electron attachment to water have shown that the single molecule does not have a negative ion state \cite{Haberland1983}, so it is unlikely that reactions of H$_2$O molecules with the primary electrons produced at the cathode are involved.  However, electron binding can occur in clusters of water molecules (H$_2$O)$_{n}^{-}$, where cluster sizes with $n \ge 2$ have been observed \cite{Haberland1984}.  Given the high electron affinity of SF$_6$ and the extremely low H$_2$O concentration, even in the high contamination data, the probability of such clusters forming at the primary ionization site should be low.  

Stable SF$_{6}^{-}$(H$_2$O)$_{n}$ clusters, with $n = 1-3$, are also known to form \cite{knighton, sieck, arnold}, thus a more likely scenario is one where water molecules interact directly with SF$_{6}^{-}$ anions as they drift towards the anode\footnote{SF$_{5}^{-}$(H$_2$O)$_{n}$ clusters should also be produced, however, we ignore them and their reactions here, because SF$_{5}^{-}$ is only produced at a few percent in our experiment (Figure~\ref{fig:avgwaveformszoom}).}.  Because these clusters drift slower than SF$_{6}^{-}$, they cannot account for the broad component in the waveform, but they can undergo further reactions with H$_2$O, producing the negative ions SOF$_4^{-}$ and F$^{-}$(HF)$_2$ with a relative probability of 4:1 \cite{arnold}.  If these ions drift faster than SF$_{6}^{-}$, as argued below, they could be responsible for most of the broad component observed in our waveforms.

With this brief overview of the chemistry of water vapor in SF$_{6}$ we can describe how some of the key features arise in the waveforms observed in our experiments.  The first is the evolution of the broad component of the waveform, which subsides with increasing $E/p$, essentially disappearing at the highest reduced fields in our measurements (e.g., Figures~\ref{fig:avgwaveformszoom}, \ref{fig:CompCleanDirty}).  This indicates that the cluster mediated reactions converting SF$_{6}^{-}$ into SOF$_4^{-}$ and F$^{-}$(HF)$_2$ become suppressed at higher $E/p$.  In our model these reactions require the stable formation and survival of the SF$_{6}^{-}$(H$_2$O)$_{n}$ clusters, which are weakly held together by hydrogen bonds that are unlikely to survive at high $E/p$.  Without these clusters the pathway to subsequent reactions is closed, leaving only SF$_{6}^{-}$ and SF$_{5}^{-}$ as observed.

Focusing on the low $E/p$ data where the effects of water vapor are most prominent, we expand our model to explain some of the  key features in the waveforms.  In our description of SF$_{6}^{-}$ and its interactions with water vapor as many as four species can be involved in transporting a negative charge from the cathode to the anode.  The drift velocity of this charge will therefore be a weighted average of each species', with the weighting determined by where exactly the conversion from SF$_{6}^{-}$ to SF$_{6}^{-}$(H$_2$O)$_{n}$, and SF$_{6}^{-}$(H$_2$O)$_{n}$ to either SOF$_4^{-}$ or F$^{-}$(HF)$_2$ occurs.  That the broad component of the waveform extends from the SF$_{6}^{-}$ peak down below the SF$_{5}^{-}$ peak means that the SOF$_4^{-}$ and F$^{-}$(HF)$_2$, and any other cluster mediated reaction products, travel faster than SF$_{6}^{-}$, with some even faster than SF$_{5}^{-}$.  Although we have no data on their mobilities in SF$_{6}$, this is reasonable given that both SOF$_4^{-}$ and F$^{-}$(HF)$_2$ are lighter than SF$_{6}^{-}$.  If we assume such a correlation between molecular mass and drift velocity (see Equation~\ref{eq:vdcross}), then F$^{-}$(HF)$_2$ would have the highest drift velocity, followed by SF$_{5}^{-}$, SOF$_4^{-}$ and SF$_{6}^{-}$, in that order.

Adopting this assumption we can explain two prominent features, the steps in amplitude at 4400 \si{\micro\second} and 4700 \si{\micro\second}, in the high contamination waveforms at low $E/p$ in Figure~\ref{fig:comp40Torr60kV}.  In our model, the former is essentially the shortest drift time in the waveform, which should correspond to F$^{-}$(HF)$_2$ being produced close to the cathode and traveling the full length of the detector.  Similarly, the second step at $\sim$4700 \si{\micro\second} should correspond to the next shortest drift time, that of SOF$_4^{-}$.  The fact that this step coincides with the SF$_{5}^{-}$ peak (Figure~\ref{fig:comp40Torr60kV}, blue curve) agrees with our assumption that two species having similar masses should have similar drift speeds.

Summarizing then, our model predicts that the charge in the region (region 1) between the SF$_{6}^{-}$ peak and the step at $\sim$4700 \si{\micro\second} should consist of a mixture of SOF$_4^{-}$ and F$^{-}$(HF)$_2$ when it arrives at the anode, while in the region (region 2) between $\sim$4700 \si{\micro\second} and $\sim$4400 \si{\micro\second} it should be only F$^{-}$(HF)$_2$.  That there is more charge in region 1 than in region 2 is expected because, as noted above, SOF$_4^{-}$ and F$^{-}$(HF)$_2$ are produced in the ratio 4:1. 

A more detailed analysis of the rich structure observed in the waveforms of the high contamination data is beyond the scope of this paper, nor is it relevant for the goals of directional dark matter detection.  For our purposes, the key features of the waveform are the SF$_{5}^{-}$ and SF$_{6}^{-}$ peaks and their properties, and the remainder of this work will describe their application to directional dark matter searches.  The data used in the following sections is the same used to produce the waveforms in Figures~\ref{fig:avgwaveforms}-\ref{fig:AvgWaveformsLowField}, which was acquired with minimum water contamination.

In hindsight, our acrylic-based TPC detector, which was designed for high reduced field operation, was not an ideal choice for operating with SF$_6$ due to its permeability to water vapor and high out-gassing rate.  Moreover, this concern extends well beyond acrylic and encompasses a broad collection of polymer-based materials that are hygroscopic.  If plastics cannot be avoided, for example because of their desired low radioactivity, then care should be taken to minimize any water vapor contamination during detector construction and data acquisition.  Besides the techniques used here to achieve this, we have also considered the use of desiccant, and gas recirculating and purification as commonly done in TPCs.

\subsection{Relative peak charge and amplitude}
\label{sec:relativeratios}

With the preceding discussion of the global features of the SF$_{6}$ waveform, we now turn our focus to the SF$_{5}^{-}$ peak.  
The importance of detecting both SF$_{5}^{-}$ and SF$_{6}^{-}$ peaks is that they enable the ability to fiducialize events along the drift direction in the TPC.  This provides a powerful tool for rejecting backgrounds in the type of rare searches of interest here, as discussed further in Section~\ref{sec:fiducial} where fiducialization is demonstrated using this tool. 
\begin{figure}[]
 \centering 
	\includegraphics[width=0.6\textwidth]{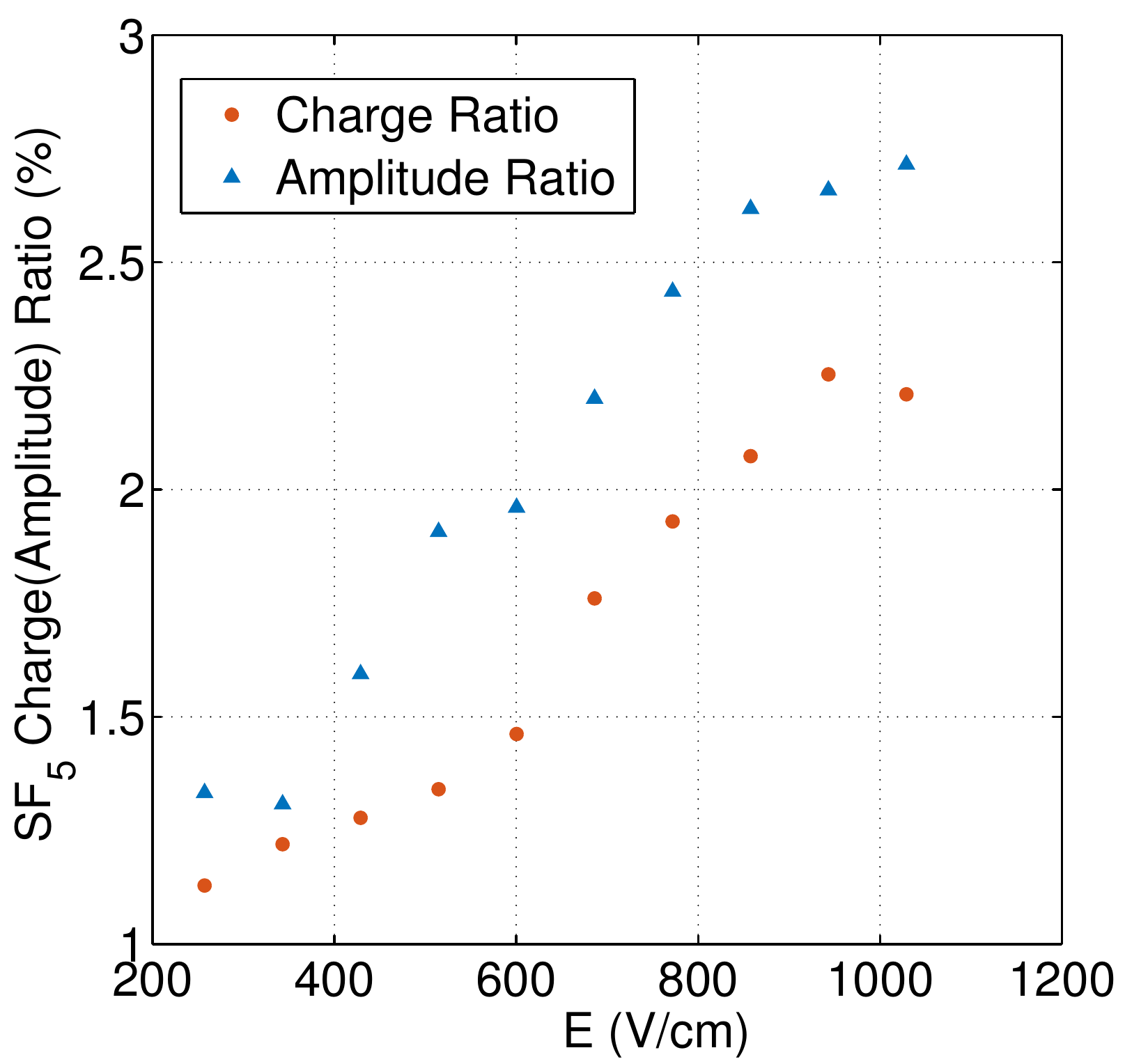}
  \caption[]{The detected charge and amplitude of the SF$_{5}^{-}$ peak relative to SF$_{6}^{-}$ in 20 Torr as a function of the electric field.  Both quantities increase with electric field but then appear to taper off at a field strength of $\sim$ 900 V$\cdot$cm$^{-1}$.  It is important to note, as discussed in the text, that these are detected quantities and not necessarily the relative amounts produced in the drift volume.  }
\label{fig:ChargeFractions}
\end{figure}

To study the behavior of the secondary SF$_{5}^{-}$ peak with field strength, we use the averaged waveforms taken at fixed $p = 20$ Torr for each of ten different electric field strengths between $257-1029$ V$\cdot$cm$^{-1}$ (Figures~\ref{fig:avgwaveforms}-\ref{fig:avgwaveformszoom}).  Using these, the amplitudes of the SF$_{5}^{-}$ and SF$_{6}^{-}$ peaks and the amount of charge contained within the peaks were computed.  The evolution of the fraction of charge in the SF$_{5}^{-}$ peak and its amplitude relative to the SF$_{6}^{-}$ peak as a function of the electric field are shown in Figure~\ref{fig:ChargeFractions}.  Both the relative charge and amplitude rise with increasing field strength but then appear to taper off at a field strength of $\sim$900 V$\cdot$cm$^{-1}$ ($E/p$ = 45  V cm$^{-1}$ Torr$^{-1}$). 

The amplitude(charge) of the SF$_{5}^{-}$ peak measured at the highest reduced field (20 Torr/1029 V cm$^{-1}$$=158$ Td) is $\sim$2.8(2.2)\% that of SF$_{6}^{-}$, which is what their relative capture cross-sections at an electron energy of $\sim 0.1$ eV would predict.  It is important to note that this is the {\it detected} ratio of SF$_{5}^{-}$ to SF$_{6}^{-}$ and is likely to be lower than what was produced at the site of ionization.  This is because of the higher electron affinity of SF$_{5}^{-}$ ($2.7-3.7$ eV), which could lead to a lower gas gain relative to SF$_{6}^{-}$ due to the greater difficulty in stripping the electron in the THGEM.

As the detectability of the small SF$_{5}^{-}$ peak is critical for fiducialization, it will require high signal-to-noise as well as investigation into possible methods to enhance it.  For example, the ratio of SF$_{5}^{-}$ to SF$_{6}^{-}$ is known to rise at higher electron energies and gas temperatures, with reports indicating that it can exceed 39\% at 593 K \cite{Miller1994}.  This is further discussed in Section~\ref{sec:peakenhance}.


\section{Reduced mobility}
\label{sec:mobility}

The drift velocities of SF$_{6}^{-}$ and SF$_{5}^{-}$ were determined by measuring the time difference between the creation of photoelectrons at the cathode using the N$_2$ laser, and the arrival of ionization at the THGEM corresponding to the respective peaks.  The 3.5 ns laser pulses generated ionization events that have essentially zero longitudinal extent.  The laser pulse also provided the trigger to the DAQ system and gave us the initial time marker, $T_0$.  We define the drift time as the time between the initial laser trigger and the arrival time of the pulse peak, $T_p$, rather than the leading edge of the ionization signal at the THGEM.  The magnitude of the drift velocity, $v_d$, is then given by
\begin{equation} \label{eq:vdT}
 v_d = \frac{L}{T_p - T_0}, 
\end{equation}
where, $L= 583 \pm 0.5$ mm, is the distance between the THGEM and the cathode.  We measured the drift velocity over a range of electric field values ($86-1029$ V$\cdot$cm$^{-1}$) and pressures (20, 30, 40 Torr).  Following convention, we report the mobilities instead of drift velocities.

The mobility, $\mu$, of a drifting ion at a specific gas density is related to the drift speed, $v_{d}$, and electric field, $E$, through the relation:
\begin{equation} \label{eq:vdmob}
v_{d} = \mu \cdot E.
\end{equation}
A standardized quantity called the reduced mobility, $\mu_{0}$, is derived from the measured mobility by the expression:
\begin{equation} \label{eq:mobred}
\mu_{0} = \frac{v_{d}}{E}\frac{N}{N_{0}},
\end{equation}
where $N_{0} = 2.687\times 10^{19}$ cm$^{-3}$ is the gas density at STP ($0^{\circ}$C and 760 Torr) and $N$ is the detector gas density at the time of measurement.

\begin{figure}[]
 \centering 
	\includegraphics[width=0.60\textwidth]{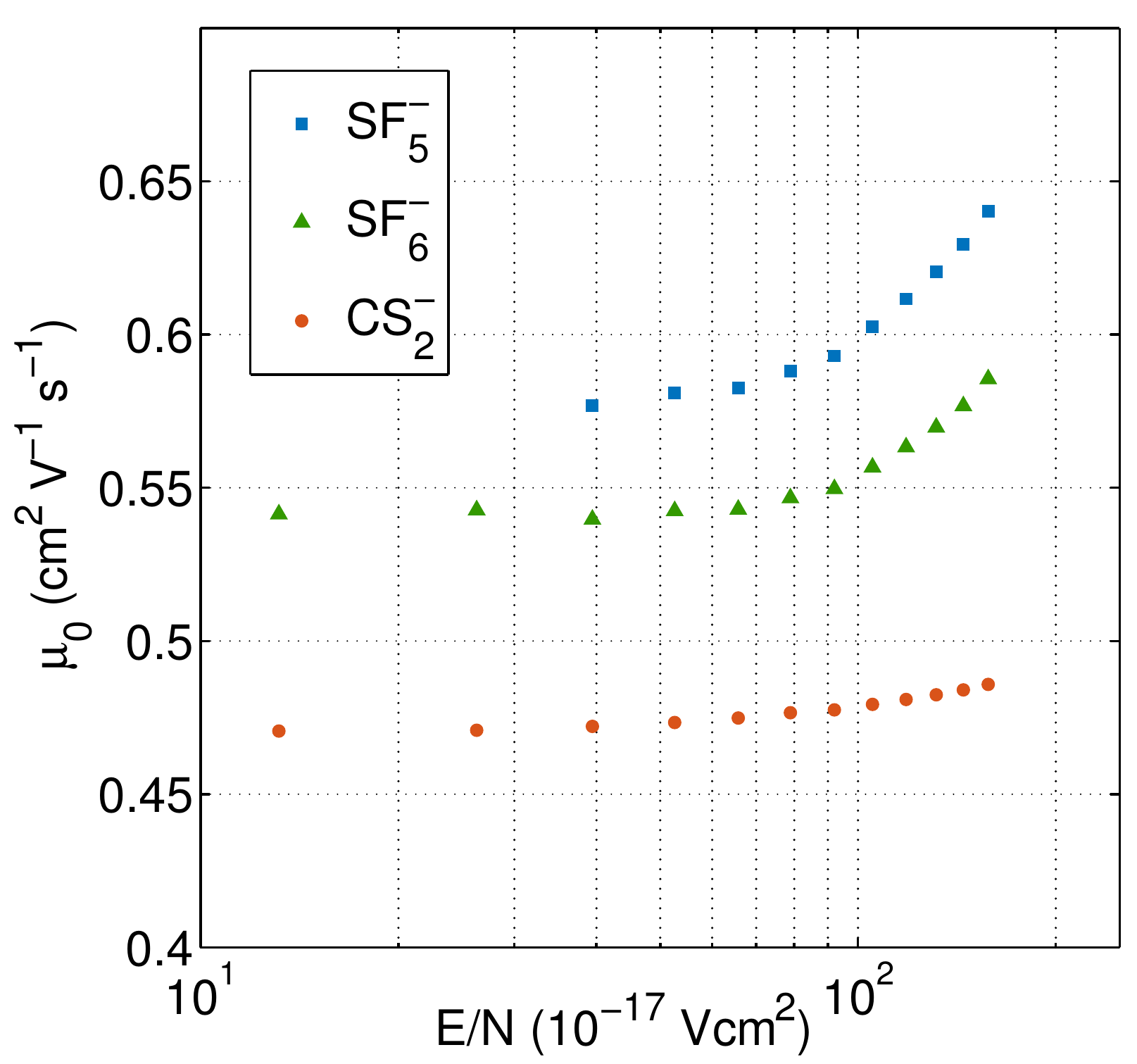}
  \caption[]{The reduced mobility as a function of reduced field for SF$_{5}^{-}$ and SF$_{6}^{-}$ in SF$_6$ and CS$_{2}^{-}$ in CS$_{2}$.  The SF$_{5}^{-}$ mobilities only go down to about 40 Td below which its peak becomes difficult to identify.  Our results for SF$_{5}^{-}$ and SF$_{6}^{-}$ are in excellent agreement with those found in Ref.~\cite{Urquijo} while the CS$_{2}^{-}$ results agree with those from Ref.~\cite{SIG2013}.  The combined uncertainty due to instrumental precision is 1\%. }
\label{fig:redmob}
\end{figure}

Our measured mobilities for CS$_{2}^{-}$, SF$_{5}^{-}$ and SF$_{6}^{-}$ are plotted in Figure~\ref{fig:redmob} as a function of the reduced field, $E/N$, in units of the Townsend\footnote{1 Td $= 10^{-17}$ V cm$^{2}$, 1 V cm$^{-1}$ Torr$^{-1}$ = 3.066 Td at $T=296$ K.}.  We find good agreement between our results for the reduced mobility of CS$_{2}^{-}$ in CS$_{2}$ and those reported by Ref.~\cite{SIG2013} in the low field regime ($<$ 50 Td), where our data overlap.  Our measurement of the reduced mobility of SF$_{6}^{-}$ in SF$_{6}$, extrapolated to zero field, is $\mu_{0}$(SF$_{6}^{-}$) $= 0.540 \pm 0.002$ cm$^{2}$V$^{-1}$s$^{-1}$, which agrees well with the result from Ref.~\cite{Urquijo}.  There is also excellent agreement over the full range of reduced fields between our dataset for SF$_{5}^{-}$ and SF$_{6}^{-}$ mobilities in SF$_{6}$ with the mass-identified measurements reported in Ref.~\cite{Urquijo}.  A comparison with other data-sets from Ref.~\cite{Patterson} and \cite{Fleming}, a majority of which do not have mass analysis, shows agreement over some ranges of reduced fields only.

The CS$_{2}^{-}$ mobility is about 13.1\% lower than the SF$_{6}^{-}$ mobility at 13 Td, but this difference rises to about 17.0\% at 158 Td which shows that SF$_{6}^{-}$ mobility increases more rapidly with reduced field than CS$_{2}^{-}$ mobility.  This is unexpected, and goes against our assumptions in Section~\ref{sec:watervapor}, because SF$_6$ is a much heavier molecule than CS$_2$ and the drift velocity for ions with mass, $m$, drifting in a gas with molecules of mass, $M$, is given by
\begin{equation} \label{eq:vdcross}
v_{d} = \left(\frac{1}{m}+\frac{1}{M}\right)^{1/2} \left( \frac{1}{3kT} \right)^{1/2} \frac{eE}{N\sigma},
\end{equation}
where $\sigma$ is the ion-gas molecule cross-section \cite{blumrolandi}.  This implies that the cross-section for SF$_{6}^{-}$:SF$_{6}$ interaction is smaller than, and changes faster with increasing field strength than that for the CS$_{2}^{-}$:CS$_{2}$ interaction.  A similar comparison between SF$_{5}^{-}$ and SF$_{6}^{-}$ shows that the mobility of the former is 6.9\% higher than the latter's at about 39 Td, and is 9.3\% larger at 158 Td.  The rise in mobility with increasing reduced field that is observed for all of the negative ions in Figure~\ref{fig:redmob} indicates that the transport processes are energy dependent.  This has important implications for diffusion at the higher reduced fields, as shown in the next section.


\section{Longitudinal diffusion}
\label{sec:diffusion}

At low field strengths where the drifting charge cloud has thermal energy, the diffusion coefficient can be approximated by its zero reduced field limit, $D(0)$\footnote{In this regime the charge cloud diffuses isotropically, so the longitudinal and transverse components, $D_L$ and $D_T$, are the same and equal to $D(0)$.}.  This is related to the mobility and gas temperature through the Nernst-Townsend-Einstein relation:
\begin{equation} \label{eq:NTE0}
\frac{D(0)}{\mu(0)} = \frac{kT}{e},
\end{equation}
where $e$ is the ion charge \cite{Mason}.  At higher field strengths, diffusion can enter the non-thermal regime where it is given approximately by the generalized Einstein relations \cite{Viehland}.  These predict that deviations from $D(0)$ will occur when the field derivative of the reduced mobility becomes non-zero, which, according to the data shown in Figure~\ref{fig:redmob}, occurs above $E/N \sim 60-70$ Td for SF$_{6}^{-}$.  In the non-thermal regime, the deviations in longitudinal diffusion, $D_L$, are proportional to this derivative and larger than those in transverse diffusion, $D_T$.  In this work we only measure longitudinal diffusion and, by comparing it with the predictions of Equation~\ref{eq:NTE0}, look for deviations from the thermal limit.

From Equation~\ref{eq:NTE0}, a starting point-like charge cloud drifting over a distance, $L$, has a longitudinal diffusion width, $\sigma_{z}$, given by
\begin{equation} \label{eq:sigmat1}
\sigma_{z}^{2} =  2D_{L}t  =  \frac{4\epsilon L}{3eE}= \frac{2kTL}{eE},
\end{equation}
where $t= L/v_{d}$ and $\epsilon = 3/2 kT$ \cite{blumrolandi}.  As our measurements are of pulse widths, we relate the diffusion in the time domain, $\sigma_{t}$, to $\sigma_{z}$
using the drift velocity:
\begin{equation} \label{eq:sigmaz}
\sigma_{z} = \sigma_{t} \cdot v_{d}.
\end{equation}
Customarily, diffusion is expressed by normalizing the measured value relative to the drift length:
\begin{equation} \label{eq:sigma0}
\sigma_{0} = \frac{\sigma_{z}}{\sqrt{L}} = \sqrt{\frac{2kT}{eE}},
\end{equation}
where $\sigma_{0}$ is typically expressed in units of \si{\micro\meter}/$\sqrt{\text{cm}}$.

\begin{figure*}[]
\centering
\subfloat[CS$_2$ 20 Torr Averaged Waveform]{
	\includegraphics[width=0.48\textwidth]{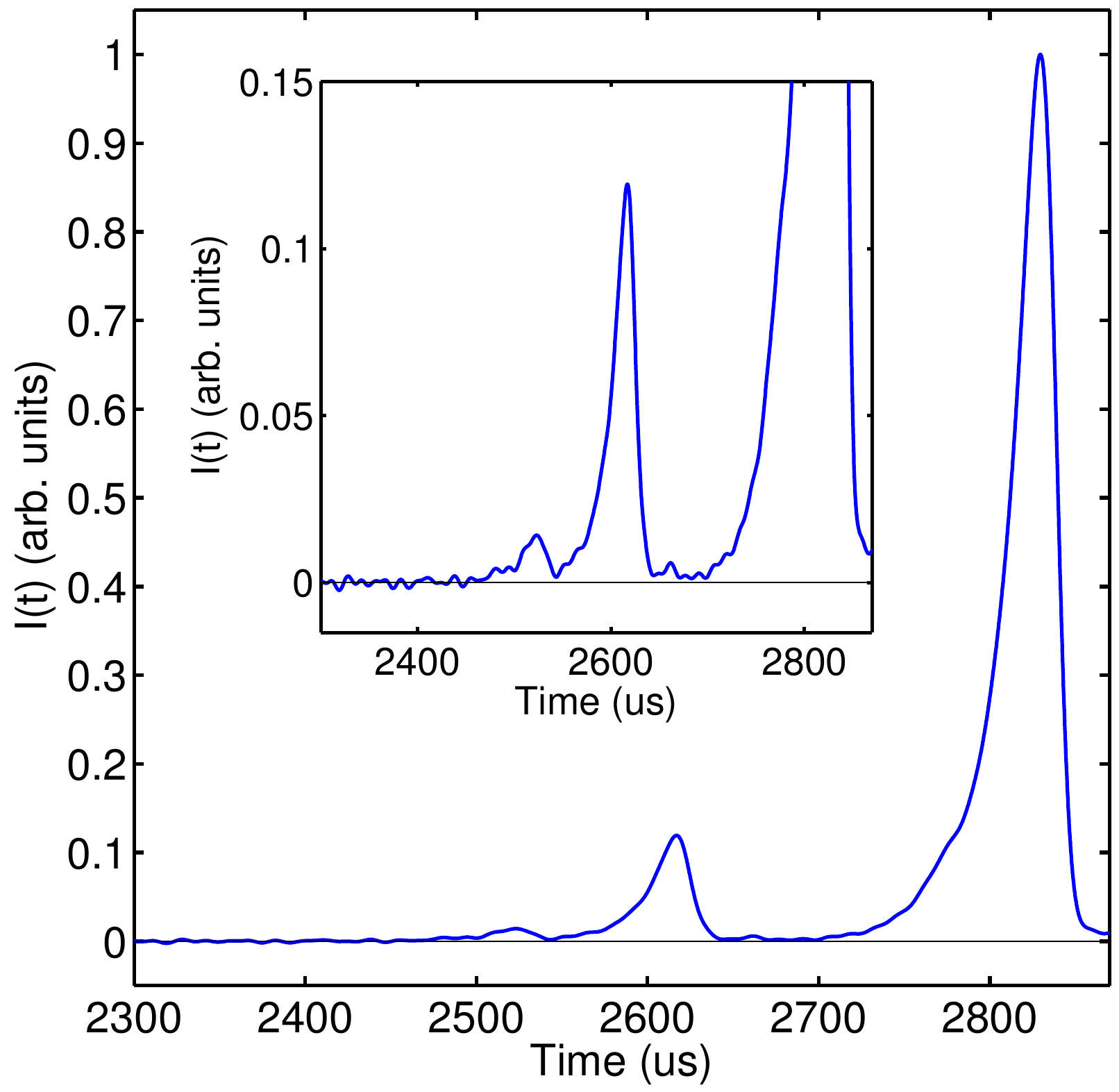}
\label{fig:CS2avgwaveform20Torr}
}
\subfloat[CS$_2$ 40 Torr Averaged Waveform]{
	\includegraphics[width=0.48\textwidth]{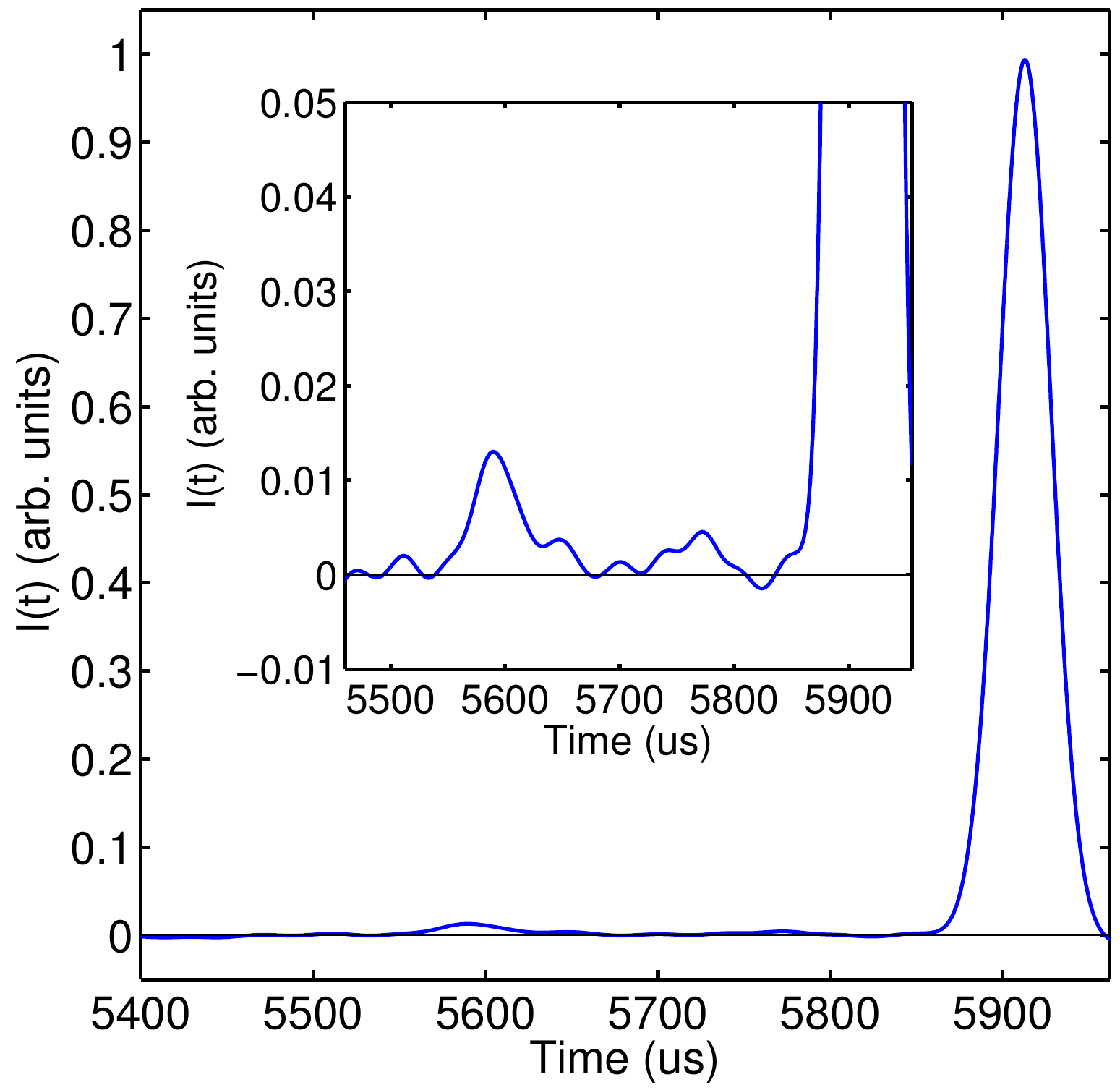}
\label{fig:CS2avgwaveforms40Torr}
}
\caption{(a) The averaged waveform for 20 Torr CS$_{2}$ at $E=1029$ V$\cdot$cm$^{-1}$ showing the presence of a large secondary peak at $\sim$2600 \si{\micro\second} and the possible appearance of two additional peaks at $\sim$2660 \si{\micro\second} and $\sim$2520 \si{\micro\second} (inset).  In addition, the distortion in the waveform shape is clearly seen in both the primary and secondary peaks at this high reduced field.  This behavior is not observed in the SF$_6$ waveforms at high reduced fields. (b) The average waveform for 40 Torr CS$_{2}$ at $E=1029$ V$\cdot$cm$^{-1}$ which shows a much smaller secondary peak and no distortion in waveform shape. }
\label{fig:CS2avgwaveforms}
\end{figure*}

\begin{figure*}[]
\centering
\subfloat[SF$_6$ diffusion]{
	\includegraphics[width=0.49\textwidth]{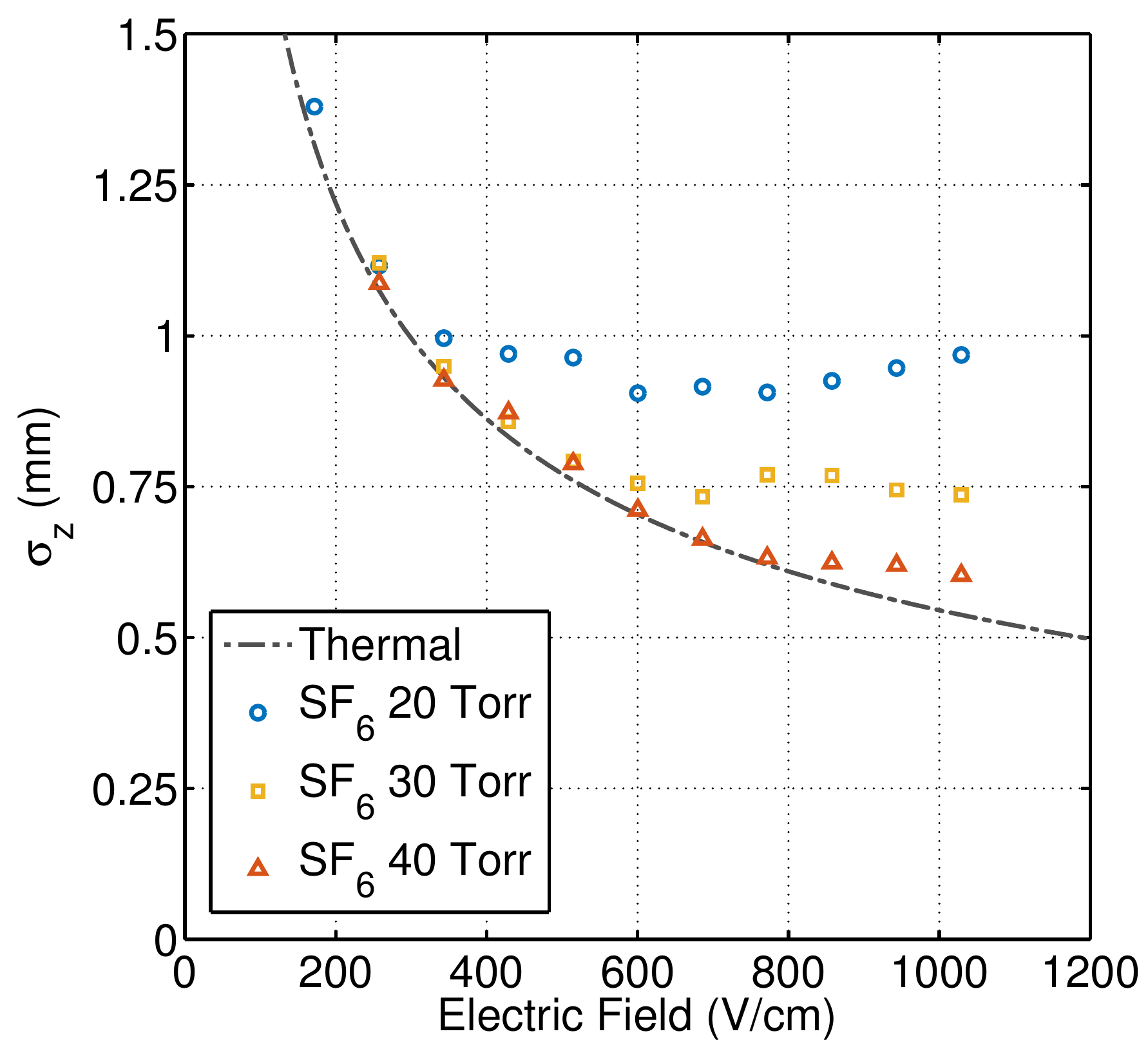}
\label{fig:SF6sigma}
}
\subfloat[CS$_2$ diffusion]{
	\includegraphics[width=0.48\textwidth]{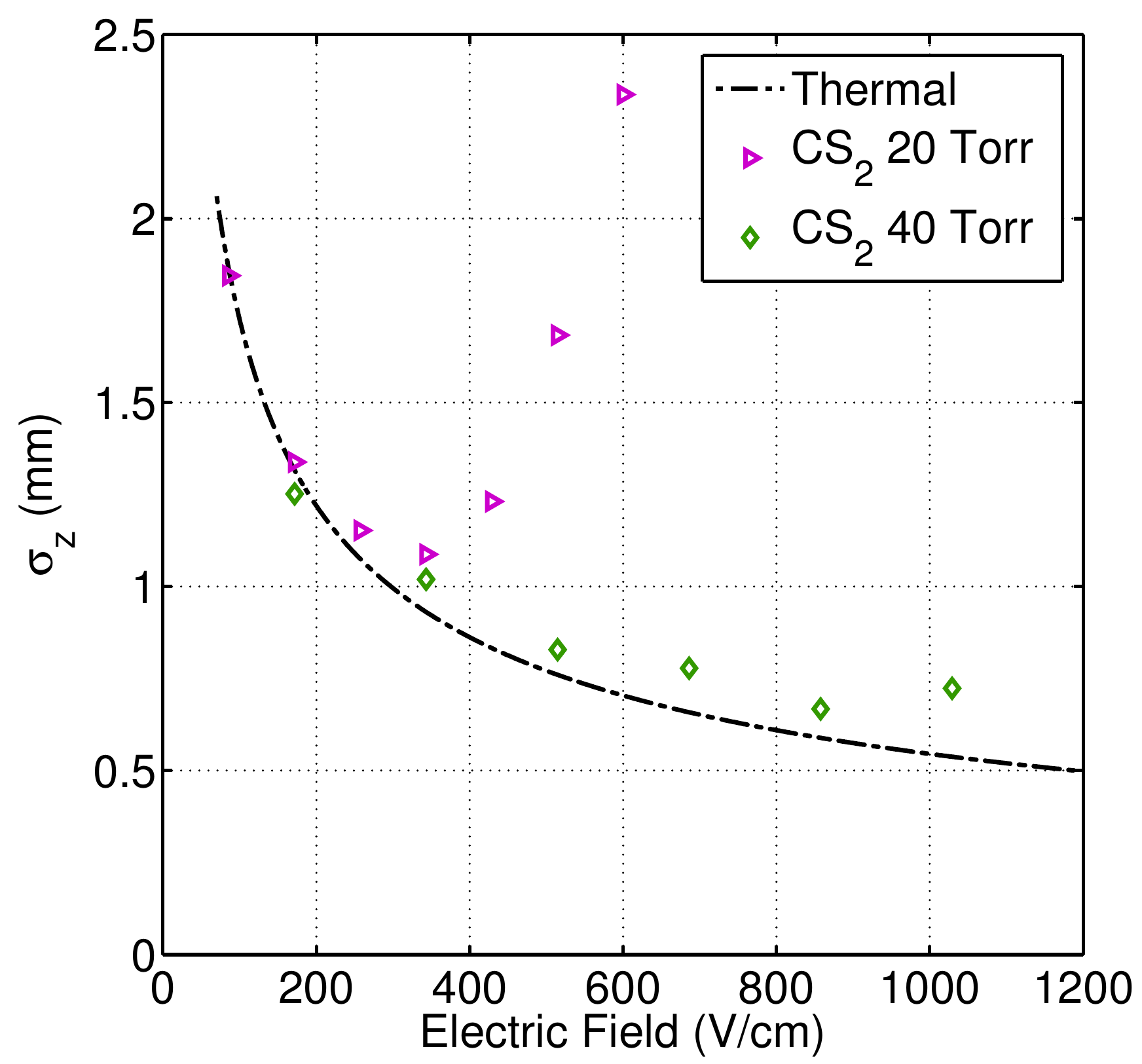}
\label{fig:CS2sigma}
}
\caption{(a) The longitudinal diffusion, $\sigma_{z}$, for 20, 30, and 40 Torr SF$_6$ as a function of electric field for a drift length of 58.3 cm.  The plotted quantity includes the broadening effects of the finite THGEM hole pitch as well as the capture process.  The dot-dashed line shows the predicted width for thermal diffusion from Equation \ref{eq:sigmat1}.  For the 40 Torr data, the measured width begins to deviate from the thermal prediction at $\sim 800$ V$\cdot$cm$^{-1}$.  Similarly, the 30 Torr and 20 Torr data deviate from the thermal diffusion curve at $\sim$600 V$\cdot$cm$^{-1}$ and 400 V$\cdot$cm$^{-1}$, respectively.  (b) The fitted pulse width for 20 and 40 Torr CS$_2$.  At 20 Torr the pulse width begins to deviate considerably from thermal at $\sim$400 V$\cdot$cm$^{-1}$.  The corresponding distortion seen in the waveform in Figure~\ref{fig:CS2avgwaveform20Torr} could, however, also be due to a longer electron capture mean-free-path at high $E/p$ (refer to the text). }
\label{fig:sigma_vs_E}
\end{figure*}

The pulses used to measure diffusion of SF$_{5}^{-}$ and SF$_{6}^{-}$ were obtained from waveforms generated using ionization produced at the cathode, a known $L$ = 58.3 cm drift distance, with the N$_2$ laser as described in Section~\ref{sec:chargegeneration}.  One thousand of these waveforms were averaged together at each pressure and electric field to increase signal-to-noise, resulting in the averaged waveforms shown in Figures~\ref{fig:avgwaveforms}, \ref{fig:avgwaveformszoom} and \ref{fig:AvgWaveformsLowField}.  As the SF$_{5}^{-}$ or SF$_{6}^{-}$ pulses are not strictly Gaussian, some care was required in extracting their widths.  The main contribution to their non-Gaussianity is from the positive ion tail on the right side, whose origin was explained in Section~\ref{sec:waveforms}.  To minimize its effect, only a fraction of the left hand side of the waveform above 10\% of the peak was used to fit to a Gaussian curve.  This fraction was determined iteratively by modeling the relative contributions of the collected charge signal and the positive ion induced signal to the pulse amplitude.  Additionally, due to the broad structure from residual water vapor contamination at low reduced field (Section~\ref{sec:waveforms}), only data with $E >171$ V$\cdot$cm$^{-1}$ at 20 Torr and $E > 257 $ V$\cdot$cm$^{-1}$ at 30 and 40 Torr were used.

Using this procedure we found $\sigma_{\text{fit}}$, which is mostly due to diffusion with small contributions from other effects. The latter are the smoothing time, $\sigma_\text{smooth}$, laser spot size, $\sigma_\text{spot}$, the spread in the electron-capture length, $\sigma_\text{capture}$, and effects at the THGEM, $\sigma_\text{THGEM}$.  We have no direct measurements of $\sigma_\text{capture}$ or $\sigma_\text{THGEM}$, so we make no attempt to correct for them here.  In our measurements, the laser spot size contribution to the longitudinal width is negligible, so we set $\sigma_\text{spot} \sim 0$.  Thus, assuming no correlation, we subtract $\sigma_\text{smooth}$ from $\sigma_{\text{fit}}$ in quadrature to get the diffusion width in time:
\begin{equation} \label{eq:sigmasub}
\sigma_{t} = \sqrt{\sigma_\text{fit}^{2} - \sigma_\text{smooth}^{2}}.
\end{equation}
Using Equation~\ref{eq:sigmaz}, we finally get $\sigma_{z}$, the longitudinal spread of the charge distribution in space due to diffusion.  The systematics on $\sigma_{z}$, mainly due to not accounting for $\sigma_\text{capture}$ and $\sigma_\text{THGEM}$, are briefly discussed below. 

The same fitting procedure was applied to our CS$_2$ data taken at 20 Torr (Figure~\ref{fig:CS2avgwaveform20Torr}) and 40 Torr (Figure~\ref{fig:CS2avgwaveforms40Torr}).  The 20 Torr, high reduced field waveform shown in Figure~\ref{fig:CS2avgwaveform20Torr} appears distorted on the left and has at least one additional peak at $\sim$ 2600 \si{\micro\second}.  These features are discussed in Section~\ref{sec:cs2_20torr}.

\subsection{$\sigma_{z}$ results}
\label{sec:diff_results}

In Figure~\ref{fig:sigma_vs_E}, the longitudinal diffusion, $\sigma_{z}$, is plotted as a function of electric field for 20, 30, and 40 Torr SF$_6$ and 20 and 40 Torr CS$_2$ data.  Overlaid is the curve for thermal diffusion calculated using Equation~\ref{eq:sigmat1}.  In the 40 Torr SF$_6$ data, $\sigma_{z}$ begins to deviate from the thermal prediction at around 800 V$\cdot$cm$^{-1}$.  Similarly, in the 30 Torr and 20 Torr SF$_6$ data, deviations from thermal diffusion occur at around 600 V$\cdot$cm$^{-1}$ and 400 V$\cdot$cm$^{-1}$, respectively.  In terms of the reduced field, the deviations all begin to occur at approximately $E/p = 20$ V$\cdot$cm$^{-1}$$\cdot$Torr or $E/N = 60$ Td.  This is close to our estimate above of $E/N \sim 60-70$ Td based on the generalized Einstein relations.

The 40 Torr CS$_2$ diffusion data shown in Figure~\ref{fig:sigma_vs_E} indicate a larger systematic than observed for SF$_6$.  This is likely due to a longer mean free path for electron capture and is discussed further below in Section~\ref{sec:diff_sys}. Assuming that this systematic is field independent, the data appear to follow thermal diffusion out to $\sim$500 V$\cdot$cm$^{-1}$ (38 Td) and perhaps even to $\sim$800 V$\cdot$cm$^{-1}$ (61 Td).  Precision measurements of $\sigma_{z}$ \cite{SIG2013} have confirmed thermal out to 23 Td and other measurements \cite{Pushkin-Ifft-2009} indicate that the low field approximation applies to CS$_2$ out to $\sim$42 Td. 

For 20 Torr CS$_2$, we observe a distortion in the waveform at high reduced fields and one or more smaller peaks begin to appear (Figure ~\ref{fig:CS2avgwaveform20Torr}), which also grow with $E/p$.  The effect of the distortion on $\sigma_{z}$ begins at $\sim$50 Td and is dramatic as seen in Figure~\ref{fig:sigma_vs_E}.  The origin of the distortion could be a deviation from thermal diffusion or a growing inefficiency in electron capture at high $E/p$, which would naturally explain the observed tail on the fast side of the waveform.  The fact that the reduced mobility has a weaker dependance on $E/p$ than SF$_6^-$ (Figure~\ref{fig:redmob}) also points to electron capture.  Measurements of the lateral diffusion should help determine which of these effects dominates.  The secondary features are discussed in Section~\ref{sec:cs2_20torr}.

\subsection{Systematics on $\sigma_{z}$}
\label{sec:diff_sys}

Here we place bounds on the two primary sources of systematic effects that contribute to our estimate of the diffusion width, $\sigma_{z}$, the spread in the electron-capture mean-free-path and the non-uniformity of the electric field near the THGEM.  Given how well matched our $\sigma_{z}$ values are to the diffusion limit at low reduced fields (Figure~\ref{fig:sigma_vs_E}), any non-diffusion contributions cannot be large.  At low reduced fields in 40 Torr CS$_2$, an upper bound on the spread in capture distance of 0.35 mm was estimated by Ref.~\cite{SIG2013}.  Based on measurements of the attachment cross-section in SF$_6$, the mean free path for attachment in our experimental apparatus should be of order $\sim$$1-10$ \si{\micro\meter} and, hence, a negligible contribution to $\sigma_{z}$.  The broadening effect due to the non-uniformity in the drift field close to the THGEM should depend on the THGEM pitch, and the fields in the holes and TPC drift region.  This can be modeled but we can provide an upper bound estimate based on diffusion in the low $E/p$ region in SF$_6$ from Figure~\ref{fig:SF6sigma}, where we expect it to be thermal.  The $\sigma_{z}$ data in this region are systematically slightly higher than the thermal prediction, thus, assigning the difference taken in quadrature to the THGEM, gives the upper bound of $\sigma_\text{THGEM} < 0.2$ mm. 

In 40 Torr CS$_2$, the systematic differences in the low $E/p$ regime (Figure~\ref{fig:CS2sigma}) are larger than in SF$_6$, which is probably due to a longer electron capture distance as discussed above.  Assuming that the contribution from the THGEM is the same for both gases, $\sigma_\text{THGEM}\sim 0.2$ mm, we can assign the remaining difference to the spread in electron capture distance in CS$_2$.  This gives $\sigma_\text{capture}\sim 0.3$ mm, which is within the upper bound for CS$_2$ from Ref.~\cite{SIG2013} given above.  In the 20 Torr CS$_2$ data we speculate that the large deviation in $\sigma_{z}$ from thermal observed above 50 Td is due to inefficient electron capture, rather than diffusion.  Measurements of lateral diffusion will test this hypothesis.  A more accurate estimate for the sum total of non-diffusion contributions, including $\sigma_\text{THGEM}$ and $\sigma_\text{capture}$, can also be determined by measuring the waveform width as a function of drift distance.  This is left for future work.

\subsection{Implications for directional searches of low mass WIMPs}
\label{sec:Low_mass_Wimps}

For dark matter searches in the low, $\sim10$ GeV/c$^2$, WIMP mass regime, the lowest possible energy thresholds are desired.  As discussed in some detail in Ref.~\cite{mayet, tatarowicz}, for TPC-based directional dark matter searches this can be achieved by lowering the pressure to lengthen the tracks, which should lead to lower \emph{directional} energy thresholds.  Reference~\cite{mayet} simulated this by using the minimum resolvable track size from directionality data taken at 100 Torr and, assuming that it is pressure independent, showed that pressures in the $\sim$5-10 Torr range would be optimal for directional low mass WIMPs searches.  Their assumption requires that the physical effects impacting track reconstruction, such as diffusion, do not worsen at lower pressures.

Our measurements of diffusion at 20 Torr for both SF$_6$ and CS$_2$ provide an important test of this assumption.  As discussed above and shown in Figure~\ref{fig:SF6sigma}, deviation from thermal diffusion in our SF$_6$ data occur at $\sim$60 Td in the $20-40$ Torr range.  This means that at lower pressures the deviation occurs at a lower drift field, where thermal diffusion is higher Figure~\ref{fig:SF6sigma}.  For example, in 20 Torr SF$_6$ the minimum longitudinal diffusion observed in our data is $\sigma_z \sim 0.9$ mm, which is quite a bit higher than the $\sigma_z \sim 0.63$ mm in 40 Torr.  This worsening of diffusion dilutes the benefit of lower pressures as described below.

For CS$_2$, the data at 20 Torr shown in Figures~\ref{fig:CS2avgwaveform20Torr} and \ref{fig:CS2sigma} are difficult to interpret with diffusion alone.  As discussed above, the long tail on the fast side of the waveform is characteristic of a long electron capture mean free path, but this requires confirmation.  The mobility data from Figure~\ref{fig:redmob}, however, do indicate that any deviations from thermal diffusion in CS$_2$ should be comparable to, if not smaller than for SF$_6$.

It is clear that further detailed studies of diffusion for both SF$_6$ and CS$_2$, which include the transverse component, are needed to better assess the low pressure regime.  Nevertheless, we can use the current data to provide a reasonable estimate of the tracking resolution as a function of pressure.  For this we define the dimensionless track resolution as $M \equiv \sigma/R \propto \sigma \cdot p$, where $R$ is the track length for a given recoil energy, and is inversely proportional to the pressure $p$.  At low reduced fields where diffusion is thermal, $\sigma$ is a function of the field only, $\sigma = \sigma\left( E \right)$, but in the non-thermal regime it also depends on the pressure as seen in Figure~\ref{fig:SF6sigma}.  Thus, so long as one remains in the thermal regime, lowering the pressure will lengthen tracks of a given energy and the resolution $M$ will scale \emph{linearly} with $p$ at a fixed $E$.  In our SF$_6$ data the transition from thermal to non-thermal diffusion occurs at $E/p \sim 60$ Td, or 20 V$\cdot$cm$^{-1}$Torr, in the $20 - 40$ Torr range.  Once in the non-thermal regime, $\sigma$ stays approximately constant with $E$ or even decreases slightly (Figure~\ref{fig:SF6sigma}).  If we conservatively take the minimum $\sigma$ to correspond to the thermal value at $E/p = 20$ V$\cdot$cm$^{-1}$Torr, then $\sigma_{\text{min}} \propto 1/\sqrt{20 \text{ V}\cdot\text{cm}^{-1}\text{Torr} \cdot p } \propto 1/\sqrt{p}$.  In this case $M \propto \sigma \cdot p \propto \sqrt{p}$, and the resolution no longer improves linearly with pressure.  Although lowering the pressure can still provide a path for directional low mass WIMP searches, the worsening of diffusion requires further study to quantify the benefits.  For example, measurements closer to the optimal $\sim$10 Torr pressures in SF$_6$ can check whether the transition from thermal is still at $E/p = 20$ V$\cdot$cm$^{-1}$Torr and whether other effects, such as the electron capture length, become significant.

\subsection{Secondary peak in CS$_2$}
\label{sec:cs2_20torr}

Finally, we return to the small, secondary peak observed in the 20 Torr CS$_2$ data shown in Figure~\ref{fig:CS2avgwaveform20Torr}.  This feature first appears at a drift field of $E = 343$ V$\cdot$cm$^{-1}$ at 20 Torr CS$_{2}$ with a drift speed that is $\sim$6.2\% faster than, and an amplitude only 0.4\% that of the primary peak.  When the drift field is increased to $E = 686$ V$\cdot$cm$^{-1}$, the secondary peak's drift speed and amplitude increase to 6.8\% and 4.6\%, respectively, relative to that of the primary peak.  Finally, at $E = 1029$ V$\cdot$cm$^{-1}$, the secondary peak is about 7.5\% faster than the primary while its amplitude has grown to about 11.7\% of the primary's peak value (Figure~\ref{fig:CS2avgwaveform20Torr}).  In the 40 Torr CS$_2$ data there is a hint of a secondary peak at the highest field, $E = 1029$ V$\cdot$cm$^{-1}$, which is a factor of $\sim$10 lower in amplitude than that of the secondary peak seen at 20 Torr.

Multiple negative ion species have been observed in CS$_2$ gas mixtures when a small amount of O$_2$ is added \cite{SI2014}.  In Ref.~\cite{SI2014}, at least three additional negative ion species (minority peaks) were reported, all with higher mobilities than CS$_2$, and peak amplitudes that grow, relative to the main CS$_2^-$ peak, with the O$_2$ fraction.  The amplitude of the largest of these three peaks is approximately a factor 2$\times$ larger than the next highest, and this ratio is maintained independent of the O$_2$ fraction or drift field, up to $E = 580$ V$\cdot$cm$^{-1}$ \cite{SI2014, drift2014}.  The only variable that affects the relative amplitudes appears to be the drift distance; increasing this lowers the amplitude of the middle peak.  To date, the physical mechanism behind the minority peaks in the CS$_2$/O$_2$ mixture is unknown.

For a number of reasons, the secondary peak seen in our 20 Torr CS$_2$ data is unlikely to be one of the minority peaks due to O$_2$ contamination: we see only one peak whereas three should clearly be visible; our secondary peak's amplitude increases by an order of magnitude with $E$ from 343 V$\cdot$cm$^{-1}$ to 686 V$\cdot$cm$^{-1}$, but no significant variation in the minority peak amplitudes was observed in Ref.~\cite{SI2014} over the range $E\sim270 - 580$ V$\cdot$cm$^{-1}$; the secondary peak is an order of magnitude smaller in our 40 Torr CS$_2$ data, which was acquired in similar conditions to the 20 Torr data.

We also note that although our acrylic TPC was a source of water vapor from out-gassing, as discussed in Section~\ref{sec:watervapor}, the permeability coefficient of water vapor in acrylic is over three orders of magnitude larger than for O$_2$.  Thus, the level of O$_2$ was probably too low to affect our data at the level seen in Figure~\ref{fig:CS2avgwaveform20Torr}, which, given all the other inconsistencies of this hypothesis, indicates a different origin for the small peak. 

A more likely hypothesis is that the peak is due to S$^-$ or CS$^-$, which are known products of the auto-dissociation of CS$_2^{-*}$ \cite{Krishnakumar, Rangwala}; i.e., in the same manner by which SF$_5^-$ is produced via Equation~\ref{eq:autodissociation}.  The cross-sections for both S$^-$ and CS$^-$ production via this mechanism are non-zero at zero electron energy and peak at 0.5 eV and 1.2 eV, respectively. The S$^-$ peak is narrower and larger by a factor $\sim$20 than that for CS$^-$.  This suggests that our secondary peak is due to S$^-$, and also explains its rapid fractional increase with $E$ described above since the S$^-$ production cross-section increases with electron energy in the $0-0.5$ eV range.


\section{Gas gain}
\label{sec:gasgain}

Previous works have shown that gas gains greater than 1000 can be achieved in electronegative gases with proportional wires \cite{CS2Gain}, GEMs \cite{GEMGain}, and bulk Micromegas (Micro Mesh Gaseous Structure) \cite{MicromegasGain}.  In contrast to electron gases where only moderate electric fields of order $100$ V$\cdot$cm$^{-1}$Torr$^{-1}$ are needed to accelerate electrons to energies close to the ionization potential of the gas, electronegative gases require much higher electric fields to initiate avalanche even though the electron affinity is usually much lower than the ionization potential \cite{Dion2010}.  For CS$_2$, measurements show that the minimum reduced field, $\left( E/p \right)_{\text{min}}$, needed to initiate avalanche is over one order of magnitude larger than for the electron drift gas P10 (10$\%$ methane in argon) \cite{Dion2010}.  A similar study can be done for SF$_6$, but we leave this for the future and instead focus on gas gain in this section.

\begin{figure*}[]
\centering
\subfloat[$^{55}$Fe energy spectrum acquired in 30 Torr SF$_6$ utilizing a 1 mm THGEM]{%
	\includegraphics[width=0.45\textwidth]{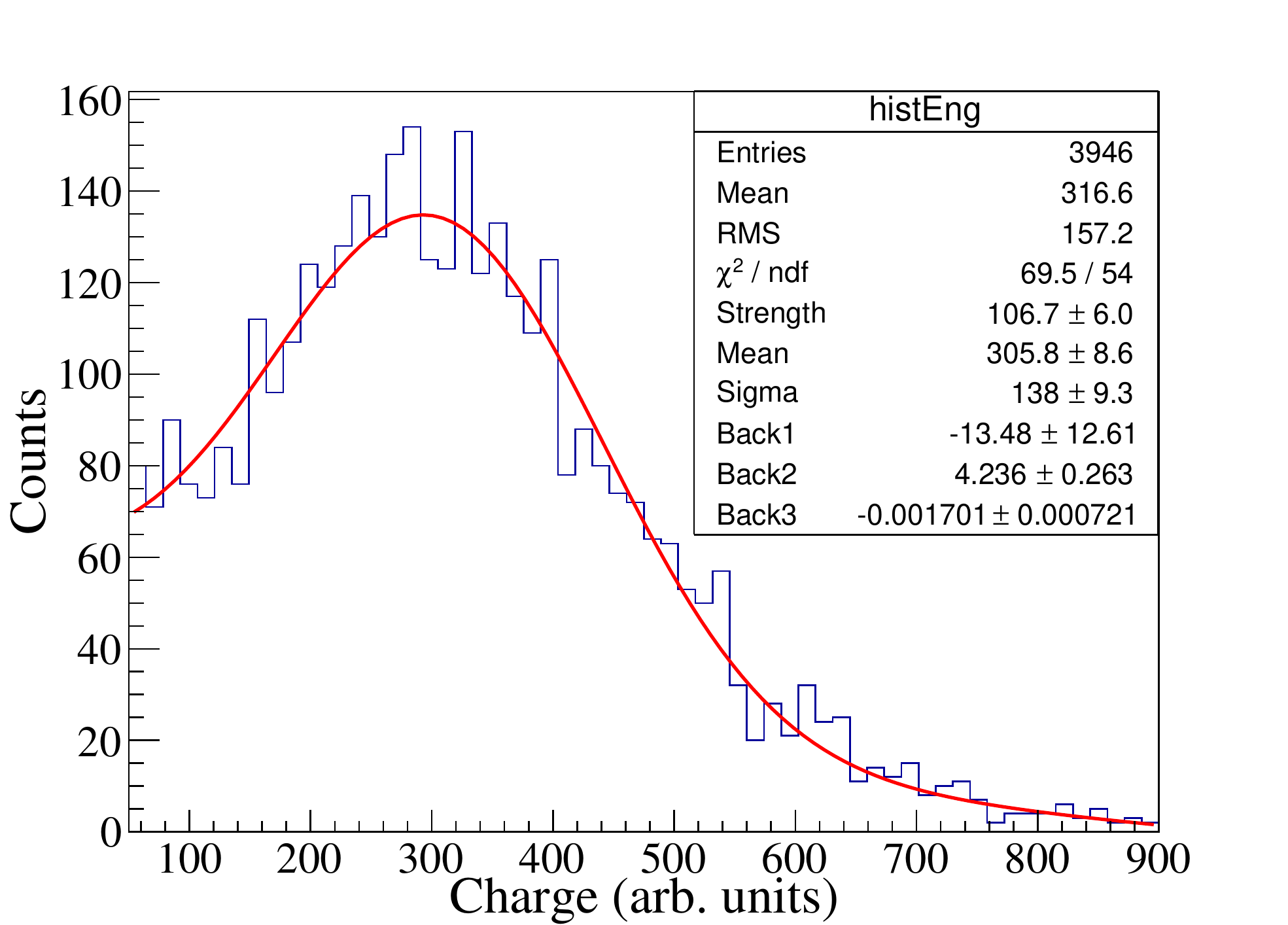}
\label{fig:fe55spect30Torr1mm}
}\hspace*{1em}
\subfloat[$^{55}$Fe energy spectrum after background subtraction]{%
	\includegraphics[width=0.45\textwidth]{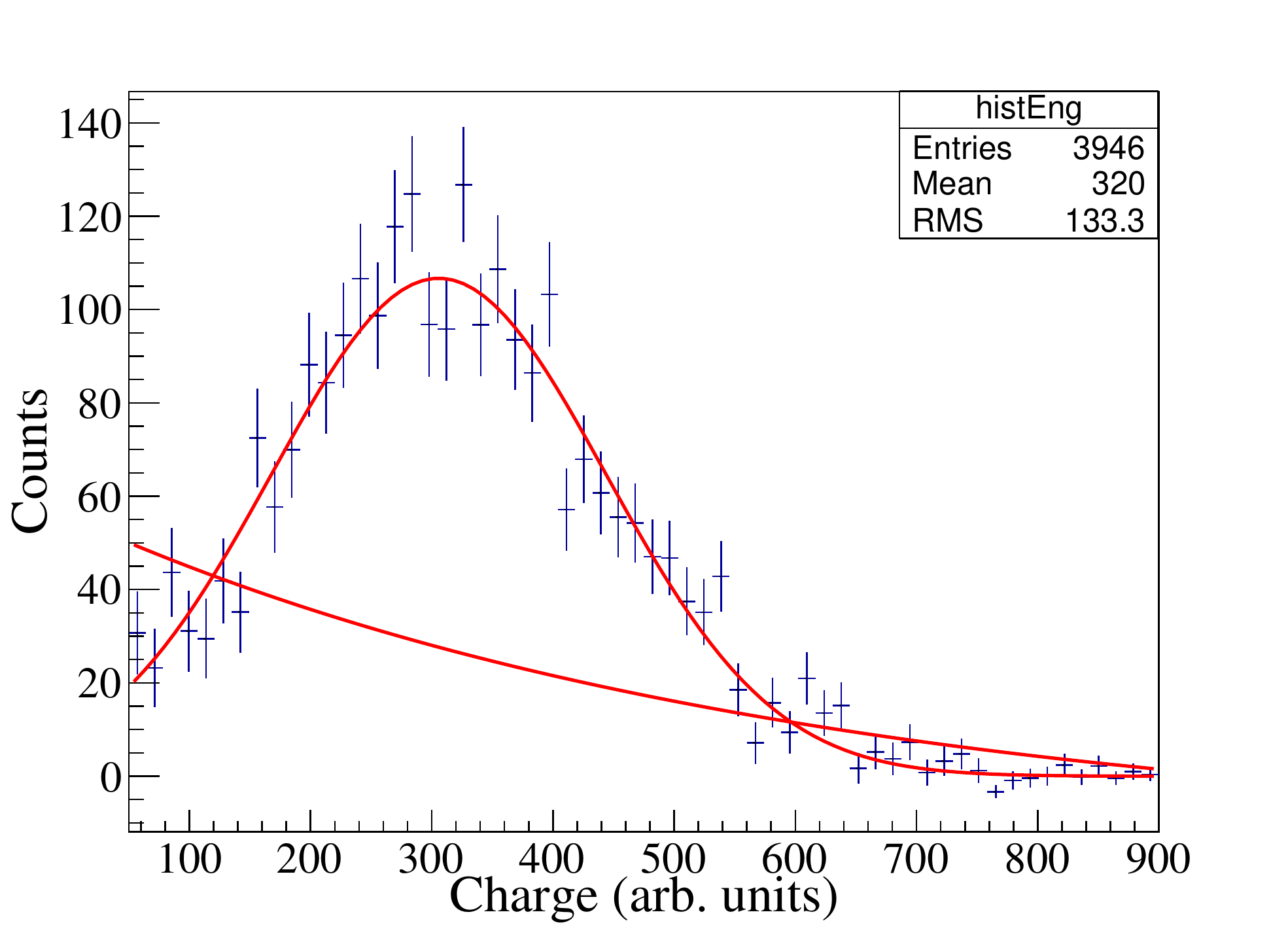}
\label{fig:fe55spect30Torr1mmbacksig}
}
\caption{ }
\label{fig:fe55spectra30Torr1mm}
\end{figure*}

\begin{figure*}[]
\centering
\subfloat[$^{55}$Fe energy spectrum acquired in 30 Torr SF$_6$ utilizing a 0.4 mm THGEM]{%
	\includegraphics[width=0.45\textwidth]{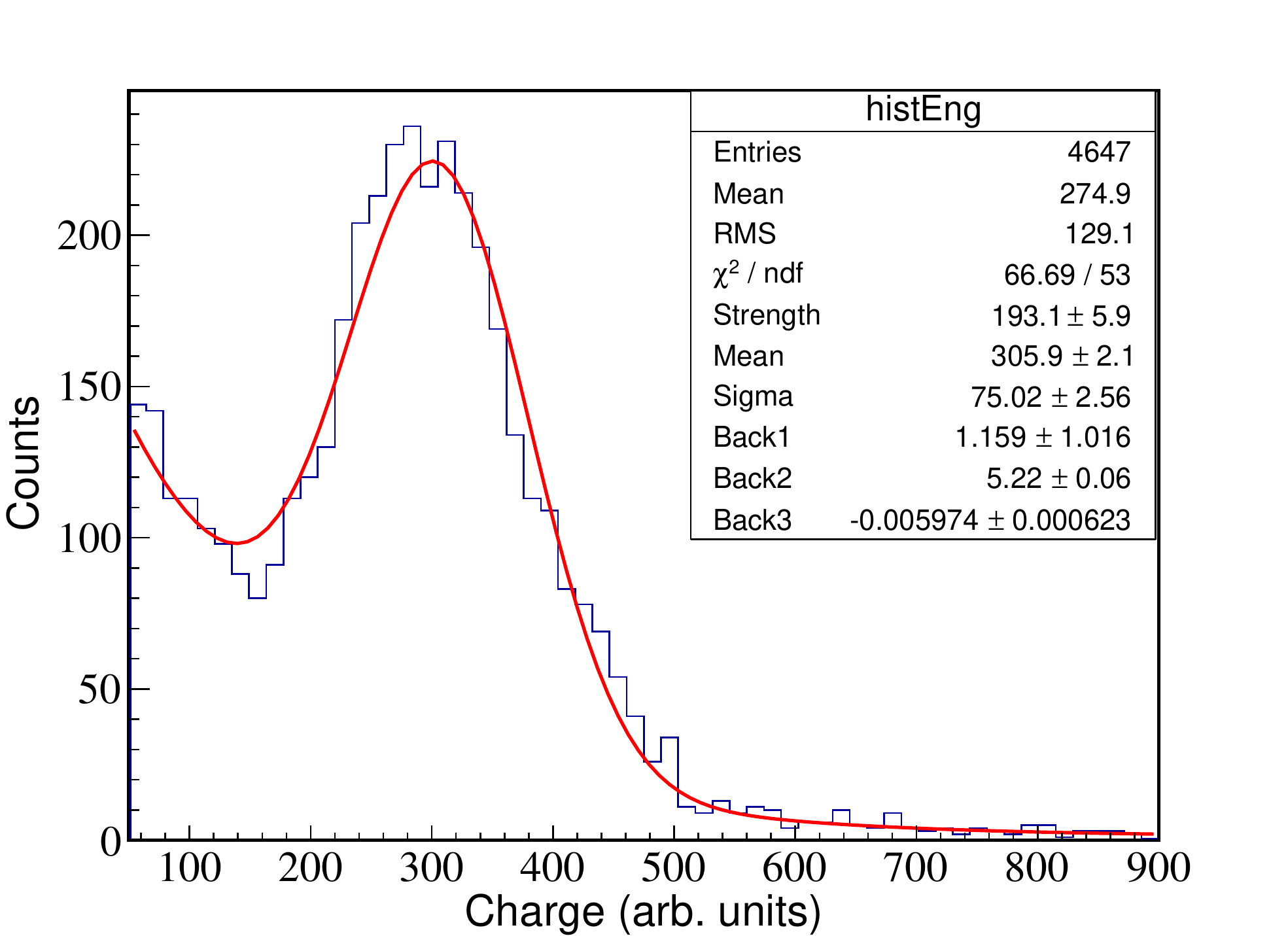}
\label{fig:fe55spect30Torr}
}\hspace*{1em}
\subfloat[$^{55}$Fe energy spectrum after background subtraction]{%
	\includegraphics[width=0.45\textwidth]{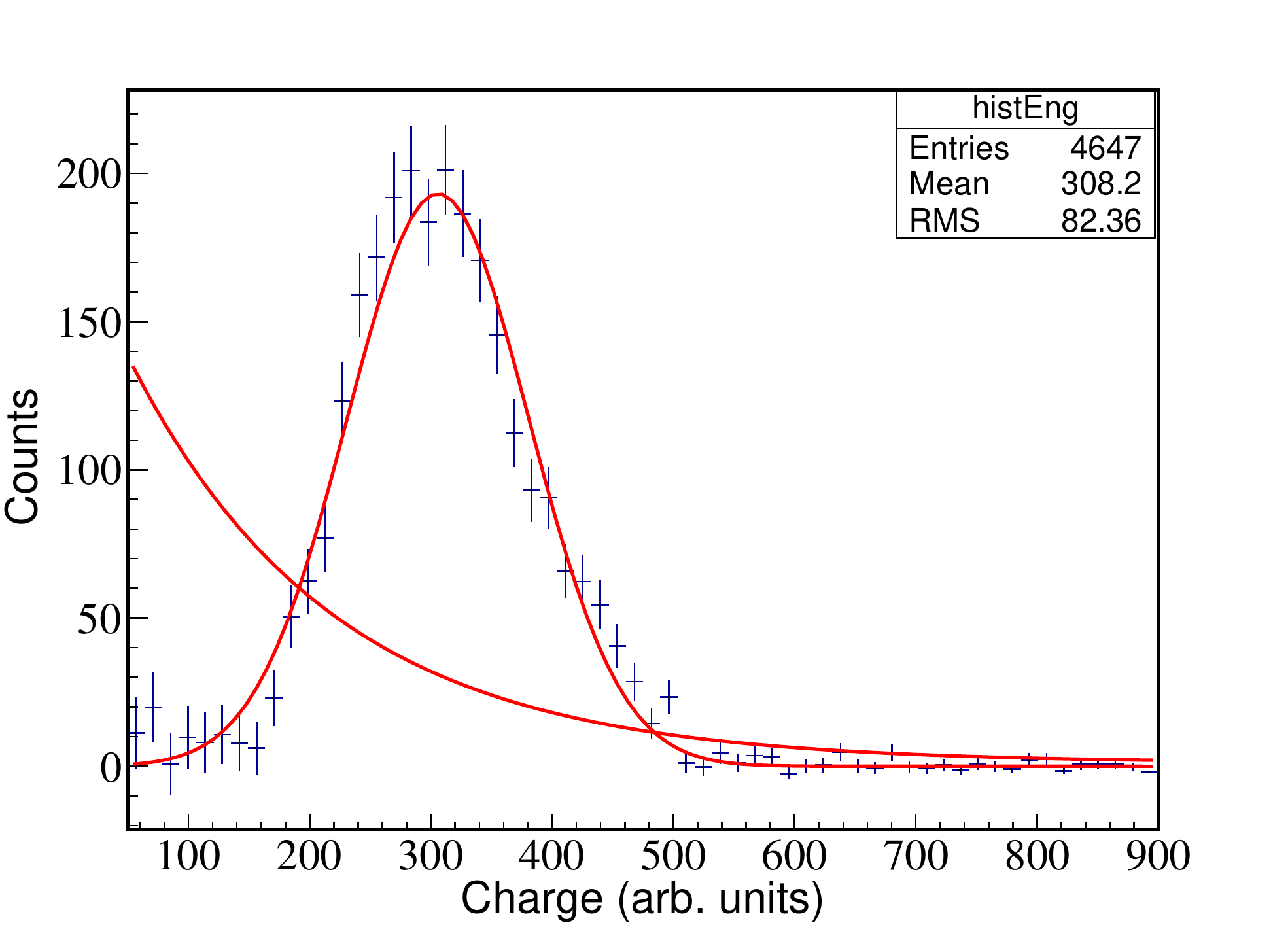}
\label{fig:fe55spect30Torrbacksig}
}
\caption{ }
\label{fig:fe55spectra30Torr}
\end{figure*}

\begin{figure*}[]
\centering
\subfloat[$^{55}$Fe energy spectrum acquired in 40 Torr SF$_6$ utilizing a 0.4 mm THGEM]{%
	\includegraphics[width=0.45\textwidth]{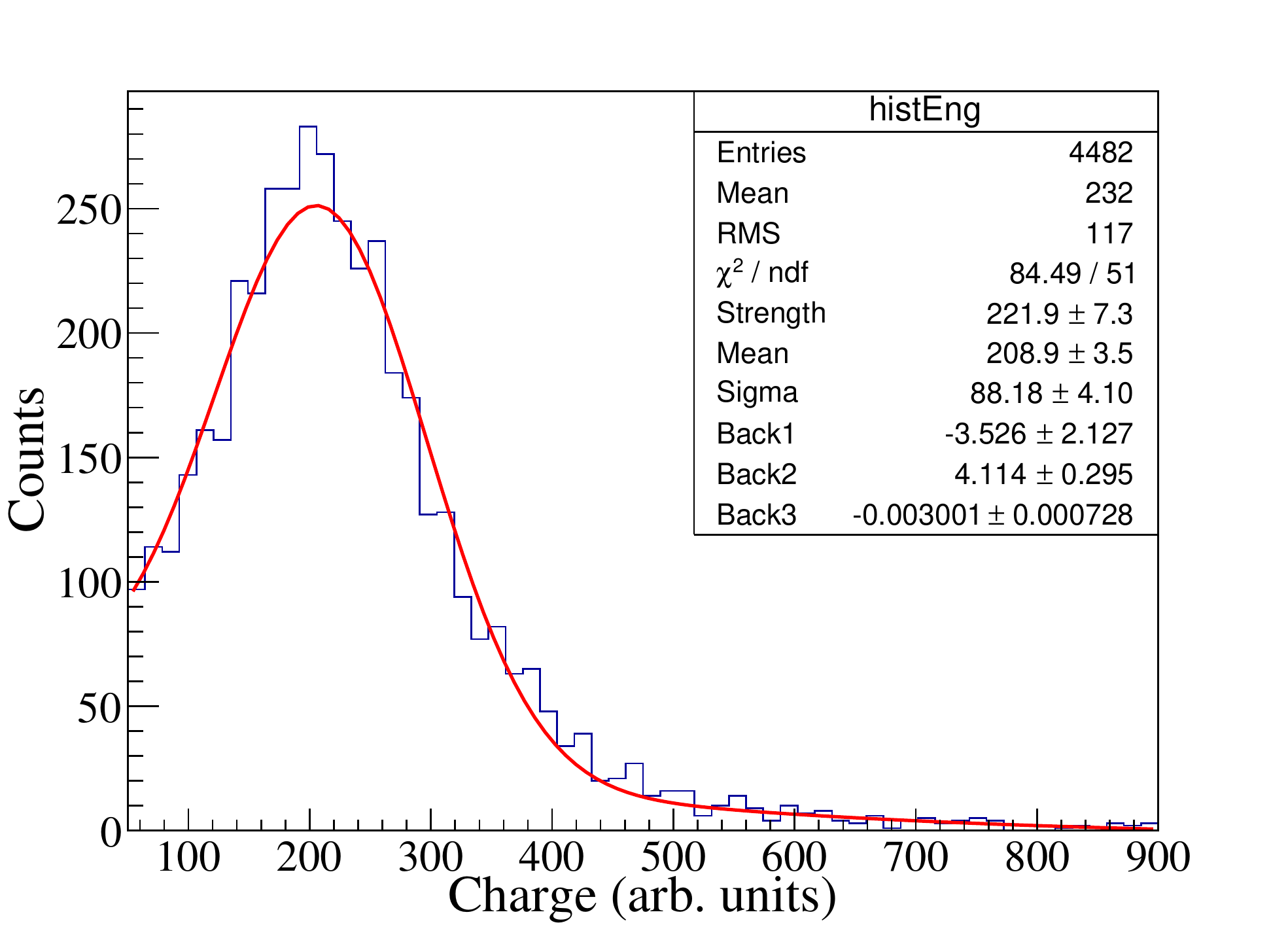}
\label{fig:fe55spect40Torr}
}\hspace*{1em}
\subfloat[$^{55}$Fe energy spectrum after background subtraction.]{%
	\includegraphics[width=0.45\textwidth]{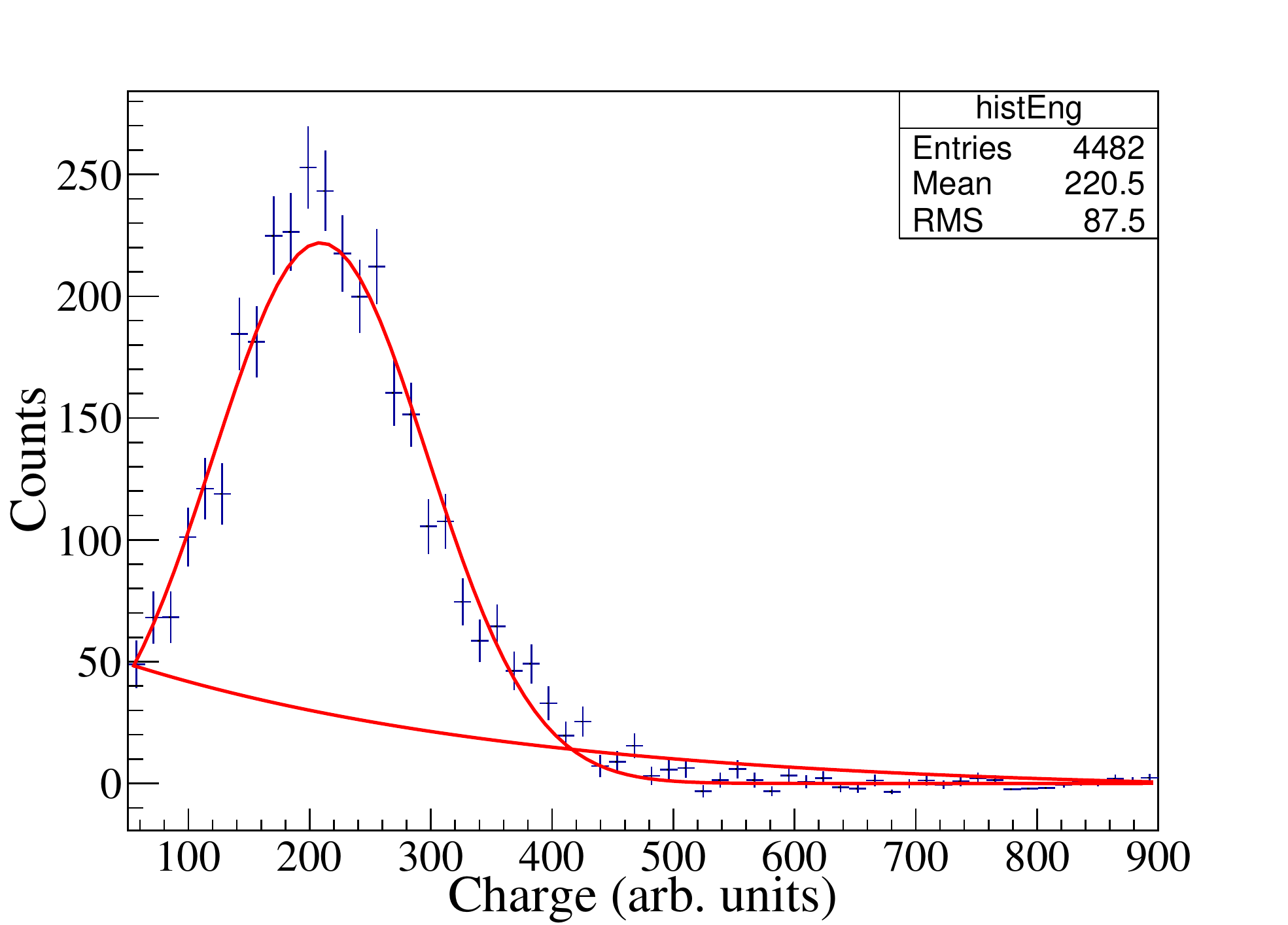}
\label{fig:fe55spect40Torrbacksig}
}
\caption{ }
\label{fig:fe55spectra40Torr}
\end{figure*}

Two THGEMs of thickness, 0.4 mm and 1 mm were used to achieve gas gain in SF$_{6}$.  Other than the thicknesses, the pitch and other THGEM parameters were the same as those described in Section~\ref{sec:detector}.  The gas gain was measured using 5.9 keV X-rays from an $^{55}$Fe source.  The number of electrons produced from the X-ray conversion are estimated using the W-factor, defined as the mean energy required to create a single electron-ion pair.  For SF$_{6}$, this value has been measured using $\alpha$ particles \cite{hilal} and a $^{60}$Co $\gamma$ source \cite{lopes}, giving W$_{\alpha} = 35.45$ eV and W$_{\gamma} = 34.0$ eV, respectively.  The slight disagreement is actually consistent with other measurements of W-factors, which find that W$_{\alpha}$ exceeds  W$_{\gamma, \beta}$ for molecular gases \cite{christophorou1971}.  Because we used an X-ray source, we adopt the W-factor from Ref.~\cite{lopes}, so the average number of primary electrons, $N_{p}$, created by an $^{55}$Fe X-ray conversion in SF$_{6}$ is
\begin{equation} \label{eq:Np}
N_{p} = \frac{E_{^{55}\text{Fe}}}{W_{\gamma}} = \frac{5.9 \text{ keV}}{34.0 \text{ eV}} \simeq 173.
\end{equation}
The effective gas gain is then given by,
\begin{equation} \label{eq:G}
G_{\text{eff}} = \frac{N_{\text{tot}}}{N_{p}},
\end{equation}
where $N_{\text{tot}}$ is the total number of charges read out with the preamplifier.  In general, this is less than the total number of charges produced in the avalanche due to inefficient charge collection, hence, the measured gain is an effective and not an absolute value.  In our case, essentially all of the electrons produced in the avalanche were collected, but there was an additional contribution to the pulse from the positive ion induced signal.  This systematic was not removed.  To determine $N_\text{tot}$ from the measured voltage pulse, $V(t)$, the standard calibration procedure of injecting a known charge into the preamplifier was used.  For this we used an ORTEC 448 Research Pulser to inject charge into the 1 pF calibration capacitor inside the ORTEC 142 preamplifier.

\begin{figure}[]
 \centering 
	\includegraphics[width=0.55\textwidth]{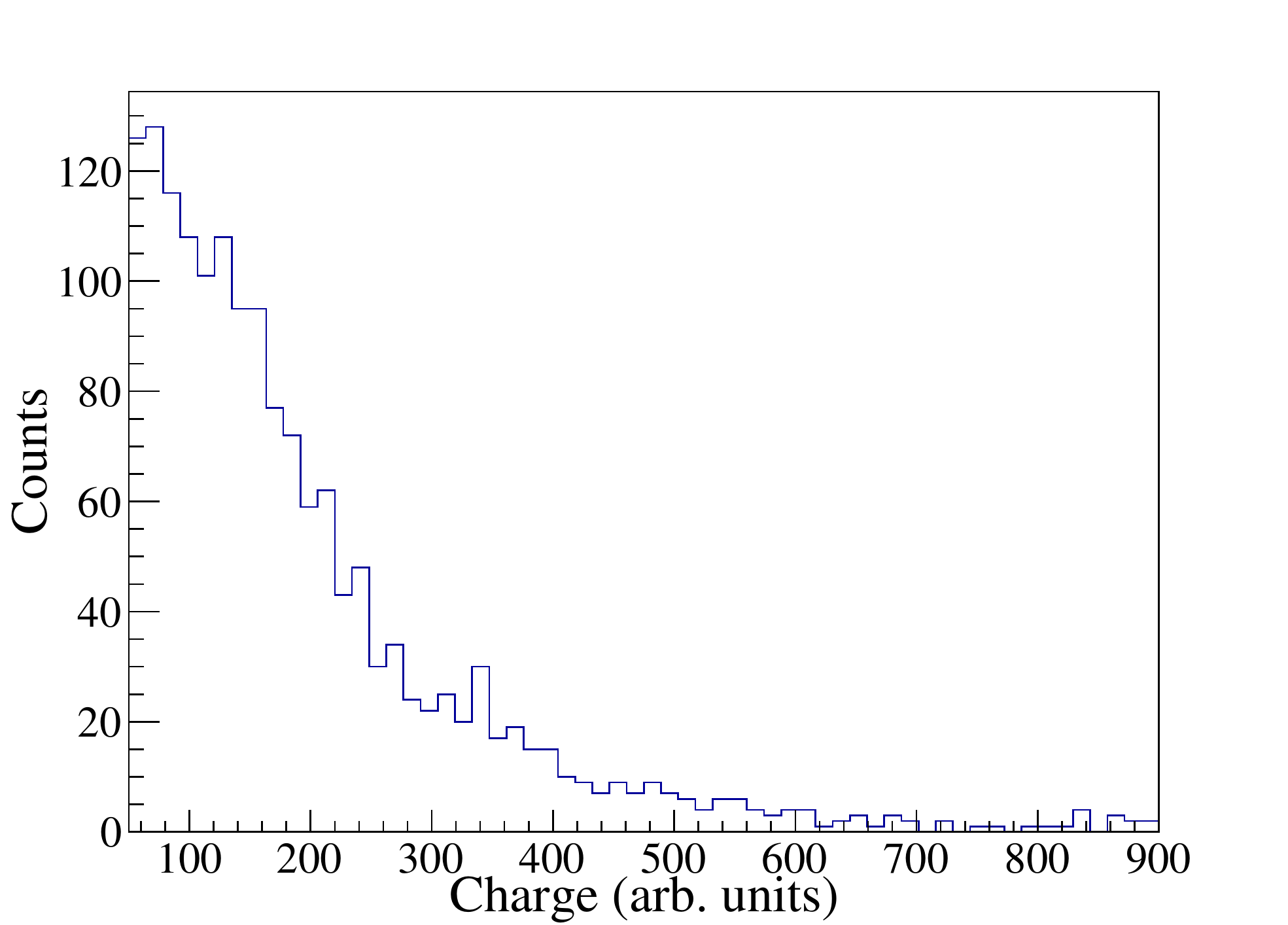}
  \caption[]{An $^{55}$Fe spectrum acquired in 60 Torr SF$_{6}$ using a 0.4 mm THGEMs.  The peak is not observable due to a combination of low gain and poor energy resolution.  However, there is a clear rate difference between $^{55}$Fe source on versus off, indicating there is indeed sufficient gas gain for detecting these low-energy events. }
\label{fig:fe55spectra60Torr}
\end{figure}

For the gain measurement at each pressure, the THGEM voltage was raised until $^{55}$Fe events were visible on the oscilloscope.  The voltage ramp continued until energetic sparks were observed and/or until the rate of micro-sparks and background events approached that of the $^{55}$Fe source.  Figures~\ref{fig:fe55spect30Torr1mm} and \ref{fig:fe55spect30Torr} show the spectra acquired in 30 Torr SF$_{6}$ using a 1 mm and 0.4 mm THGEM, respectively.  The spectrum taken with the 1 mm THGEM (Figure~\ref{fig:fe55spect30Torr1mm}) is much broader, indicating a worse energy resolution, than that taken with the 0.4 mm THGEM spectrum (Figure~\ref{fig:fe55spect30Torr}).  Figures~\ref{fig:fe55spect40Torr} and \ref{fig:fe55spectra60Torr} show the spectra acquired in 40 Torr and 60 Torr, respectively, both using the 0.4 mm THGEM.  For the 60 Torr spectrum the maximum stable gas gain was not sufficient to clearly resolve the peak above background.  However, there was a clear rate difference above the trigger threshold when the $^{55}$Fe source was switched on and off, indicating that the tail of the $^{55}$Fe distribution is contained in the spectrum.  At 20 and 100 Torr a similar rate difference was observed between source on and off using the 0.4 mm THGEM, but spectra were not acquired due to instability. 

None of the spectra are Gaussians, but contain an extra exponential component due to micro-sparks and background events.  To better identify the background and signal components and quantify their shapes, the spectra were fit with a Gaussian signal component, and an exponential plus constant for the background component.  The fitted total spectrum and the separated signal and background components are shown in Figures~\ref{fig:fe55spect30Torr1mm} and \ref{fig:fe55spect30Torr1mmbacksig} for the 30 Torr data acquired with the 1 mm THGEM.  The reduced chi-square ($\chi^2/$ndf) of the fit is 1.29.  Similar fits are shown in Figures ~\ref{fig:fe55spect30Torr} and \ref{fig:fe55spect30Torrbacksig} for the 30 Torr data, and Figures~\ref{fig:fe55spect40Torr} and \ref{fig:fe55spect40Torrbacksig} for the 40 Torr data, both acquired with the 0.4 mm THGEM.  The reduced chi-squares for these fits are 1.26 and 1.66, respectively.

The mean of the Gaussian fit was used to derive the effective gas gain and the width gave the energy resolution, both of which are tabulated in Table~\ref{table:THGEM_Gains} for each experimental configuration.  Other important parameters that describe the operating conditions for the different gain measurements are also listed there to aid in interpreting our results.  Of these, the reduced field inside the THGEM holes, $E_{h}/p$, will be most useful in understanding the differences in the energy resolution and gas gains shown in Table~\ref{table:THGEM_Gains}.  The electric field, $E_{h}$, in the THGEM was approximated by $\Delta$$V/d$, where $\Delta V$ is the voltage across the THGEM and $d$ is its thickness. 

The spectra shown in Figures~\ref{fig:fe55spectra30Torr1mm} - \ref{fig:fe55spectra60Torr}, with the corresponding gas gains and energy resolutions summarized in Table~\ref{table:THGEM_Gains}, can be understood with some knowledge of the physical processes governing the avalanche process in negative ion gases.  These involve stripping the electron from the negative ion, which initiates the avalanche, and the recapture of electrons by SF$_{6}$ in the avalanche, both of which can negatively impact gas gain and energy resolution.  The stripping will occur at some depth, $z$, inside the THGEM hole that is determined by the electron detachment mean-free-path, $\lambda_\text{detach}$, a function of the reduced field.  A large $\lambda_\text{detach}$, relative to the THGEM thickness, $d$, will lead to a larger average depth, $z$, where the avalanche begins, resulting in lower gas gains, larger gain fluctuations, and worse energy resolution.

In addition, the avalanche process in negative ion gases will suffer from a competition with recapture on the neutral molecule or its fragments produced in the THGEM holes (e.g., by auto-dissociation).  Although the cross-sections for attachment in SF$_{6}$ fall with electron energy, the higher electron energies in the THGEM will favor auto-dissociation to SF$_{5}^-$, SF$_{4}^-$, SF$_{3}^-$ and F$^-$\footnote{The electron affinities of SF$_{5}$, SF$_{4}$, SF$_{3}$, and F are $2.7-3.7$ eV \cite{Christodoulides}, 1.50 eV \cite{MillerSF4}, 1.84 eV \cite{MillerSF4}, and 3.4012 eV \cite{Blondel}, respectively.} over collisional stabilization to SF$_{6}^-$, and these fragments all have higher electron affinities than SF$_{6}^-$ (1.06 eV) \cite{christophorou2001}.  Regardless of the details, if recapture occurs the avalanche is halted momentarily until the electron can be stripped again, which further suppresses the gain and worsens energy resolution.  As the cross-sections for attachment, dissociation, and ionization of SF$_{6}$ and its fragments depend on the electron energy, the distinctive spectral shapes, energy resolutions, and gas gain must originate from the dependence on the reduced field in the THGEM.

With this overview, we can attempt to understand the spectra shown in Figures~\ref{fig:fe55spectra30Torr1mm} - \ref{fig:fe55spectra60Torr} (also refer to Table~\ref{table:THGEM_Gains}).  A comparison of the 30 Torr spectra taken with the 0.4 mm and 1 mm THGEMs shows a factor $\sim$2 worse energy resolution in the 1 mm THGEM.  This difference is clearly due to the 2$\times$ lower reduced field, $E_{h}/p$, in the 1 mm THGEM, which, as discussed above, will lead to a larger $\lambda_\text{detach}$ and higher probability of recapture, both of which will lead to the large gain fluctuations that result in poor energy resolution.  If the 1 mm THGEM could have sustained a larger $\Delta V$, leading to a higher $E_{h}/p$ in the holes, a potentially much larger gas gain and better energy resolution could have resulted.

Next, we look at the differences between the 30 and 40 Torr spectra, both taken in the 0.4 mm THGEM.  The energy resolution in 40 Torr is almost 2$\times$ worse, nearly as poor as for the 30 Torr data taken in the 1 mm THGEM.  Here again, it is due to the lower reduced field in the 40 Torr case, $E_{h}/p = 550$ kV$\cdot$cm$^{-1}$, relative to that for the 30 Torr case, $E_{h}/p = 683$ kV$\cdot$cm$^{-1}$.  The fact that the $E_{h}/p$ lies closer to the 30 Torr, 0.4 mm case then to the 40 Torr, 1 mm case, indicates that either $\lambda_\text{detach}$ or the attachment probability depend strongly on energy.  Which of these variables dominates in the effects we see here is not known at this time.  We note, however, that although the reduced fields differ, the electric fields are comparable for the two cases, $E_\text{h} \sim 20$ kV$\cdot$cm$^{-1}$, which supports our claim that the relevant processes are governed by $E_{h}/p$. 

The low gas gains at the higher $60 - 100$ Torr pressures were also due to low $E_{h}/p$, which we were unable to sustain at the levels achieved at low pressures.  In the 60 Torr 0.4 mm THGEM data, we could only reach $E_{h}/p$ = 425 kV$\cdot$cm$^{-1}$, which was insufficient to raise all $^{55}$Fe events above the trigger threshold.  This caused the peak to fall below the lower edge of Figure~\ref{fig:fe55spectra60Torr} and resulted in the broadening of the peak, which is also apparent from the figure.  To achieve gain at higher pressures, multiple THGEMs should work and other MPGD devices, such as thin GEMs and Micromegas, should be attempted as well.  The latter two could also achieve much higher reduced fields, albeit over a shorter avalanche region, which could help with improving the energy resolution.  These are interesting questions for future studies.

\begin{table}
\caption[THGEM gas gains]{THGEM parameters and results }
\label{table:THGEM_Gains}
\centering
\begin{tabular}{p{1.2cm}p{1.2cm}p{1.2cm}p{2.2cm}p{3.4cm}p{1.1cm}p{1.5cm}}
\toprule
$d$ (mm) & $p$ (Torr) &  $\Delta V$ (V) & $E_{h}$ (kV$\cdot$cm$^{-1}$) &  $E_{h}/p$ (V$\cdot$cm$^{-1}$Torr$^{-1}$)  & $G_{\text{eff}}$ & $\sigma/\mathcal{E}$ (\%) \\
\midrule
0.4 	&	30 	&	820		&	20.50  	& 683 	& 3000 	& 25  	\\
1.0 	&	30	&	1005	&	10.05 	& 335 	& 3000 	& 45		\\
0.4 	&	40	&	880		&	22.00		& 550 	& 2000 	& 42 		\\
0.4  	& 60 	& 1020 	& 25.50  	& 425 	& - 		& -  		\\ 
\bottomrule
\end{tabular}
\end{table}


\section{Event fiducialization}
\label{sec:fiducial}

\subsection{$^{252}$Cf data}
\label{sec:Cf252data}

We showed in Figure~\ref{fig:avgwaveforms} of Section~\ref{sec:waveforms} that at high drift fields, the waveform of the charge arriving at the anode consists mainly of the two SF$_5^-$ and SF$_6^-$ peaks.  Having two or more species of charge carriers with differing mobilities is critical for event fiducialization in gas-based TPCs employed in dark matter and other rare event searches.  The ability to fiducialize in these experiments allows for the identification and removal of the most pernicious backgrounds, which originate at or near to the inner surfaces of the detector.  While identifying the event location in the readout plane (X,Y) of a TPC is straightforward, locating the event along its drift direction (Z) is challenging.  Unlike in accelerator-based experiments, the time of interaction (T$_0$) in a gas-based TPC used for rare searches is not available, so Z-fiducialization had proven difficult.  The recent discovery of minority charge carriers in CS$_2$ + O$_2$ mixtures \cite{SI2014}, has changed this by allowing the differences in their mobility to be used to derive the Z coordinate of the event (e.g., see Equation~\ref{eq:Z}).  This has transformed the DRIFT dark matter experiment \cite{drift2014}, which, until this discovery, had operated for close to a decade with backgrounds from radon progeny recoils at the TPC cathode that severely impacted the dark matter search \cite{DRIFT2007, DRIFT2014a, DRIFT2014b, DRIFT2015a, DRIFT2015b}.
 
The differences in the SF$_5^-$ and SF$_6^-$ mobilities in pure SF$_6$ are used to measure the $Z$ coordinate of the event through the relation:
\begin{equation} \label{eq:Z}
Z = \frac{v_{s}\cdot v_{p}}{v_{s} - v_{p}}\Delta T,
\end{equation}
where $v_{p}$ and $v_{s}$ are the drift speeds of the negative ions in the primary (SF$_6^-$) and secondary (SF$_5^-$) peaks, respectively, and $\Delta T$ is the time separation of the peaks.  Note that the anode (THGEM) is at $Z$ = 0, and the cathode at $Z$ = 58.3 cm.

\begin{figure*}[]
\centering
\subfloat[$\Delta T$ laser calibration pulses]{
	\includegraphics[width=0.45\textwidth]{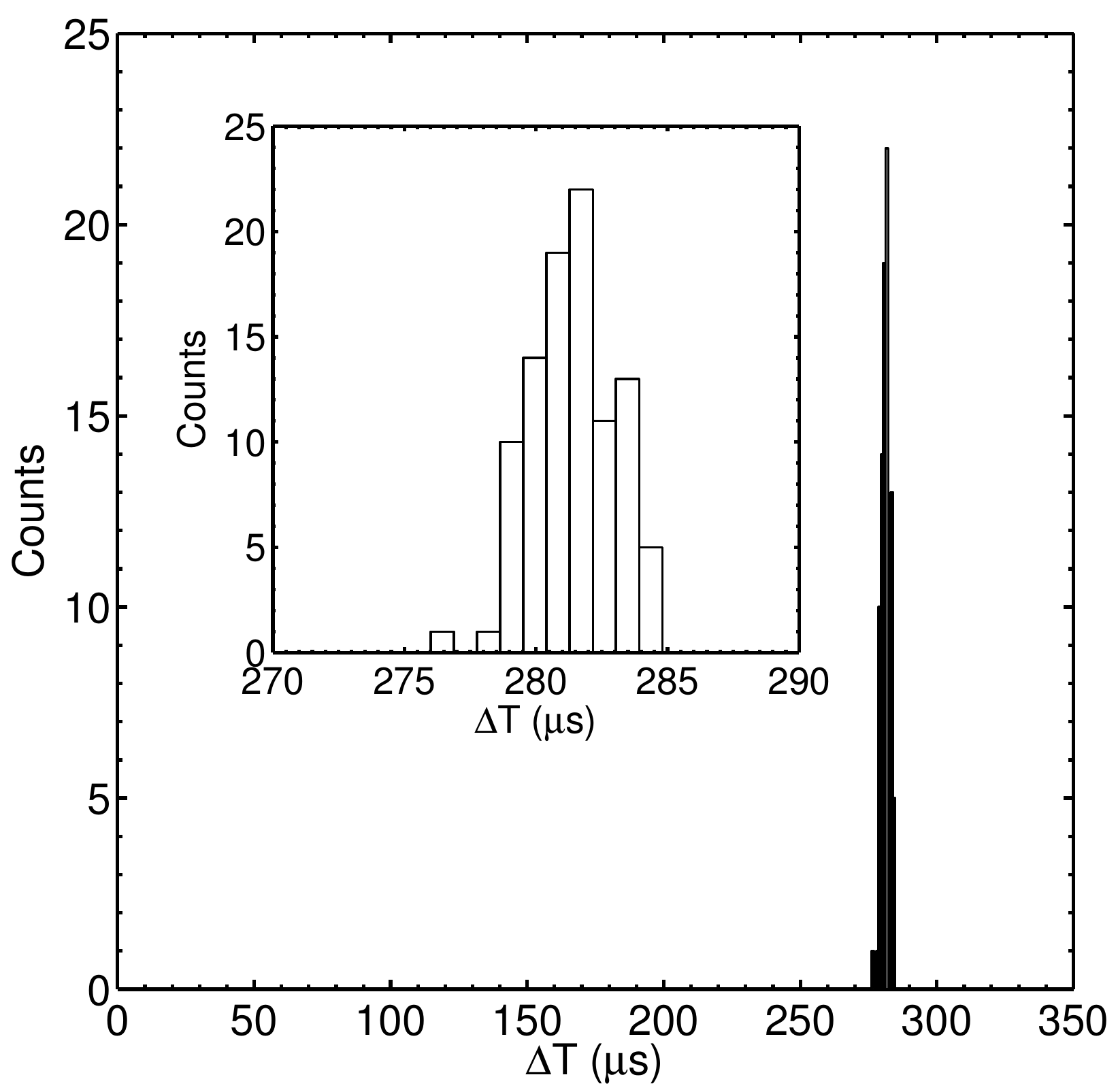}
\label{fig:LaserT}
}\hspace*{1em}
\subfloat[$\Delta T$/Z $^{252}$Cf events]{
	\includegraphics[width=0.45\textwidth]{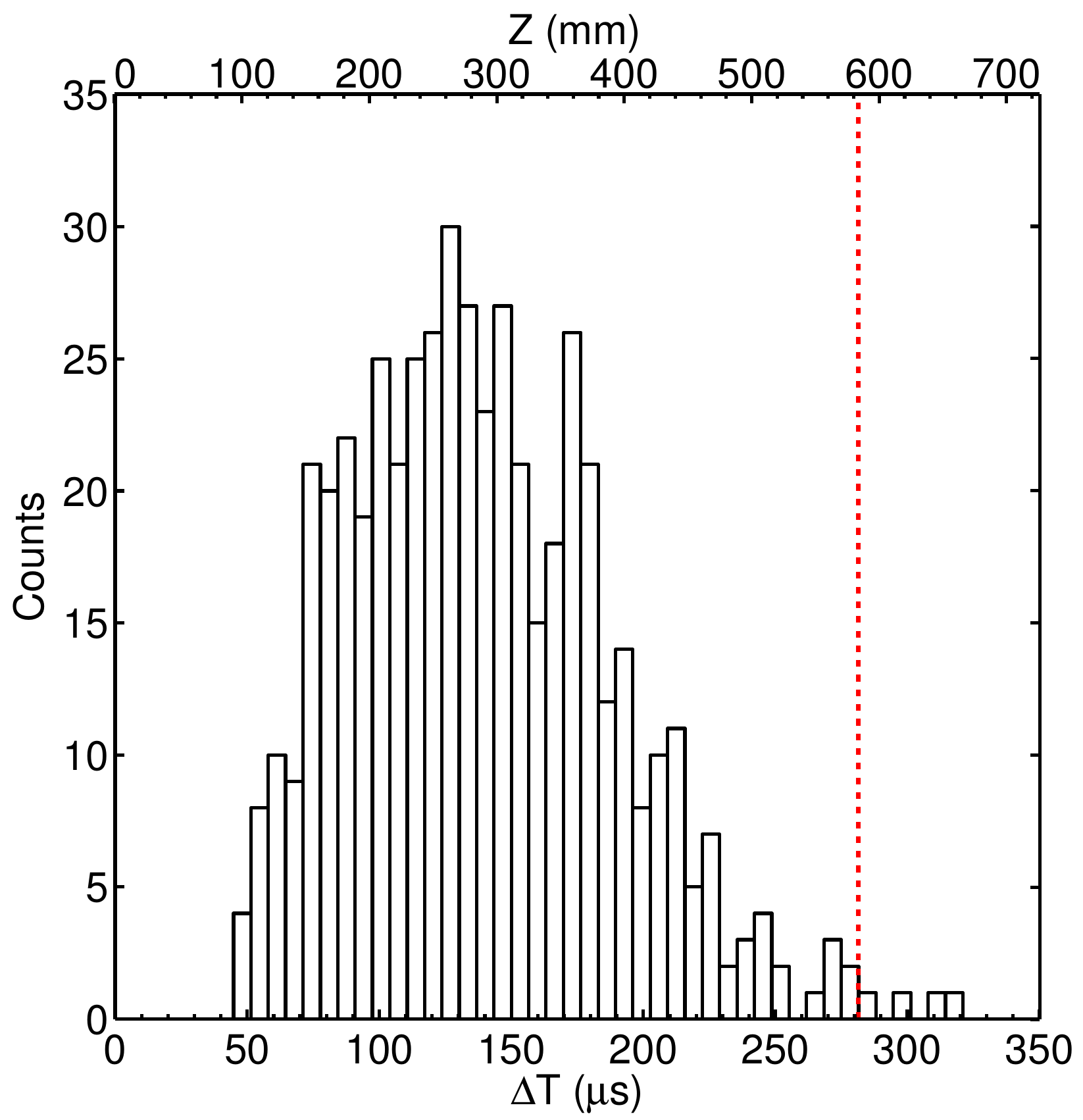}
\label{fig:Cf252T}
}
\caption{(a) Distribution of the time difference between secondary, SF$_5^-$, and primary, SF$_6^-$ peaks ($\Delta T$) for the laser calibration pulses obtained in 30 Torr SF$_6$ and $E= 1029$ V$\cdot$cm$^{-1}$.  (b) The same distribution for events that passed analysis cuts from the $^{252}$Cf data shows a broad distribution of Z locations.  The dotted vertical line shows the position of the cathode at Z = 583.5 mm.  The events with Z locations greater than the cathode location are those with misidentified peaks.  There are no events below 100 mm because the two peaks are not separable for drift distances less than this.}
\label{fig:Zmm}
\end{figure*}

\begin{figure*}[]
\centering
\subfloat[$^{252}$Cf event]{
	\includegraphics[width=0.32\textwidth]{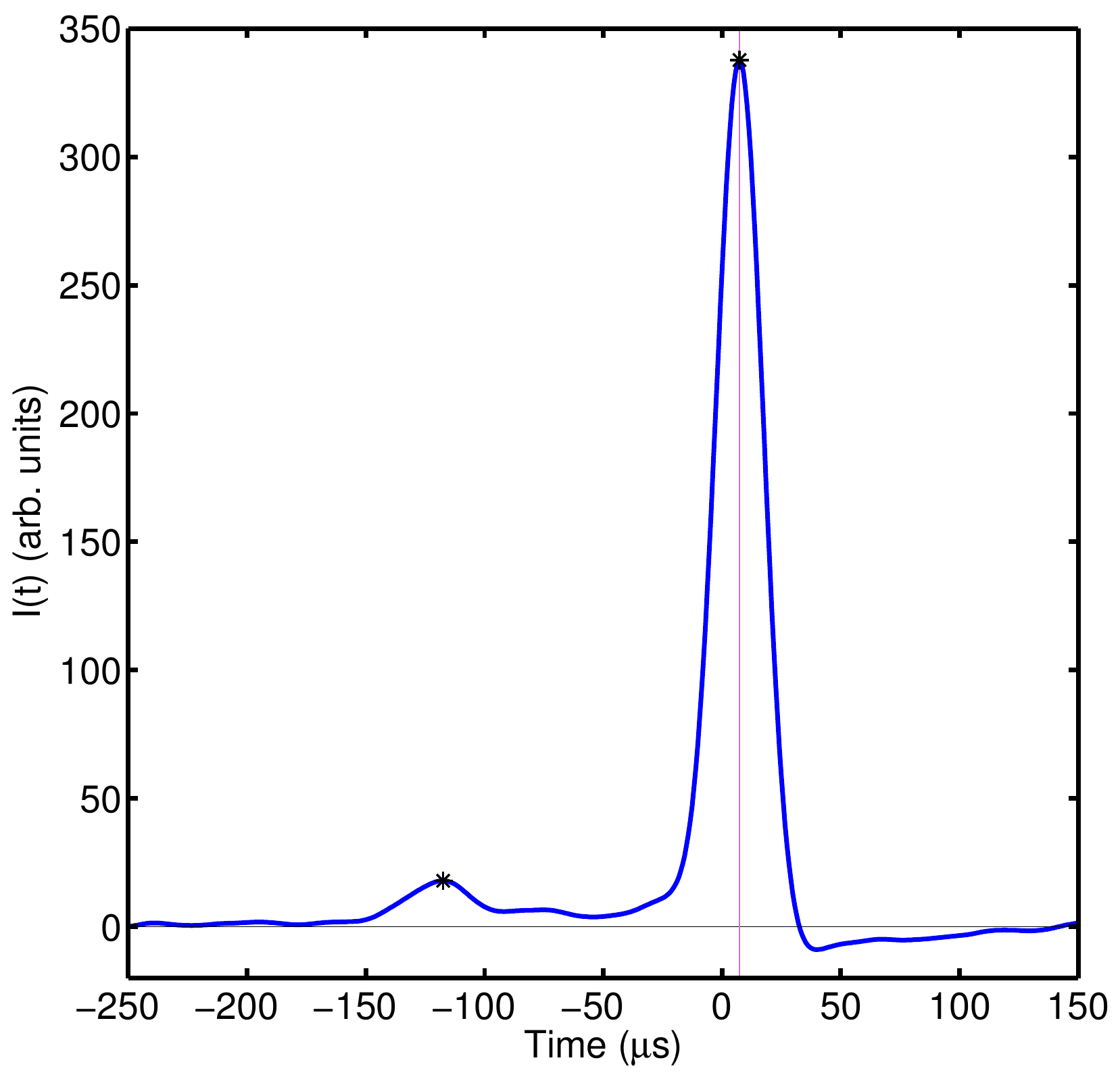}
\label{fig:252Cfevent-1002}
}
\subfloat[$^{252}$Cf event]{
	\includegraphics[width=0.31\textwidth]{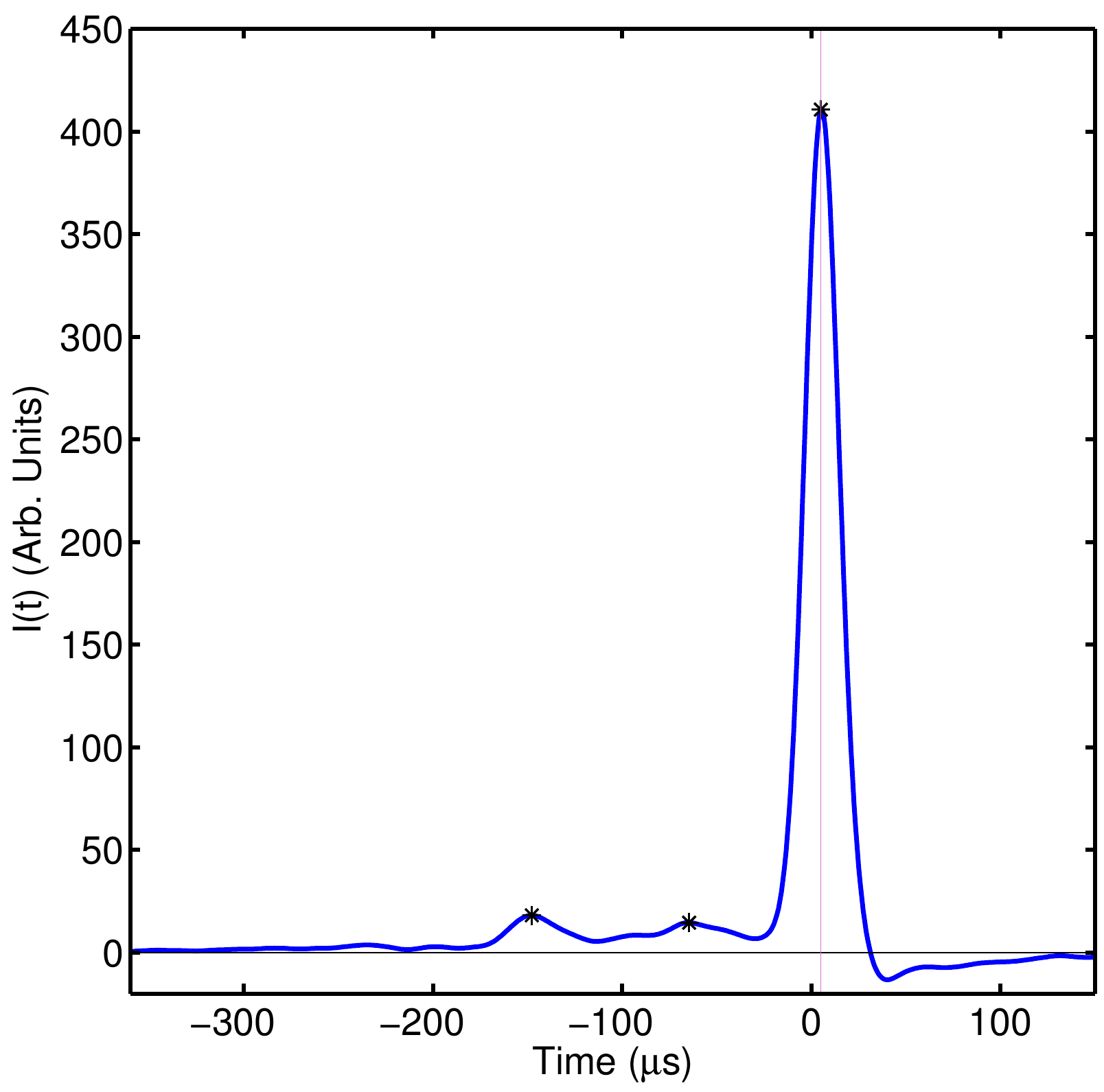}
\label{fig:252Cfevent-1153}
}
\subfloat[$^{252}$Cf event]{
	\includegraphics[width=0.31\textwidth]{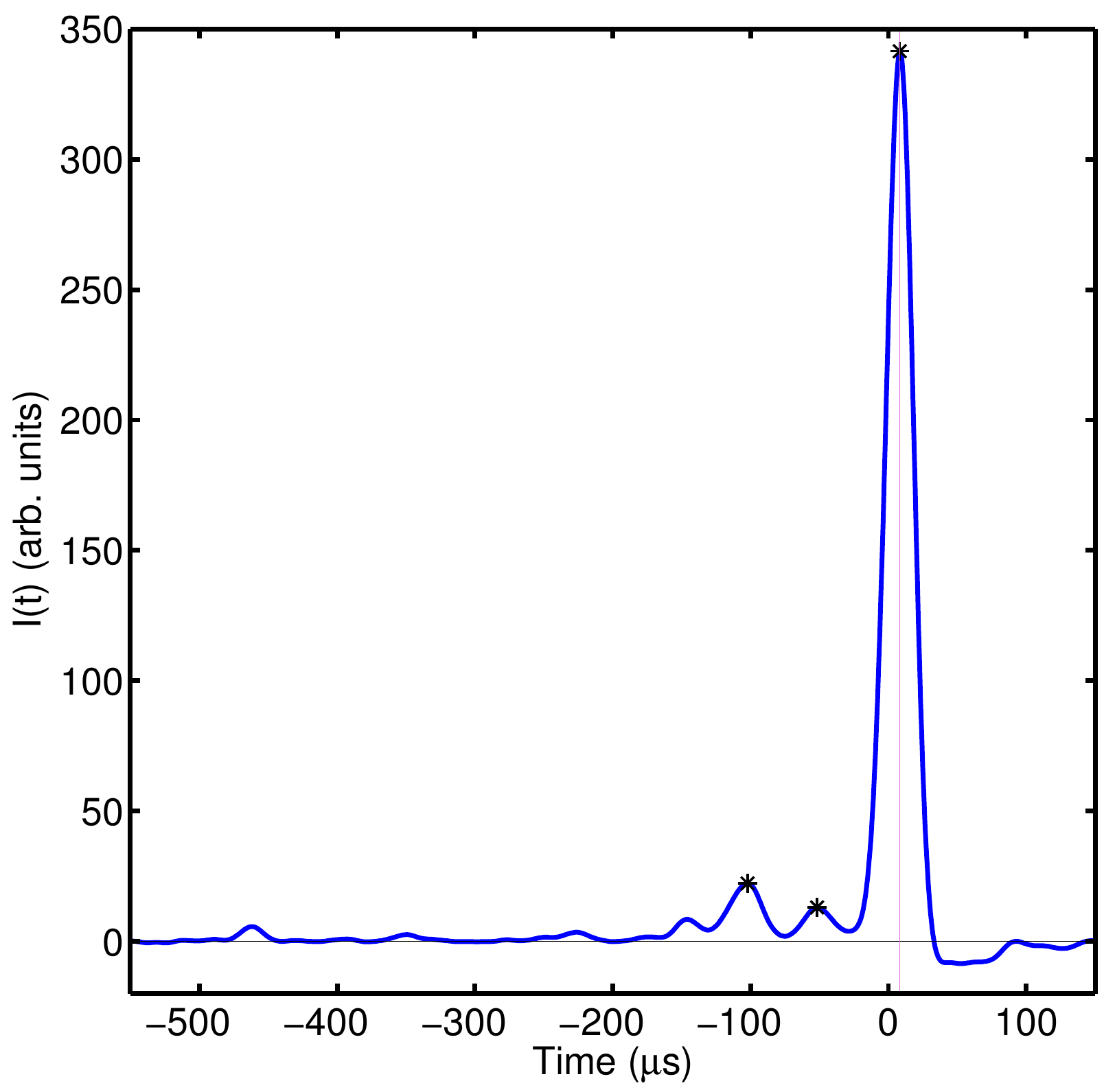}
\label{fig:252Cfevent-1243}
}
\caption{(a) An event from the $^{252}$Cf run in 30 Torr SF$_6$ and $E= 1029$ V$\cdot$cm$^{-1}$ showing two distinct peaks.  The black markers identify the locations of peaks detected by peak finding algorithm.  The magenta vertical line passing through the black marker passes through the location of the primary, SF$_6^-$, peak.  (b) An event from the same data run with three detected peaks.  (c) An event with as many as five peaks; three are detected by the peak finding algorithm.}
\label{fig:Cf252events}
\end{figure*}

To test how well one can determine the location of events in SF$_6$ using this method, we used a $^{252}$Cf source to generate ionization events at different locations in the detection volume.  The $^{252}$Cf source was placed near the outside surface of the vessel and about 20 cm from the cathode.  The detector was operated at 30 Torr with $E = 1029$ V$\cdot$cm$^{-1}$ where the highest gas gains were achieved (Section~\ref{sec:gasgain}).  This was important for identifying the small SF$_{5}^{-}$ peak in low energy recoils, which produce less ionization than the nitrogen laser illuminating the cathode.  Preceding the $^{252}$Cf run, an energy calibration was done with an internally mounted $^{55}$Fe source.  In addition, to calibrate $\Delta T$ we pulsed the laser onto the cathode to generate ionization from a known, fixed $Z$ location.  This provided further confirmation that fiducialization works in SF$_6$ and was also used to quantify the $Z$ resolution.

The SF$_{5}^{-}$ and SF$_{6}^{-}$ peaks were found through an automated process using a derivative based peak finding algorithm.  Although the algorithm performs efficiently for a large data-set, the derivative based approach tends to give false peak detections for noisy data.  To reduce the chance of false peak detections affecting the accuracy of $Z$, we only accepted events that have two and only two identified peaks, one corresponding to SF$_5^-$ and the other to SF$_6^-$.  This greatly reduced the efficiency of our analysis, but our aim here was only to demonstrate event fiducialization in SF$_6$, with work on increasing the efficiency left for future work.  In addition, only events with energy $>$ 60 keVee were accepted so that the SF$_5^-$ peaks were more easily identified, and also to better aid discrimination against electronic recoils due to the gamma-rays from the $^{252}$Cf source.

The distribution of the time difference, $\Delta T$, between the SF$_5^-$ and SF$_6^-$ peaks for the laser calibration data is shown in Figure~\ref{fig:LaserT}.  The distribution has a mean of 281.3 \si{\micro\second} (583.5 mm) and FWHM of about 3.5 \si{\micro\second} (7.3 mm), demonstrating the fundamental accuracy and precision of fiducialization in SF$_6$.  The distribution of the same timing parameter from the $^{252}$Cf run is shown in Figure~\ref{fig:Cf252T}.  The mean and shape of the distribution grossly agree with expectations based on the location of the source, which results in a larger solid angle intersecting the detector volume on the anode side.  Note that there are no events seen with $Z < 10$ cm because the SF$_{5}^{-}$  and SF$_{6}^{-}$ peaks cannot be resolved individually at low $Z$ by our simple peak finding algorithm. 

A sample event from the $^{252}$Cf exposure with a relatively well-defined SF$_5^-$ peak is shown in Figure~\ref{fig:252Cfevent-1002}, demonstrating the feasibility of fiducialization on an event by event basis.  Also note that the relative amplitude of the SF$_5^-$ and SF$_6^-$ peak in this event is 5.3\%, higher than the laser generated ionization data from Section~\ref{sec:relativeratios} at the same reduced field, and for some events in our dataset, the relative amplitude exceeded 8\%.  This can be explained if the energies of electrons produced by nuclear recoils are significantly higher than those produced by laser illumination of the cathode, and higher than the energy gained from the drift field before capture.  In addition to a larger relative SF$_5^-$ peak, higher electron energies also increase the probability of other species being produced, for example SF$_4^-$ and F$^-$ (briefly discussed in Section~\ref{sec:gasgain}).
 
Examples of events potentially demonstrating this effect are shown in Figures~\ref{fig:252Cfevent-1153} and \ref{fig:252Cfevent-1243}.  These events possess more than two peaks, indicating that other negative ion species besides SF$_5^-$ and SF$_6^-$ are being produced due to the initial energies of liberated electrons.  This adds a complication into the analysis to determine the event location, which requires further study.  On the other hand, the sensitivity of the relative strength of the SF$_5^-$, SF$_6^-$, and other peaks to electron energies could open up possibilities beyond fiducialization.  One potential application is for discriminating between electron and nuclear recoils.  If the distribution of electron energies created by an electron recoil is characteristically distinct from the one created by a nuclear recoil, than the relative charge in the peaks could be used to identify the type of particle that created the ionization.

\subsection{Secondary peak enhancement}
\label{sec:peakenhance}

The efficiency with which one can fiducialize in SF$_6$ is largely dictated by how well the relatively small SF$_5^-$ peak is detected.  Here we consider a few possible approaches that might enhance its relative abundance.
  
The first, motivated by the behavior of the minority peaks in CS$_2$ $+$ O$_2$ gas mixtures \cite{SI2014}, is to add a small amount ($<$ 1 Torr to a few Torr) of O$_2$ into SF$_6$. We attempted this and, not surprisingly, saw no significant change in the relative abundance of SF$_5^-$.  Another approach that is motivated by the energy dependence of the SF$_{5}^{-}$ and SF$_{6}^{-}$ production cross-sections, which favors a larger SF$_{5}^{-}$/SF$_{6}^{-}$ ratio at higher electron energies, is to operate at higher reduced drift fields.  One drawback of this, depending on how high one needs to increase $E/p$, is that it could increase diffusion to unacceptable levels (Figure~\ref{fig:SF6sigma}).

The most straightforward approach would be to increase the gas gain, thereby increasing the overall signal-to-noise for detecting the SF$_{5}^{-}$ peak.  As the gains in our measurements with a single THGEM are already at or close to the maximum, two or more THGEMs as well as other MPGD amplification devices should be attempted.  As discussed at length in Section~\ref{sec:gasgain}, amplification devices with the highest possible reduced fields are desired to counteract the physical effects that compete with avalanche production in a negative ion gas.  This is especially important for SF$_{5}^{-}$, which, due to its high electron affinity, would benefit from high $E/p$ to efficiently strip the electron and initiate the avalanche.

There also exists an interesting alternative method to increase the production of SF$_5^{-}$ in SF$_6$.  A study of the production cross-section for SF$_5^{-}$ by auto-dissociation has shown that the first peak at $\sim$ 0.0 eV is very sensitive to temperature \cite{ChenSF6temp}.  Increasing the temperature from 300 K to 880 K increases the relative cross-section for the formation of SF$_5^{-}$ by about two orders of magnitude for electron energies $\sim$ 0.0 eV, while the cross-section hardly varies for electron energies of $\sim$0.38 eV near the second peak.

Since increasing the gas temperature effectively raises the vibrational and rotational excitation energy of the SF$_6$ molecules, this led Ref.~\cite{ChenSF6temp} to consider the possibility of photo-enhancing the SF$_5^{-}$ production via the processes:
\begin{equation}\label{eq:laserSF6}
\mathrm{ n(h\nu)_{\text{laser}} + SF_6 \rightarrow (SF_{6}^{*})_{\text{laser}}  }
\end{equation}  

\begin{equation}\label{eq:laserSF6SF5}
\mathrm{ (SF_{6}^{*})_{\text{laser}} +e^- \rightarrow (SF_{5}^{-})_{\text{laser}} + F. }
\end{equation} 
Using a CO$_2$ laser ($9.4 - 10.6$ \si{\micro\meter}) to vibrationally excite SF$_6$ molecules, they observed an enhancement in SF$_5^{-}$ that was radiation wavelength dependent and different for $^{32}$S and $^{34}$S isotopes.  It should be noted that infrared excitation should not result in the photodetachment of the SF$_{6}^{-}$ anion as measurements have shown that the threshold for this process is at 3.16 eV (392 nm) \cite{Datskos}.  Nevertheless, implementing this idea or increasing the gas temperature for large TPCs presents practical challenges that must be weighed against any benefit.  These are experimental questions that require further investigation.


\section{Conclusion}
\label{sec:conclusion}

For the first time it has been shown that gas gain is achievable in a low pressure gas TPC detector with SF$_6$ as the bulk gas. This has allowed us to make a series of measurements that have demonstrated the negative ion drift behavior of SF$_6$ and have led to the discovery of additional features, which make it an ideal target for spin-dependent directional dark matter experiments. Using THGEMs operating in $20 - 100$ Torr SF$_6$, we were able to detect signals from low energy $^{55}$Fe events with gas gains between $2000 - 3000$.  We found that the energy resolution depends on the reduced field in the amplification region, indicating that electron detachment and/or re-attachment are competing with the avalanche process. 

In addition, we also found a number of interesting features in the signal waveforms.  The first resulted from complex interactions of SF$_6$ with water vapor, which was out-gassing from our acrylic TPC vessel.  Another was the discovery of a secondary peak due to SF$_{5}^{-}$, which drifts faster and arrives earlier than the main negative ion species, SF$_{6}^{-}$.  With these two negative ion species drifting in SF$_6$ we demonstrated the ability to fiducialize events along the drift direction, which is critical for background rejection in the rare searches of interest here. 

Mobility measurements of both SF$_{5}^{-}$ and SF$_{6}^{-}$ were made up to high reduced fields, as were those of the negative ion CS$_{2}^{-}$.  These all agree well with published data in regions of $E/p$ where there is overlap.  However, we did observe an additional peak in the CS$_{2}^{-}$ waveform at high reduced fields where no published data exist.  We speculate that this is due to an additional species, either S$^-$ or CS$^-$, produced by a similar mechanism to that for SF$_{5}^{-}$. 

Finally, the diffusion properties of all three negative ion species, SF$_{5}^{-}$, SF$_{6}^{-}$ and CS$_{2}^{-}$, were also measured to high reduced fields.  These confirmed that all three species drift with thermal diffusion at low $E/p$, as expected, but deviate from it at high reduced fields beyond some critical value of $E/p$.  This deviation from thermal diffusion has important implications for directional low mass WIMP searches where low pressure operation is required.  

The work described here has laid the groundwork for future studies on the use of SF$_6$ in TPCs.  The mechanism of gas amplification in SF$_6$ needs better understanding so that gas gain and energy resolution can be improved.  Other amplification devices should be tested as should operation at higher gas pressures.  The latter, if successful, could open up possibilities for applications that require high pressure operation with gases similar to SF$_6$, such as SeF$_6$ which is of interest for neutrinoless double-beta decay searches.  Investigations on increasing the SF$_{5}^{-}$ fraction are also needed to improve the efficiency for fiducialization.  Finally, measurements of diffusion and other properties important for track reconstruction should extend to the lower pressures of interest for directional low mass WIMP searches.


\section*{ Acknowledgements}
\noindent This material is based upon work supported by the NSF under Grant Nos. 1103420 and 1407773.


\end{document}